\documentclass[paper=a4, fontsize=12pt, abstract=on]{article}
\usepackage{amsmath,amssymb,amsfonts,mathrsfs,hyperref,microtype, amsthm}
\usepackage{geometry}
\usepackage{pstricks}
\usepackage[utf8]{inputenc}
\usepackage[T1]{fontenc}
\usepackage[framed,numbered]{matlab-prettifier}
\usepackage{stackengine}
\usepackage[pdftex]{graphicx}
\usepackage{eufrak}
\usepackage{graphicx,color}
\usepackage[T1]{fontenc}
\usepackage{rotating}
\usepackage{booktabs}
\usepackage{mathtools}
\usepackage{float}
\usepackage{breqn}
\usepackage{graphicx}
\usepackage{caption}
\usepackage{multirow}
\usepackage{authblk}
\usepackage{comment}
\usepackage{siunitx}
\usepackage{upgreek}
\usepackage{mathrsfs}
\usepackage{tikz}
\usetikzlibrary{patterns}
\usepackage{subcaption}
\usepackage{enumitem}
\usepackage[english]{babel}
\usepackage{pgfplots}
\usepgfplotslibrary{groupplots} 
\usepackage{tocbibind}  
\usepackage{appendix}

\newtheorem{theorem}{Theorem}[section]
\newtheorem{remark}{Remark}[section]
\newtheorem{definition}{Definition}[section]
\newtheorem{lemma}{Lemma}[section]
\newtheorem{proposition}{Proposition}[section]
\newtheorem{corollary}{Corollary}[section]

\numberwithin{equation}{section}
\theoremstyle{definition}
\newtheorem{example}{Example}[section]

\def\E{\mathbb{E}}

\def\F{{\rm I\kern-0.16em F}}
\def\B{{\rm I\kern-0.16em B}}
\def\C{{\rm I\kern-0.46em C}}
\def\G{{\rm I\kern-0.50em G}}
\def\D{{\rm I\kern-0.50em D}}

\newcommand{\R}{\mathbb{R}}
\newcommand{\N}{\mathbb{N}}

\DeclareMathAlphabet{\mathscrbf}{OMS}{mdugm}{b}{n}

\newcommand{\boldgamma}[1]{\boldsymbol{ #1}}

\usepackage{color} 
\newcommand{\gr}[1]{{\color{blue} #1}}

\def\ind{\mathrel{\hbox{\rlap{%
				\hbox to 7.5pt{\hrulefill}}\raise6.6pt\hbox{\eka\char'167}}}}
\usepackage{stackengine}

\usepackage{mathtools}

\usepackage{accents}
\newcommand\thickbar[1]{\accentset{\rule{.6em}{1.5pt}}{#1}}
\newcommand\thickbarr[1]{\accentset{\rule{.5em}{1.2pt}}{#1}}
\newcommand\barbelow[1]{\stackunder[1pt]{$#1$}{\rule{.6em}{1.5pt}}}
\newcommand\barbeloww[1]{\stackunder[1pt]{$#1$}{\rule{.5em}{1.1pt}}}

\DeclareMathOperator{\Span}{span}

\newcommand\bigsubsetneqq[1][1.20]{%
	\mathrel{\vcenter{\hbox{\scalebox{#1}{$\subsetneqq$}}}}}

\date{ }

\title{ {\bf Generalized Families of Fractional Stochastic Dominance}}

\author{
	Ehsan Azmoodeh\thanks{University of Liverpool, Department of Mathematical Sciences, L69 7ZL Liverpool, United Kingdom E-mail: \texttt{ehsan.azmoodeh@liverpool.ac.uk}} \, and \,  
	Ozan H\"ur \thanks{Scientific and Technological Research Council of Türkiye (TÜBİTAK), Pk. 16, Mamak Ankara 06261, Türkiye. Email:  \texttt{ozan.hur@tubitak.gov.tr}}
}

\begin{document}
	\maketitle
	\begin{abstract}
		\noindent Introduced by Müller et al. in their seminal paper \cite{muller}, fractional stochastic dominance (SD) offers a nuanced approach to ordering distributions. In this paper, we propose a fundamentally new framework by replacing the fixed parameter $\gamma \in [0,1]$ in fractional SD with a function $\boldsymbol{\upgamma}: \R \to [0,1]$. This yields two novel families, multi-fractional stochastic dominance (MFSD) and functional fractional stochastic dominance (FFSD). They enable the ranking of a broader range of distributions and incorporate a more informative utility class, including those with local non-concavities whose steepness varies depending on the location. Furthermore, our framework introduces the concept of partial greediness, which dynamically captures how behaviour of decision makers adapts to changes in wealth. We also extend this framework to encompass almost stochastic dominance. We provide the mathematical foundations of our generalized framework and study how it offers a novel tool for ordering distributions across various settings.
		\end {abstract}
\noindent \textbf{MSC2020 subject classifications}: Primary 60E15; Secondary 90B50; 91B06; 91B16

\tableofcontents

\section{Introduction}
Stochastic orderings have served as a fundamental and versatile framework for comparing distributions of random variables, capturing a wide range of relationships relevant to diverse fields.  Among these methods, stochastic dominance (SD) is particularly important in decision making under uncertainty, as it allows for ranking alternatives according to the risk preferences of decision makers \cite{Muller_Book, stochastic_orders, denuit}.

At its core, SD involves two key components: the integral condition, which is the ordering rule based on distribution functions, and the utility class, which reflects the shared preferences of decision makers. One of the most well-known concepts in stochastic dominance, \textit{first-order stochastic dominance} (FSD), ranks two alternatives based on the distribution functions $ F $ and $ G $ using the integral condition $ F(x) \geq G(x) $ for all $ x \in \mathbb{R} $, denoted as $ F \preceq_{\textsc{FSD}} G $. This is equivalent to the condition $ \mathbb{E}_F[u] \leq \mathbb{E}_G[u] $ for all $u$ within the utility class, consisting of all increasing functions for which the expectations exist. Similarly, the other widely used stochastic dominance relation, \textit{second-order stochastic dominance} (SSD), states that $ G $ dominates $ F $ in SSD, denoted as $ F \preceq_{\textsc{SSD}} G$, if the integral condition $ 0 \leq \int_{-\infty}^{t} (F(x) - G(x)) \, dx $ holds for all $ t \in \mathbb{R} $. This condition is equivalent to $ \mathbb{E}_F[u] \leq \mathbb{E}_G[u] $, where $ u $ belongs to the class of all increasing and concave utility functions.

A closer look immediately reveals a substantial gap between FSD and SSD. From a mathematical perspective, the integral condition of FSD is highly restrictive, as even a single violation at any point prevents the distributions from being ranked. This results in FSD having low discriminatory power. On the other hand, SSD does not hold when the functions fail to satisfy global concavity, as in the case of those with local non-concave segments. From a decision theoretic perspective, these functions reflect the preferences of decision makers with mixed risk attitudes—for example, individuals who buy insurance while also engaging in gambling (see \cite{FriedmanSavage,Markowitz}).\\

To address the limitations above, a landmark paper \cite{muller} introduced \textit{ fractional stochastic dominance}, which bridges FSD and SSD by incorporating a fixed parameter $\gamma \in [0,1]$. 
\begin{definition}[\textbf{Fractional $(1+\gamma)$\textsc{-SD}}] \label{def:mullers_gamma}
	For fixed $\gamma\in [0,1]$, we say $G$ dominates $F$ in the sense of fractional $(1+\gamma)$\textsc{-SD}, denoted by $F \preceq_{(1+\gamma)\textsc{-SD}} G$, if 
	\begin{equation}\label{eq:muller_IntegralCondition}
		\int_{-\infty}^{t} \left(  F(x) - G(x) \right)_{-}dx  \le \gamma  \int_{-\infty}^{t} \left(F(x) - G(x) \right)_{+} dx, \quad \forall t\in \mathbb{R}.  
	\end{equation}
\end{definition}

The utility counterpart of the relation $F \preceq_{(1+\gamma)\textsc{-SD}} G$ is as follows. For a given $\gamma\in [0,1]$, $F \preceq_{(1+\gamma)\textsc{-SD}} G$ if and only if $ \mathbb{E}_F[u] \leq \mathbb{E}_G[u] $ for every $u\in \mathscr{U}^{*}_{\gamma}$, where 
\begin{equation}
	\mathscr{U}^{*}_{\gamma}  : =\Bigg\{   u:\R \to \R\, \,  :  \,  
	0\leq \gamma \frac{u(x_4) -u(x_3)}{x_4-x_3} \leq \frac{u(x_2)-u(x_1)}{x_2-x_1}, \forall \, x_1 < x_2 \le x_3 <x_4  \Bigg\}.     \label{eq:MullerUtilitySpacesNonSmooth}
\end{equation}	
Specifically, $\gamma=0$ corresponds to FSD, while $\gamma=1$ aligns with SSD. For values of $\gamma \in (0,1)$, the resulting fractional $(1+\gamma)$\textsc{-SD} relaxes the integral condition of FSD, offering more informative criteria for ranking. Moreover, it expands the SSD utility class, consisting of a broader range of functions, including those with local non-concavities. This pivotal contribution has since served as a foundation for further research, inspiring a series of follow-up studies \cite{FractionalSD,fractional_distorted,further_properties,mullerTechnicalNoteRanking2022,fractional_rdu,mullerMultivariateAlmostStochastic2023}, among others.\\

In this paper, motivated by the significant interest and the contributions of the fractional $(1+\gamma)$\textsc{-SD} framework, we extend its capabilities by replacing the fixed coefficient $\gamma$ with a function $\boldsymbol{\upgamma}: \R \to [0,1]$. This introduces two distinct generalized families of fractional SD, depending on whether the extension takes place in the integral condition \eqref{eq:muller_IntegralCondition} or the utility class \eqref{eq:MullerUtilitySpacesNonSmooth}. The rest of this paper is organized as follows. In Section  \ref{sec:genarilzation_integral},  we introduce \textit{multi-fractional stochastic dominance} (MFSD) by extending the integral condition \eqref{eq:muller_IntegralCondition}.  We derive the corresponding utility class and prove an associated “if and only if” theorem. We examine the consequences of this generalization and show that MFSD can effectively order distributions that cannot be ordered via fractional SD. In section \ref{sec:generilization_utilty}, we provide another generalized family of  \textit{functional fractional stochastic dominance} (FFSD) obtained from applying the functional extension on the utility side \eqref{eq:MullerUtilitySpacesNonSmooth}.  We analyse the properties and implications, presenting an "if and only if" theorem that connects the FFSD utility class with the integral condition. Additionally, we mathematically establish the key distinctions between FFSD and MFSD through their respective utility classes, supported by examples based on piecewise continuously differentiable  functions. So far, the first two sections focus more on the mathematical aspects. Section \ref{sec:economical} explores functional extensions within the decision theoretic framework, focusing on the concept of \textit{partial greediness} in both FFSD and MFSD. This novel measure quantifies how risk taking behaviour varies with wealth. This variability enables the development of classification tools for both utility spaces. Additionally, we reveal a key relationship between the function $\boldsymbol{\upgamma}$, partial greediness, and average risk aversion, providing a decision theoretic interpretation for $\boldsymbol{\upgamma}$. In Section \ref{sec:func_ASD}, we follow a similar approach as in previous sections to construct the generalized family of \textit{functional almost stochastic dominance}. This framework extends the concept of \textit{almost stochastic dominance} \cite{almost_SD}. Section \ref{sec:conclusion} concludes the paper. All technical details and proofs are provided in the appendix.

\section{Generalization through integral condition}\label{sec:genarilzation_integral}
In this section, we introduce the multi-fractional stochastic dominance framework, explore its key features, and examine how the generalization via integral condition \eqref{eq:muller_IntegralCondition} extends the fractional $(1+\gamma)$-\textsc{SD} framework.

\subsection{MFSD: Definition}\label{sec:definition}
We begin by defining multi-fractional stochastic dominance.

\begin{definition}[\textbf{Multi-fractional $(1+\boldgamma{\upgamma})$\textsc{-SD}}] \label{def:Extension_gammaSD}
	Let $\boldgamma{\upgamma}:  \R  \to [0,1]$ be an arbitrary \textbf{non-decreasing} function. Let $F$ and $G$ be two arbitrary distribution functions. We say that $G$ dominates $F$ in the sense of the multi-fractional $(1+\boldgamma{\upgamma})$\textsc{-SD}, denote by $F \preceq^{\textsc{mf}}_{(1+\boldgamma{\upgamma})\textsc{-SD}} G$ if
	\begin{equation}\label{eq:Extension_gammaSD_IntegralCondition}
		\int_{-\infty}^{t} \left(  F(x) - G(x) \right)_{-}dx  \le \boldgamma{\upgamma}(t)  \int_{-\infty}^{t} \left(F(x) - G(x) \right)_{+} dx, \quad \forall t\in \mathbb{R}.  
	\end{equation}
\end{definition} 

\begin{remark}
	\label{rem:RightAfterMainDefinition}
	\begin{itemize}
		\item[(i)] Clearly, the multi-fractional $(1+\boldsymbol{\upgamma})$\textsc{-SD} coincides with the fractional $(1+\gamma)$\textsc{-SD} as soon as $\boldgamma{\upgamma}(x) =\gamma \in [0,1]$ for every $x$.  Furthermore, when for some $\gamma \in(0,1)$, function $\boldgamma{\upgamma}(t) \leq \gamma$ for every $t\in \R$, then the multi-fractional $(1+\boldsymbol{\upgamma})$\textsc{-SD} implies the fractional $(1+\gamma)$\textsc{-SD} and, similarly, if function $\boldgamma{\upgamma}(t) \geq \gamma$ for every $t\in \R$, then the fractional $(1+\gamma)$\textsc{-SD} implies the multi-fractional $(1+\boldsymbol{\upgamma})$\textsc{-SD}.
		
		\item[(ii)] The non-decreasing nature of the function $\boldsymbol{\upgamma}$ is not technically required but is crucial for defining a meaningful order relation. To elaborate this further, consider $\boldsymbol{\upgamma}$ as a non-constant, non-increasing function. In this case, any value of $\boldsymbol{\upgamma}$ other than its minimum will not influence the ordering of the distribution functions. This is because smaller values of $\boldsymbol{\upgamma}$ impose stronger dominance rules. Consequently, the smallest value of $\boldsymbol{\upgamma}$ dictates the order relation across the entire support of the distributions.
		
		\item[(iii)] By our assumptions, function $\boldsymbol{\upgamma}$ is non-decreasing and bounded, hence it has at most countably many discontinuity of the first kind and clearly, $\boldsymbol{\upgamma} (t) \le \boldsymbol{\upgamma}(t^+)$ for each $t$.  Therefore, due to continuity of the integral operator, without of loss of generality, one can always assume that function $\boldsymbol{\upgamma}$ is always right continuous. This desirable feature allows us to add one extra layer of randomness to multi-fractional stochastic dominance framework by considering $\boldsymbol{\upgamma}$ as a distribution function (see Example \ref{ex:MFSDUtilityStrictSubsetFSD}).  We expect that such a desirable viewpoint provides us extra strengths that can be useful in certain applications.

	\end{itemize}
\end{remark}

As previously mentioned, function  $\boldsymbol{\upgamma}$ enables the capture of local relations between given distributions, which then allows the notion of multi-fractional $(1+\boldsymbol{\upgamma})$\textsc{-SD} to provide us with a \textbf{local interpolation} phenomenon, as opposed to the global one provided by fractional $(1+\gamma)$\textsc{-SD} (or fractional degree SD introduced in \cite{FractionalSD}). We clarify this in the following example.

\begin{example}[\textbf{Local interpolation}]\label{ex:local_interpolation} 
	Let $\gamma\in(0,1)$ be a constant. Consider the following non-decreasing function
	\begin{equation*}
		\boldgamma{\upgamma}(t)=\begin{cases}0 & \text { if } t \le t_1 \\ \gamma & \text { if } t_1 < t \leq t_2 \\ 1 & \text { if } t>t_2 \end{cases}
	\end{equation*}
	where $t_1<t_2<t_3$ are real numbers. Assume that $F \preceq^{\textsc{mf}}_{ (1+ \boldgamma{\upgamma})\textsc{\textsc{-SD}}} G$.  Then,  $\boldgamma{\upgamma}\left(t_1\right)=0$ entails that $F \preceq_{\textsc{FSD}} G$ on the portion $\left(-\infty, t_1\right]$ of the full supports. Similarly, the condition $\boldgamma{\upgamma}\left(t_2\right)=\gamma$ yields that $F \preceq_{(1+\gamma)\textsc{\textsc{-SD}}} G$ on the interval $\left(-\infty, t_2\right]$ (this directly follows from the non-decreasing property of function $\boldgamma{\upgamma}$). Lastly, $\boldgamma{\upgamma}\left(t\right)=1$ for all $t\in(t_2,\infty)$ implies that $F \preceq_{\textsc{SSD}} G$. Hence, with appropriately chosen non-decreasing function $\boldgamma{\upgamma}$, one can simultaneously order given $F$ and $G$ from first (strongest) to second (weakest) order on different portions (capture with function $\boldgamma{\upgamma}$) of the supports.
\end{example}

Now, we discuss yet another feature of MFSD framework. It is known that the assumption of equal means for the distribution functions ($\mu_F=\mu_G$) renders them indistinguishable in terms of the fractional $(1 + \gamma)$-\textsc{SD}, for any constant $\gamma \in [0, 1)$ \footnote{We exclude the limiting case $\gamma=1$ due to the fact that it corresponds to (less interesting) SSD.}. Nonetheless, as demonstrated in the following example, the multi-fractional $(1+\boldsymbol{\upgamma})$\textsc{-SD} can effectively order distributions with equal means, provided that an appropriate function i.e. non-constant and satisfying $\boldgamma{\upgamma}(t)=1$ for some $t\in \R$, is chosen. 

\begin{example}\label{ex:identical_means}
	Consider $\delta_{\mu} \sim G$ (dirac measure at point $\mu\in \R$), distribution function 
	\begin{equation*}
		F(x)=\begin{cases}
			0 & \mbox{ if } x < \mu-\varepsilon \\
			1/2 & \mbox{ if } \mu-\varepsilon \le x < \mu+\varepsilon\\
			1 & \mbox{ if } \mu+\varepsilon \le x 
		\end{cases}  \, \text{and} \, \, \boldsymbol{\upgamma}(t)=
		\begin{cases}
			0 & \mbox{ if } \mu-\varepsilon\le t \le \mu \\
			\frac{t-\mu}{ \varepsilon} & \mbox{ if } \mu< t <\mu+\varepsilon\\
			1 & \mbox{ if } \mu+\varepsilon\le t
		\end{cases} 
	\end{equation*}
	where $\mu>\varepsilon>0$. The straightforward calculation shows that the integral condition \eqref{eq:Extension_gammaSD_IntegralCondition} holds. Hence,  $F \preceq^{\textsc{mf}}_{(1+\upgamma)\textsc{\textsc{-SD}}} G$ .	However, $F \npreceq_{ (1+ \gamma)\textsc{-SD}} G$ for any constant $\gamma \in [0,1)$ since the inequality \eqref{eq:muller_IntegralCondition} fails at $t=\mu+\varepsilon$. 
	In contrast, we have  $\boldgamma{\upgamma}(\mu+\varepsilon)=1$.
\end{example}

We can readily construct an additional example that extends Example \ref{ex:identical_means} to a scenario beyond identical means. However, to avoid redundancy, we will not pursue this further. The following example, though rather technical, examines the non-trivial relationship between fractional and multi-fractional SD rules. While we provide the example here for context, the explicit constructions are provided in the supplementary material to avoid disrupting the flow.

\begin{example}\label{ex:3}
	The purpose of the example is to show that for any given constant $\gamma \in (0,1)$ and every (non-decreasing) function $\boldgamma{\upgamma}$ satisfying $\boldgamma{\upgamma}(t^0) < \gamma < \boldgamma{\upgamma}(t_0)$ for some $t^0<t_0$, there exists distribution functions $F$ and $G$ so that either one the followings holds: 
	\begin{itemize}
		\item[(i)] $F \preceq^{\textsc{mf}}_{ (1+ \boldgamma{\upgamma})\textsc{\textsc{-SD}}} G$ but $F \npreceq_{(1+\gamma)\textsc{\textsc{-SD}}} G$
		\item[(ii)] $F \npreceq^{\textsc{mf}}_{ (1+ \boldgamma{\upgamma})\textsc{\textsc{-SD}}} G$ but $F \preceq_{(1+\gamma)\textsc{\textsc{-SD}}} G$.\end{itemize}
	
	The full details of the construction are provided in the supplementary material.
	
\end{example}

The conclusion is that in both non-trivial scenarios, one can always construct two distribution functions $F$ and $G$ so that there are not distinguishable by the fractional $(1+\gamma)$\textsc{-SD} but possible to order them with the multi-fractional $(1+\boldsymbol{\upgamma})$\textsc{-SD} and vice versa. 

\subsection{The MFSD universe}\label{sec:MFSDUniverse}

In this section, we formalize all prior observations within a structured framework. For every constant $\gamma \in [0,1]$ and, non-decreasing function $\boldgamma{\upgamma }: \R \to [0,1]$, let introduce


\begin{align*}
	\mathscrbf{E}_{\textsc{f}}(\gamma)  & :   =    	  \Big\{   \left(F,G\right) \in  \mathscrbf{D}(\R)  \times \mathscrbf{D}(\R)      \, : \,  F \preceq_{(1+\upgamma)\textsc{\textsc{-SD}}} G      \Big    \},\\
	\mathscrbf{E}_{\textsc{mf}}(\boldgamma{\upgamma})  & : =    \Big\{      \left(F,G\right) \in  \mathscrbf{D}(\R)  \times \mathscrbf{D}(\R)   \, : \,  F \preceq^{\textsc{mf}}_{ (1+ \boldgamma{\upgamma})\textsc{\textsc{-SD}}} G          \Big \}	
\end{align*}
where $\mathscrbf{D}(\R)$ is the set of all distribution functions supported on $\R$. 
Also,  we denote $\mathscrbf{E}_{\textsc{f}}(1) = \mathscrbf{E}_{\textsc{ssd}}$. Note that, clearly we have the set identity $  \mathscrbf{E}_{\textsc{mf}}(\boldgamma{\upgamma}) = \mathscrbf{E}_{\textsc{f}}(\gamma)$  as soon as $\boldsymbol{\upgamma} \equiv \gamma$ is the constant function, and furthermore, when $ \boldgamma{\upgamma} (t )  \ge \gamma$  $(\boldgamma{\upgamma} (t )  \le \gamma)$ for every $t \in \R$, the set inclusion  $	 \mathscrbf{E}_{\textsc{f}}(\gamma)      \,  \subsetneq  \,   \mathscrbf{E}_{\textsc{mf}}(\boldgamma{\upgamma})$ $(  \mathscrbf{E}_{\textsc{mf}}(\boldgamma{\upgamma})  \,  \subsetneq \,  \mathscrbf{E}_{\textsc{f}}(\gamma))$ holds. 
Lastly, Example \ref{ex:3} shows that for every constant $\gamma \in (0,1)$ and every (non-decreasing) function $\boldgamma{\upgamma}$ satisfying $\boldgamma{\upgamma}(t^0) < \gamma < \boldgamma{\upgamma}(t_0)$ for some $t^0<t_0$, we have 
\begin{equation*}
	\mathscrbf{E}_{\textsc{mf}}(\boldgamma{\upgamma})  \setminus \mathscrbf{E}_{\textsc{f}}(\gamma)   \neq \emptyset \quad  \text {  and }   \quad  \mathscrbf{E}_{\textsc{f}}(\gamma)     \setminus   \mathscrbf{E}_{\textsc{mf}}(\boldgamma{\upgamma}) \neq \emptyset .
\end{equation*}

Before proceeding further, for notational convenience throughout this paper, we use the following notation for the limits of a non-decreasing function $\boldsymbol{\upgamma}$. We define  $$\thickbar{\boldsymbol{\upgamma}}:=\lim_{t\to +\infty}\boldgamma{\upgamma}(t)=\sup_{t\in\mathbb{R}}\boldsymbol{\upgamma}(t) \quad \text{and} \quad  \barbelow{\boldsymbol{\upgamma}} := \lim_{t\to -\infty}\boldgamma{\upgamma}(t)=\inf_{t \in \R} \boldsymbol{\upgamma}(t).$$
The next result aims to summarise the non-trivial relations between the corresponding universes.
\begin{theorem}\label{thm:universal_relation}
	Let $\boldgamma{\upgamma}:\mathbb{R}\rightarrow [0,1]$ be an arbitrary non-decreasing function. Then, the following statements hold:
	
	\begin{itemize}
		\item[(a)] The inclusions $
		\mathscrbf{E}_{\textsc{f}}(\barbelow{\boldsymbol{\upgamma}}) \subseteq \mathscrbf{E}_{\textsc{mf}}(\boldgamma{\upgamma}) \subseteq \mathscrbf{E}_{\textsc{f}}(\thickbar{\boldsymbol{\upgamma}}) 
		$
		holds and are strict as soon as $\boldgamma{\upgamma}$ is a non-constant function.
		
		\item[(b)] The corresponding universes associated to fractional and multi-fractional stochastic dominances coincide with $\mathscrbf{E}_{\textsc{ssd}}$, namely that 
		
		\begin{align}\label{eq:unviersal_inclusion}
			\bigcup_{   \gamma   \in[0,1]     }  \mathscrbf{E}_{\textsc{f}}(\gamma)    \quad  = \quad
			\bigcup_{    \mathclap{\substack{  \boldgamma{\upgamma}:\R \to [0,1]     \\
						\text{non-decreasing}   }  } }  \,  \mathscrbf{E}_{\textsc{mf}}(\boldgamma{\upgamma}) \quad = \quad \mathscrbf{E}_{\textsc{ssd}}.
		\end{align}
		In addition, if the \textbf{non-interesting} case  i.e. $\boldgamma{\upgamma}\equiv\gamma=1$ is excluded from both sides, then the relation \eqref{eq:unviersal_inclusion} reduces to
		\begin{equation}\label{eq:unviersal_inclusion2}
			\bigcup_{   \gamma   \in[0,1)     }  \mathscrbf{E}_{\textsc{f}}(\gamma)    \quad \bigsubsetneqq \quad 
			\bigcup_{    \mathclap{\substack{  \boldgamma{\upgamma}:\R \to [0,1]     \\
						\text{non-decreasing} \\
						\boldgamma{\upgamma} \not\equiv 1  }  } }  \,  \mathscrbf{E}_{\textsc{mf}}(\boldgamma{\upgamma})\quad = \quad \mathscrbf{E}_{\textsc{ssd}}.
		\end{equation}

	\end{itemize}	
\end{theorem}

\begin{remark}
	Part (b) of Theorem \ref{thm:universal_relation} demonstrates a significant feature of multi-fractional stochastic dominance in the sense that by removing the constant function $\boldgamma{\upgamma}\equiv\gamma=1$ (corresponding to SSD case), still the universe of multi-fractional stochastic dominance remains as large as that of SSD whilst this is not the case for fractional stochastic dominance. 
	\end{remark}

	\subsection{Utility class}\label{sec:utilityclass}
	In this section, first we introduce the set of utility functions that generates the multi-fractional SD. Then, we prove the "if and only if" theorem between the introduced utility class and the integral condition \eqref{eq:Extension_gammaSD_IntegralCondition}. As mentioned previously, the utility class corresponding to MFSD is not readily available. Therefore, the definition requires preliminary steps to be established. As our starting point, we present the following simple lemma and the subsequent proposition, inspired by the fractional SD utility space $\mathscr{U}^*_{\upgamma}$ and by  \cite[Theorem 2.4]{muller} (specifically, direction (b)).
	
	\begin{lemma}\label{lem:NiceUtilityBelongClass}
		Let $\boldgamma{\upgamma}: \mathbb{R} \rightarrow [0,1]$ be an arbitrary non-decreasing function.   For each $t \in \R$, we consider the following functions space:
		\begin{equation}
			\begin{aligned}	\label{eq:OneSidedUtilityClass}
				\mathscr{U}^{t}_{\boldgamma{\upgamma}(t)}: =\Bigg\{   u:&\R \to \R\, \, :  \,  u(x)=u(t), \, \forall \, x \ge t \text{ and, }  \\
				&\forall \, x_1 < x_2 \le x_3 <x_4,  \, 0\leq \boldgamma{\upgamma}(t) \frac{u(x_4) -u(x_3)}{x_4-x_3} \leq \frac{u(x_2)-u(x_1)}{x_2-x_1} \Bigg\}.
			\end{aligned}
		\end{equation}
		
		Let $F$ and $G$ be two arbitrary distribution functions. For every $t \in \R$, define the \textbf{continuous} (utility) function $u_t$ via its right-derivative as 
		\begin{equation}\label{eq:Nice_UtilityFunction}
			u_t^{\prime}(x)= 
			\begin{cases}
				\boldgamma{\upgamma}(t) & \text { if } G(x) \le F(x) \text { and } x \leq t \\ 
				1 & \text { if } F(x) < G(x) \text { and } x \leq t \\ 
				0 & \text { if } t<x.
			\end{cases}
		\end{equation}
		Then, $u_t \in \mathscr{U}^{t}_{\boldgamma{\upgamma}(t)}\,$ for every $t \in \R$.
	\end{lemma}

	\begin{proposition}\label{prop:Utility_Implies_IntegralCondition}
		Let $F$ and $G$ be two arbitrary probability distributions having \textbf{finite means} and,  $\boldgamma{\upgamma}: \mathbb{R} \rightarrow [0,1]$ be a non-decreasing function.  Then,  relation  
		\begin{equation}\label{eq:Extension_UtilityCondition}
			\E_F[u_t]  \le \E_G \left[u_t\right],  \quad \forall \, t \in \R
		\end{equation}
		implies that $F \preceq^{\textsc{mf}}_{(1+\boldgamma{\upgamma})\textsc{-SD}} G$.
	\end{proposition}

	The conclusion is that any desirable utility class for MFSD should contain the family of those base-type functions $\{u_t : t \in \R\}$ given in \eqref{eq:Nice_UtilityFunction}.  Observe that these functions enjoy two important characteristics: 
	\begin{itemize}
		\item[(i)] They are constant over the region $(t,+\infty)$ and hence their derivatives vanish (this in particular implies that no contribution in the expected utility through the integration by parts formula).
		\item[(ii)] On the complementary range $(-\infty,t)$ they should manifest the fractional $(1+\gamma)$\textsc{-SD} utilities with $\boldsymbol{\upgamma}(t)$ instead of a fixed parameter $\gamma$.  
	\end{itemize}
	
	Based on the discussion thus far, we take the function spaces $\mathscr{U}^{t}_{\boldgamma{\upgamma}(t)}$ for $t \in \mathbb{R}$ as the building blocks of the MFSD utility class. Before proceeding, let us summarize the some of the key mathematical properties that are crucial for the subsequent discussion.

	\begin{proposition} \label{prop:u_t_properties}
		Let $\boldgamma{\upgamma}:\mathbb{R}\rightarrow [0,1]$ be an arbitrary non-decreasing function and let $t \in \R$.
		\begin{itemize}
			\item[(a)] Every $u\in \mathscr{U}^{t}_{\boldgamma{\upgamma}(t)}$ is bounded from above (in fact $u(x) =u(t)$ for all $x \ge t$) and non-decreasing function hence differentiable almost everywhere.
			
			\item[(b)] Suppose $ \boldgamma{\upgamma}(t) \neq 0 $. Then, every function $ u\in \mathscr{U}^{t}_{\boldgamma{\upgamma}(t)} $ is continuous on $\R$ and  Lipschitz continuous on any interval of the form $ [a,\infty) $. Consequently, absolutely continuous on every compact interval of  $\R$.
			\item[(c)] Let $\boldsymbol{\upgamma}(t)<1$ and  \begin{equation}\label{eq:both_sided_u_t}
				\widetilde{\mathscr{U}}^{t}_{\boldsymbol{\upgamma}(t)}:=\{u\in \mathscr{U}^{t}_{\boldsymbol{\upgamma}(t)}: u_-'(x) \text{ and } u_+'(x)  \text{ exist for all }  x \in\R \}, 
			\end{equation}
			then $\widetilde{\mathscr{U}}^{t}_{\boldsymbol{\upgamma}(t)} \subsetneq \mathscr{U}^{t}_{\boldsymbol{\upgamma}(t)}$.
			\item[(d)] The class $\mathscr{U}_{\boldgamma{\upgamma}(t)}^t$ is a convex cone and closed under topology of pointwise convergence.
			\item[(e)] Let $ s \le t$. Then,  the following set-inclusions are in place: \begin{equation*}
				\mathscr{U}^s_{\boldsymbol{\upgamma}(t)}    \subseteq \mathscr{U}^t_{\boldsymbol{\upgamma}(t)} \text{ and } \mathscr{U}^s_{\boldsymbol{\upgamma}(t)}  \subseteq \mathscr{U}^s_{\boldsymbol{\upgamma}(s)}.\label{eq:inclusion_gamma}
			\end{equation*}
			\item[(f)]	Limit $\displaystyle{\lim_{t\rightarrow -\infty }\mathscr{U}^t_{\boldsymbol{\upgamma}(t)}}$ as $t \to -\infty$ exists, and moreover,  \begin{equation*} 
				\displaystyle{\lim_{t\rightarrow -\infty }\mathscr{U}^t_{\boldsymbol{\upgamma}(t)}}=:\mathscr{U}^{-\infty}_{\barbeloww{\boldsymbol{\upgamma}}} =\{\text { constant functions }\}.
			\end{equation*}
			\item[(g)] Limit $\displaystyle{\lim_{t\rightarrow \infty }\mathscr{U}^t_{\boldsymbol{\upgamma}(t)}}$ as $t \to +\infty$ exists, and moreover, 
			\begin{equation*}\label{eq:Limit+Set} 
				\displaystyle{\lim_{t\rightarrow \infty }\mathscr{U}^t_{\boldsymbol{\upgamma}(t)}}=\mathscr{U}^{+\infty}_{\thickbarr{\boldsymbol{\upgamma}}}
			\end{equation*}
			where $\mathscr{U}^{+\infty}_{\thickbarr{\boldsymbol{\upgamma}}} := \{ u \, :\, 0 \le  \thickbarr{\boldsymbol{\upgamma}} \frac{u(x_4) -u(x_3)}{x_4-x_3} \leq \frac{u(x_2)-u(x_1)}{x_2-x_1}, \, \forall \,  x_1 < x_2 \le x_3 <x_4\}=\mathscr{U}^{*}_{\thickbarr{\boldsymbol{\upgamma}}}.$ 
		\end{itemize}
	\end{proposition}
	
	For each $t \in \R$, the class $\mathscr{U}^{t}_{\boldgamma{\upgamma}(t)}$ contains (utility) functions with limited variety of characteristics. Therefore, by gathering spaces $\mathscr{U}^{t}_{\boldgamma{\upgamma}(t)}$ for all $t\in [-\infty,\infty]$ one can create a larger class which contains a broader range of utility functions with different attributes.  Define
	\begin{equation}\label{eq:UtilityClassFounded}
		\mathscr{U}^{\text{Union}}_{\boldsymbol{\upgamma}}:=\bigcup_{t\in [-\infty,\infty]}\mathscr{U}^t_{\boldsymbol{\upgamma}(t)}.
	\end{equation}
	
	In above, for the technical reasons, when $\boldsymbol{\upgamma}(t)=0$, we restrict the utility space $\mathscr{U}^{t}_{\boldsymbol{\upgamma}(t)}$ to consist of non-decreasing \textbf{continuous} functions. As it follows from Proposition \ref{prop:u_t_properties} (d), although for each $t \in [-\infty,\infty]$, the class $\mathscr{U}^t_{\boldsymbol{\upgamma}(t)}$ is a convex cone, however, $\mathscr{U}^{\text{Union}}_{\boldsymbol{\upgamma}}$ is not necessarily closed under (arbitrary) addition. Given that the sum of utility functions naturally produces a new utility function, and considering the linearity of expected utilities, the class $\mathscr{U}^{\text{Union}}_{\boldsymbol{\upgamma}}$ alone is not a useful candidate for the utility class of the multi-fractional $(1+\boldsymbol{\upgamma})$\textsc{-SD}. However, it is simple to overcome the latter issue by considering all the possible positive (finite) linear combinations. Hence, we finally introduce the following final (utility) class.

	\begin{definition}[\textbf{Utility class of MFSD}] \label{def:MFSDUtilityClass}
		Let $\boldgamma{\upgamma}: \mathbb{R} \rightarrow[0,1]$ be an arbitrary non-decreasing function. A (utility) class for the multi-fractional $(1+\boldsymbol{\upgamma})$\textsc{-SD} is defined as 
		\begin{equation}\label{eq:MFSD-UtilityClass_nondif}
			\mathscr{U}^{\textsc{mf}}_{\boldsymbol{\upgamma}} := \Span_{\mathbb{R}_+}  \Big \{u  \, :\, u\in \mathscr{U}^{\text{Union}}_{\boldsymbol{\upgamma}}    \Big\} = \Big\{    u = \sum_{\text{finite}} \lambda_k u_k, \, \lambda_k \ge 0, \, u_k \in    \mathscr{U}^{\text{Union}}_{\boldsymbol{\upgamma}}   \Big\}.
		\end{equation}	
	\end{definition}

	Now, we are ready to state the main theorem of this section.

	\begin{theorem} \label{thm:Main_Theorem_IFF}
		Let $\boldgamma{\upgamma}: \mathbb{R} \rightarrow[0,1]$ be an arbitrary non-decreasing function. For every two distribution functions $F$ and $G$, the following statements are equivalent:
		\begin{itemize}
			\item[(a)]	
			$F \preceq^{\textsc{mf}}_{(1+\boldgamma{\upgamma})\textsc{\textsc{-SD}}} G$.
			\item[(b)]  $\E_F[u]\leq \E_G[u]$  for every $u \in \mathscr{U}^{\textsc{mf}}_{\boldsymbol{\upgamma}}$.
		\end{itemize}
	\end{theorem}

	Following the main theorem, we provide the subsequent remark to emphasize several important points.
	
	\begin{remark}
		\begin{itemize}
			\item[(i)] Clearly, if the non-decreasing function $\boldgamma{\upgamma}\equiv\gamma \in [0,1]$ be constant, then $\mathscr{U}^{\textsc{mf}}_{\boldsymbol{\upgamma}}= \mathscr{U}^{*}_{\gamma}$. In addition, for a non-decreasing function $\boldsymbol{\upgamma}$, the inclusions 
			\begin{equation*}
				\mathscr{U}^{*}_{\thickbarr{\boldsymbol{\upgamma}}}    \subseteq   \,	\mathscr{U}^{\textsc{mf}}_{\boldsymbol{\upgamma}}  \, \subseteq \mathscr{U}^{*}_{\barbeloww{\boldsymbol{\upgamma}}}
			\end{equation*}
			hold and, both inclusions being strict as soon as $\boldsymbol{\upgamma}$ is a non-constant (see Theorem \ref{thm:universal_relation}, part (a)).  From the utility theory perspective, the second set inclusion means that $\mathscr{U}^{\textsc{mf}}_{\boldsymbol{\upgamma}}$ provides us a more informative class compare to  $\mathscr{U}^{*}_{\barbeloww{\boldsymbol{\upgamma}}}$. However, the first inclusion implies that the utility class $\mathscr{U}^{\textsc{mf}}_{\boldsymbol{\upgamma}}$ represents the preferences of a broader class of decision makers.		
			\item[(ii)] Due to the linear structure of $\mathscr{U}^{\textsc{mf}}_{\boldsymbol{\upgamma}}$ and, for every $t$, observing that $\mathscr{U}^{t}_{\boldsymbol{\upgamma}(t)} \subsetneq  \mathscr{U}^{*}_{\boldsymbol{\upgamma}(t)} $, while multi-fractional  $(1+\boldsymbol{\upgamma})$\textsc{-SD} establishes stochastic dominance over $\mathscr{U}^{\textsc{mf}}_{\boldsymbol{\upgamma}}$ that might possibly there is no dominance (in the sense of fractional SD) in any $\mathscr{U}^*_\gamma$ for all $\gamma< \thickbar{\boldsymbol{\upgamma}}$. 
			\item[(iii)]  For a given parameter $\gamma$, one can choose a function $\boldsymbol{\upgamma}:\R\to [0,\gamma]$ to further expand the set from $\mathscr{U}^*_\gamma$ to $\mathscr{U}^{\textsc{mf}}_{\boldsymbol{\upgamma}}$. This will ensure that individuals in $\mathscr{U}^*_\gamma$ also agree on the ordering of random variables under $(1+\boldsymbol{\upgamma})$\textsc{-SD}, since $\mathscr{U}^*_\gamma \subsetneq\mathscr{U}^{\textsc{mf}}_{\boldsymbol{\upgamma}}$. Similarly, choosing a function $\boldsymbol{\upgamma}:\R\to [\gamma,1]$ will result in a smaller set of decision makers, $\mathscr{U}^{\textsc{mf}}_{\boldsymbol{\upgamma}} \subsetneq \mathscr{U}^*_\gamma$, who unanimously agree on the ordering of random variables in the $(1+\gamma)$\textsc{-SD} sense. Hence, MFSD can be utilised to further interpolate between FSD and $(1+\gamma)$\textsc{-SD} by choosing a function $\boldsymbol{\upgamma}:\R\to [0,\gamma]$, or between $(1+\gamma)$\textsc{-SD} and SSD by choosing a function $\boldsymbol{\upgamma}:\R\to [\gamma,1]$.
		\end{itemize} 
	\end{remark}	
	
	Next, we continue with a few illustrating examples to gain better understanding of the utility class.
	
	\begin{example}\label{prop:StepFunctionGeneralForm}
		Let  $\gamma_1 < \gamma_2 <\cdots <\gamma_n $ be $n$ real numbers belonging to the interval $[0,1]$. Consider function 
		\begin{equation}\label{eq:GammaStepFunction}
			\boldsymbol{\upgamma} (t) = 
			\begin{cases}
				\gamma_1 & \mbox{ if }    t \le t_1\\
				\gamma_2 & \mbox { if } t_1 < t \le  t_2\\
				\vdots\\
				\gamma_{n-1} & \mbox { if } t_{n-2} < t \le t_{n-1}\\
				\gamma_n & \mbox { if }   t > t_{n-1}.\\
			\end{cases}	
		\end{equation}	
		Then, we have 
		\begin{align*}
			\mathscr{U}^{\text{Union}}_{\boldsymbol{\upgamma}} =    \bigg( \bigcup_{k=1}^{n-1} \mathscr{U}^{t_k}_{\boldsymbol{\upgamma}(t_{k})} \bigg)    \bigcup             \mathscr{U}^{+\infty}_{\thickbarr{\boldsymbol{\upgamma}}} =\bigg( \bigcup_{k=1}^{n-1} \mathscr{U}^{t_k}_{\gamma_k} \bigg)  \bigcup  \mathscr{U}^{*}_{\gamma_n}
		\end{align*}
		where function class $\mathscr{U}^{*}_{\gamma_n}$ corresponds to the fractional $(1+\gamma_n)$-\textsc{SD}. 
	\end{example}

	The purpose of the following example is to provide yet another capability of multi-fractional SD through its utility space.
	
	\begin{example}\label{ex:MFSDUtilityStrictSubsetFSD}
		In certain occasions, one may require a given family of utility functions sharing particular features belongs to a single utility space corresponding to a given SD rule. Next, we construct a parametrized family $\mathscr{U}_{\Theta}$ of utility functions such that the only single utility space $\mathscr{U}^{*}_{\gamma}$, containing $\mathscr{U}_{\Theta}$, associated to the fractional $(1+\gamma)$\textsc{-SD} corresponds to first order stochastic dominance, namely that $\gamma=0$ whereas one can capture $\mathscr{U}_\Theta$ with a multi-fractional utility space $\mathscr{U}^{\textsc{mf}}_{\boldsymbol{\upgamma }}$ that is a strict subset of the FSD utility space $\mathscr{U}_{0}$ and hence resulting in a stochastic dominance rule in which a larger set of distribution functions can be ordered. 
		Let $\theta^* >0$ be fixed. Let $\Theta = (\theta^*,\infty)$ and consider a function $\widetilde{\boldsymbol{\upgamma}} : \Theta \to (0,1]$ so that (i) it is non-decreasing, (ii) $\widetilde{\boldsymbol{\upgamma}} (\theta) < \theta $ for every $\theta \in \Theta$, and (iii) $\lim_{\theta \to \theta^*} \widetilde{\boldsymbol{\upgamma}} (\theta) =0$. 	For every $\theta \in \Theta $, we define a utility function $u_\theta$ via its right-derivative as 
		\begin{equation*}
			u'_\theta (x) =
			\begin{cases}
				2 \widetilde{\boldsymbol{\upgamma}} (\theta)& \mbox{ if }   x \le \widetilde{\boldsymbol{\upgamma}} (\theta),\\
				1 + \widetilde{\boldsymbol{\upgamma}} (\theta)& \mbox{ if }  \widetilde{\boldsymbol{\upgamma}} (\theta) < x \le \theta,\\
				\widetilde{\boldsymbol{\upgamma}} (\theta) & \mbox{ if }  x > \theta.
			\end{cases}
		\end{equation*}	 
		Then, it is simple to observe that $u_\theta \in \mathscr{U}^*_{\widetilde{\boldsymbol{\upgamma}}(\theta)}$ (see \eqref{eq:MullerUtilitySpacesNonSmooth} for definition) for each $\theta \in \Theta$. Let  $\mathscr{U}_{\Theta } : =\left \{       u_\theta \, :\, \theta \in \Theta          \right \}$. Now, for a moment assume that for a given constant $\gamma \in [0,1]$ the set inclusion $\mathscr{U}_{\Theta} \subseteq \mathscr{U}^*_\gamma$  holds. Then, property (iii) above dictates that $\gamma =0$ must be.  In other words, the only possibility to capture all the utility functions $u_\theta$ in a single utility space $\mathscr{U}^*_\gamma$ associated to the fractional $(1+\gamma)$\textsc{-SD} corresponds to the first stochastic dominance. On the other hand, define the non-decreasing function 
		\begin{equation*}
			\boldsymbol{\upgamma} (\theta)=
			\begin{cases}
				0 & \mbox { if } \theta \le \theta^*,\\
				\widetilde{\boldsymbol{\upgamma}}(\theta) & \mbox { if }  \theta > \theta^*.
			\end{cases}
		\end{equation*}
		Now, we claim that $\mathscr{U}_{\Theta} \subseteq \mathscr{U}^{\textsc{mf}}_{\boldsymbol{\upgamma}} \subset \mathscr{U}_{0}$ where clearly the second inclusion is strict. For each $\theta \in \Theta$, define utility functions $v_\theta$ and $w_\theta$ via their right-derivatives as $v'_\theta (x) = \widetilde{\boldsymbol{\upgamma}}(\theta)$, and 
		
		\begin{equation*}
			w'_\theta (x) =
			\begin{cases}
				\widetilde{\boldsymbol{\upgamma}}(\theta) & \mbox{ if } x \le \widetilde{\boldsymbol{\upgamma}}(\theta),\\
				1 & \mbox{ if } \widetilde{\boldsymbol{\upgamma}}(\theta) < x \le \theta,\\
				0  & \mbox { if }  x > \theta.
			\end{cases}
		\end{equation*}
		Then, clearly, $v_\theta, w_\theta \in \mathscr{U}^{\textsc{mf}}_{\boldsymbol{\upgamma }}$, and therefore, $u_\theta = v_\theta + w_\theta \in \mathscr{U}^{\textsc{mf}}_{\boldsymbol{\upgamma }}$ for each $\theta$ in which proves the validity of the first inclusion. 
		To finalize this example through visualization, we shall specify $\widetilde{\boldsymbol{\upgamma}}(\theta)$. We set $\theta^*=1$ and choose $\widetilde{\boldsymbol{\upgamma}}(\theta)=1-e^{(1-\theta)}$. It is apparent that it is a non-decreasing function, where $1-e^{(1-\theta)}< \theta$ for every $\theta \in \Theta$, $\lim_{\theta \to \theta^*} 1-e^{(1-\theta)}=0$, and $\lim_{\theta \to 1} 1-e^{(1-\theta)}=1$. Now, consider and exponential random variable $E \sim \text{Exp}(1)$ and let $X = E+1$ with the distribution function given as 
		\begin{equation*}
			F_X(\theta) =	\boldsymbol{\upgamma} (\theta)=
			\begin{cases}
				0 & \mbox { if } \theta < 1,\\
				1-e^{(1-\theta)} & \mbox { if }  \theta \ge 1.
			\end{cases}
		\end{equation*}
		Then, derivative of the utility function $u'_\theta$ becomes
		\begin{equation*}
			u'_\theta (x) =
			\begin{cases}
				2 (1-e^{(1-\theta)})& \mbox{ if }   x \le 1-e^{(1-\theta)},\\
				2-e^{(1-\theta)}& \mbox{ if }  1-e^{(1-\theta)} < x \le \theta,\\
				1-e^{(1-\theta)} & \mbox{ if }  x > \theta.
			\end{cases}
		\end{equation*}	  
		
		\begin{figure}[H]
			\centering
			\tikzset{every picture/.style={line width=0.75pt}} 
			
			\begin{tikzpicture}[scale=0.5]
				\begin{groupplot}[
					group style={
						group size=2 by 2,  
						horizontal sep=2cm,  
						vertical sep=2cm     
					},
					]
					\nextgroupplot[ 
					axis lines = left,
					xlabel = $\theta$,
					ylabel = {$\boldsymbol{\upgamma}(\theta)$},
					xtick={-1,-3,-5,1,3,5,7},
					xticklabels={$-1$,$-3$,$-5$,$\theta^*$,$3$,$5$,$7$}
					]
					\addplot [
					domain=1:9, 
					samples=50, 
					color=red,
					line width=1.3pt
					]
					{1-e^(1-x)};
					
					\addplot [
					domain=-5:1, 
					samples=50, 
					color=red,
					line width=1.4pt
					]
					{0};
					
					\nextgroupplot[ 
					ymin= 0,
					ymax= 2,
					axis lines = left,
					xlabel = {$x$},
					ylabel = {$u'_{1.01}(x)$},
					]
					
					\addplot [
					domain=-5:0.009, 
					samples=50, 
					color=blue,
					line width=1.3pt
					]
					{2*(1-e^(1-1.01))};
					
					\addplot [
					domain=0.0091:1.01, 
					samples=50, 
					color=blue,
					line width=1.3pt
					]
					{2-e^(1-1.01)};
					
					\addplot [
					domain=1.01:9, 
					samples=50, 
					color=blue,
					line width=1.3pt
					]
					{1-e^(1-1.01)};
					
					\nextgroupplot[ 
					ymin= 0,
					ymax= 2,
					axis lines = left,
					xlabel = {$x$},
					ylabel = {$u'_{1.1}(x)$},
					]
					
					\addplot [
					domain=-5:0.09, 
					samples=50, 
					color=blue,
					line width=1.3pt
					]
					{2*(1-e^(1-1.1))};
					
					\addplot [
					domain=0.091:1.1, 
					samples=50, 
					color=blue,
					line width=1.3pt
					]
					{2-e^(1-1.1)};
					
					\addplot [
					domain=1.01:9, 
					samples=50, 
					color=blue,
					line width=1.3pt
					]
					{1-e^(1-1.1)};

					\nextgroupplot[ 
					ymin= 0,
					ymax= 2,
					axis lines = left,
					xlabel = {$x$},
					ylabel = {$u'_{3}(x)$},
					]
					
					\addplot [
					domain=-5:0.86, 
					samples=50, 
					color=blue,
					line width=1.3pt
					]
					{2*(1-e^(1-3))};
					
					\addplot [
					domain=0.861:3, 
					samples=50, 
					color=blue,
					line width=1.3pt
					]
					{2-e^(1-3)};
					
					\addplot [
					domain=3.01:9, 
					samples=50, 
					color=blue,
					line width=1.3pt
					]
					{0.86};
				\end{groupplot}
			\end{tikzpicture}
			\caption{Graphs of $u'_{\theta}$ for values $\theta=1.01$, $\theta=1.1$ and $\theta=3$ (blue) and $\boldsymbol{\upgamma}$ (red)}
		\end{figure}
		
	\end{example}

	To highlight the comprehensiveness of our proposed utility class in modeling diverse risk preferences, it is worth to emphasize the inclusion of both star-shaped utility functions \cite{landsbergerLotteriesInsuranceStarshaped1990, landsbergerTaleTwoTails1990} and fourth-degree polynomial functions \cite{benishayFourthDegreePolynomialUtility1987} within our utility class $\mathscr{U}^{\textsc{mf}}_{\boldsymbol{\upgamma}}$. These specific utility functions are of particular interest due to their ability to capture local convexities. Such convex segments represent intervals where individuals might exhibit risk-seeking behavior, a phenomenon that traditional concave utility functions fail to capture.
	
	\subsection{Further mathematical aspects }\label{sec:mathematical_aspects}
	
	In this section, we gather several basic properties of multi-fractional stochastic dominance. First, we investigate a few interesting properties of the
	utility class $u\in \mathscr{U}^{\textsc{mf}}_{\boldsymbol{\upgamma}}$. 
	
	\begin{proposition}\label{prop:utility_class_properties}
		Let $\boldgamma{\upgamma}: \mathbb{R} \rightarrow[0,1]$ be an arbitrary non-decreasing function. Then utility class $\mathscr{U}^{\textsc{mf}}_{\boldsymbol{\upgamma}}$ enjoys the following properties:
		\begin{itemize}
			\item[(a)] Any function $u \in \mathscr{U}^{\textsc{mf}}_{\boldsymbol{\upgamma}}$ is non-decreasing. In addition, it contains all linear non-decreasing functions. 
			\item[(b)] It is invariant under \textbf{positive translations} that is, if $u\in \mathscr{U}^{\textsc{mf}}_{\boldsymbol{\upgamma}}$ and $c\ge 0$ then $u_c\in \mathscr{U}^{\textsc{mf}}_{\boldsymbol{\upgamma}}$ where $u_c(x)=u( x+c)$. 
		\end{itemize}
	\end{proposition}

	\begin{proposition} \label{prop:basic_properties}
		Let $\boldgamma{\upgamma}: \mathbb{R} \rightarrow[0,1]$ be a non-decreasing function.	Assume that $X\sim F$ and $Y\sim G$ are  two arbitrary distributions. Then, the following statements are in order:
		\begin{itemize}
			\item[(a)] Further assume that $F$ and $G$ have finite left supports,  denoted as $\ell_F$ and  $\ell_G$. Then, $F \preceq^{\textsc{mf}}_{(1+ \boldgamma{\upgamma})\textsc{-SD}} G$ implies that $\ell_F \le \ell_G$.
			\item[(b)] If $F \preceq^{\textsc{mf}}_{ (1+ \boldgamma{\upgamma})\textsc{\textsc{-SD}}} G$, then $\mu_G \ge \mu_F$.
			\item[(c)] Let $\boldgamma{\upgamma}_1,\boldgamma{\upgamma}_2: \mathbb{R} \rightarrow[0,1]$ be non-decreasing functions such that $\boldgamma{\upgamma}_1(t)\le \boldgamma{\upgamma}_2(t)$ for all $t\in\R$. Then, relation $				F \preceq^{\textsc{mf}}_{ (1+ \boldgamma{\upgamma_1})\textsc{-SD}} G $ implies that $ F \preceq^{\textsc{mf}}_{ (1+ \boldgamma{\upgamma_2})\textsc{-SD}} G$.
			\item[(d)] For all $c\ge 0$, $				X\preceq^{\textsc{mf}}_{ (1+ \boldgamma{\upgamma})\textsc{-SD}} X+c$. 
			\item[(e)] If $a \le b$, then $\delta_a \preceq^{\textsc{mf}}_{(1+ \boldsymbol{\upgamma})\textsc{-SD}} \delta_b$, where $\delta_x$ is a Dirac measure at point $x$.
		\end{itemize} 
	\end{proposition}

	\begin{remark}\label{rem:inclusion_gamma1_gamma_2}
		It is worth to point out that for any non-decreasing functions $\boldgamma{\upgamma}_1,\boldgamma{\upgamma}_2: \mathbb{R} \rightarrow[0,1]$  with $\boldgamma{\upgamma}_1(t)\le \boldgamma{\upgamma}_2(t)$ for every $t \in \R$, we have
		
		\begin{equation} \label{eq:inclusion_gamma1_gamma_2}
			\mathscr{U}^{\textsc{mf}}_{\boldgamma{\upgamma}_2}   \, \subseteq  \mathscr{U}^{\textsc{mf}}_{\boldgamma{\upgamma}_1} \quad \mbox{ and } \quad \mathscrbf{E}_{\textsc{mf}}(\boldgamma{\upgamma}_2) \supseteq \mathscrbf{E}_{\textsc{mf}}(\boldgamma{\upgamma}_1).
		\end{equation}
	\end{remark}	
	
	As mentioned earlier in Remark \ref{rem:RightAfterMainDefinition} (iii), incorporating the function $\boldgamma{\upgamma}$ as a CDF introduces an additional layer of randomness, enriching our framework. In line with this, we now present a probabilistic characterization of Proposition \ref{prop:basic_properties} (c).
	
	\begin{proposition} Let $F,G,H_1,H_2$ be distribution functions such that $H_2 \preceq_{\textsc{FSD}} H_1$. Then, relation $F \preceq^{\textsc{mf}}_{ (1+ \boldsymbol{H_1})\textsc{-SD}} G$ implies that $	F \preceq^{\textsc{mf}}_{ (1+ \boldsymbol{H_2})\textsc{-SD}} G.$
	\end{proposition}

	\begin{proposition}[\textbf{Stability properties}]\label{prop:StabilityProperties} 
		For an arbitrary non-decreasing function $\boldsymbol{\upgamma}$, the multi-fractional $(1+\boldsymbol{\upgamma})$\textsc{-SD} is closed under the following operations:
		\begin{itemize}
			\item [(a)] Positive location transformations: for all $c\ge 0$,
			\begin{equation*}
				X \preceq^{\textsc{mf}}_{ (1+ \boldgamma{\upgamma})\textsc{-SD}} Y \quad \text{ implies that }   \quad X+c \preceq^{\textsc{mf}}_{ (1+ \boldgamma{\upgamma})\textsc{-SD}} Y+c.
			\end{equation*}

			\item[(b)] Convolutions with non-negative, independent random variables:
			\begin{equation*}
				X \preceq^{\textsc{mf}}_{ (1+ \boldgamma{\upgamma})\textsc{-SD}} Y  \quad \text{ implies that }   \quad    X+Z \preceq^{\textsc{mf}}_{ (1+ \boldgamma{\upgamma})\textsc{-SD}} Y+Z
			\end{equation*}
			where $Z$ is non-negative and independent of $X$ and $Y$.
			\item[(c)] Mixing: 
			\begin{equation*}
				[X|\Theta=\theta]\preceq^{\textsc{mf}}_{ (1+ \boldgamma{\upgamma})\textsc{-SD}} [Y|\Theta=\theta], \quad \forall \theta \in \text{supp}(\Theta)  \quad \text{ implies that }   \quad  X \preceq^{\textsc{mf}}_{ (1+ \boldgamma{\upgamma})\textsc{-SD}} Y.
			\end{equation*}
			
			\item[(d)] Let $(\boldgamma{\upgamma}_n:n\ge 1)$ be a sequence of non-decreasing functions so that $\boldsymbol{\upgamma}_n \stackrel{p.wise}{\longrightarrow} \boldsymbol{\upgamma}$  as $n \to \infty$. Then, $\boldgamma{\upgamma}$ is non-decreasing and, moreover $X \preceq^{\textsc{mf}}_{ (1+ \boldgamma{\upgamma}_n)\textsc{-SD}} Y$, for every $n \in \mathbb{N}$ implies that 
			$X \preceq^{\textsc{mf}}_{ (1+ \boldgamma{\upgamma})\textsc{-SD}} Y$. 
		\end{itemize}
	\end{proposition}

	Inspired by \cite[Theorem 4.8]{muller} we provide the following result which extends the invariance property to more general transformations.
	\begin{proposition}\label{prop:gamma_MFgamma}
		If $X \preceq^{\textsc{mf}}_{ (1+ \boldgamma{\upgamma}\times\gamma)\textsc{-SD}} Y$ for a non-decreasing function $\boldgamma{\upgamma}: \mathbb{R} \rightarrow[0,1]$  and the constant $\gamma \in [0,1]$,  then we have: 
		\begin{itemize}
			\item[(a)]
			$ \phi(X) \preceq^{\textsc{mf}}_{ (1+ \boldgamma{\upgamma})\textsc{-SD}} \phi(Y)$ for all $\phi \in \mathscr{U}^*_{\gamma}$ with $\phi(x)\ge x$,
			\item[(b)] $\psi(X) \preceq_{ (1+ \gamma)\textsc{-SD}} \psi(Y)$ for all $\psi \in \mathscr{U}^{\textsc{mf}}_{\boldsymbol{\upgamma}}$.
		\end{itemize}
	\end{proposition}

	\begin{remark}
		Recall that when $\gamma=1$, the utility class $\mathscr{U}^*_\gamma$ consists of increasing concave functions.  Then, in the special case of Proposition \ref{prop:gamma_MFgamma} when $\gamma=1$, it implies that: (i) the multi-fractional $(1+\boldsymbol{\upgamma})$-\textsc{SD} is invariant under any increasing concave transformations $\phi$ that satisfy $\phi(x)\ge x$, (ii) for every $\psi \in \mathscr{U}^{\textsc{mf}}_{\boldsymbol{\upgamma}}$, the relation $\psi (X) \preceq_{\textsc{SSD}} \psi(Y)$ holds.
	\end{remark}

	\section{Generalization through utility class} \label{sec:generilization_utilty}
	In this section, we study the generalization of fractional $(1+\gamma)$-\textsc{SD} through the utility class in \eqref{eq:MullerUtilitySpacesNonSmooth}, which gives rise to generalized families of functional fractional stochastic dominance (FFSD).
	
	\subsection{FFSD: Definition and the integral condition}\label{sec:FFSD_definition}
	In Section \ref{sec:genarilzation_integral}, we extended the fractional  $(1+\gamma)$-\textsc{SD} using an integral condition by replacing the constant parameter $\gamma$ with its functional conjugate  $\boldsymbol{\upgamma}$. Due to the nature of the integral condition \eqref{eq:muller_IntegralCondition}, this extension was relatively straightforward, involving the association of the function $\boldsymbol{\upgamma}$ with a single point. However, the extension through the utility class introduces a degree of complexity, as the set condition of  $\mathscr{U}^{*}_{\gamma}$ given in \eqref{eq:MullerUtilitySpacesNonSmooth} involves four points. Therefore, we require a new framework to enable a functional extension via the function $\boldgamma{\upgamma}$ of a single variable. We construct this with the following requirements in mind:  1) The fractional SD should be representable such that, for any $\gamma \in [0,1]$, the set condition given in \eqref{eq:MullerUtilitySpacesNonSmooth} is both a necessary and sufficient to verify whether a given function $u$ belongs to the fractional SD utility space within the new framework (see Proposition  \ref{prop:one_is_enough}). 2) The framework must be meaningful in decision analysis (see Section \ref{sec:economical}).

	Based on the discussion above, we build our framework through the one-sided derivatives of utility functions. This approach not only serves our purposes but also provides a tractable and meaningful structure. Hence, we introduce the following.
	
	\begin{definition}\label{def:prop_+_-} We say a function $u$ has $\boldsymbol{D}_\pm$ property if $ u $ is both-sided differentiable, i.e. left $ u_-'(x) $ and right $ u_+'(x) $ derivatives exist at each point $ x \in \mathbb{R}$. For a given  $\gamma \in [0,1]$, we  define 
		\begin{equation}
			\widetilde{\mathscr{U}}^{*}_{\gamma} := \mathscr{U}^{*}_{\gamma} \cap \mathscr{U}(\boldsymbol{D}_\pm) 
		\end{equation}	
		where
		\begin{equation}\label{eq:both_sided_U}\mathscr{U}(\boldsymbol{D}_\pm):=\{ u : \mathbb{R} \to \mathbb{R} : u \text{ satisfies the } \boldsymbol{D}_\pm \text{ property} \}.
		\end{equation}
	\end{definition}
	We present the following proposition that confirms the representability of the fractional SD as discussed in Remark \ref{rem:framework}.
	
	\begin{proposition}\label{prop:one_is_enough} Let $u$ be a function whose both-sided derivatives exist at each point $x \in \R$. Then, for a given $\gamma \in [0,1]$, $u\in \widetilde{\mathscr{U}}^{*}_{\gamma}$ if and only if
		\begin{equation*}
			0\le\gamma u_+'(y) \le u_+'(x)  \text{ for all } x \le y.
		\end{equation*}
		A similar conclusion is valid for the left-derivative $u'_-$.
	\end{proposition}
	Building on our development so far, we provide the following derivative-relation property for the functional extension of  $\widetilde{\mathscr{U}}^{*}_{\gamma}$. 
	
	\begin{definition}[$\boldsymbol{D}_\pm(\boldsymbol{\upgamma})$\textbf{-property}]\label{def:D_pm(gamma)Property}
		Let $\boldsymbol{\upgamma}:\R\to[0,1]$ be an non-decreasing arbitrary function. We say that function $u:\R \to \R$ satisfies in $\boldsymbol{D}_\pm(\boldsymbol{\upgamma})$ if it is both-sided differentiable, i.e. it possesses $\boldsymbol{D}_\pm$ property and, moreover \textbf{at least} one of the following conditions hold 
		\begin{equation}\label{eq:D_pm(gamma)Property}
			(i) \quad 0 \le \boldsymbol{\upgamma} (y) u_+'(y) \le   u_+'(x), \, \forall \, x \le  y
			\quad (ii) \quad  0 \le \boldsymbol{\upgamma}(y) u'_-(y) \le u'_{-}(x), \, \forall  x\le y.
		\end{equation}	
	\end{definition}
	Clearly, both conditions (i) and (ii) become equivalent and simplify to $0 \le \boldsymbol{\upgamma}(y) u'(y) \le u'(x), \forall x\le y$ at the differentiability points of $u$. We will revisit this relationship in Proposition \ref{prop:piece_wise}. We provide the following utility space for a given non-decreasing function $\boldsymbol{\upgamma}:\mathbb{R}\to[0,1]$:
	\begin{equation}\label{eq:FFSD_utiltiy}
		\widetilde{\mathscr{U}}^{\textsc{ff}}_{\boldsymbol{\upgamma}}   := \Big\{   u: \R \to \R  \, :\,    \boldsymbol{D}_\pm(\boldsymbol{\upgamma}) \text{ property holds}    \Big\}.              
	\end{equation}	
	
	Now, we are ready to define functional fractional stochastic dominance.
	
	\begin{definition}[\textbf{Functional fractional  $(\boldgamma{\upgamma})$\textsc{-SD}}] \label{def:FFSD} Let $\boldgamma{\upgamma}:\R \to [0,1]$ be an arbitrary non-decreasing function. Let $F$ and $G$ be two distribution functions. We say that $G$ dominates $F$ in the sense of the functional fractional $(\boldgamma{\upgamma})$\textsc{-SD}, denote by $F \preceq^{\textsc{ff}}_{\boldgamma{\upgamma}\textsc{-SD}} G$ if 
		\begin{equation}\label{eq:ff_utiltiy_condition}
			\E_F  \left[  u \right]  \le \E_G \left[ u \right], \quad \forall \, u \in \widetilde{\mathscr{U}}^{\textsc{ff}}_{\boldsymbol{\upgamma}}
		\end{equation}	
		provided that expectations exist.
	\end{definition}
	
	The following result establishes the relationship between both MFSD and FFSD.
	
	\begin{proposition}\label{prop:mf_one_sided} Let $\boldgamma{\upgamma}:\R \to [0,1]$ be arbitrary non-decreasing function. Then, the inclusion
		$ \widetilde{\mathscr{U}}^{\textsc{mf}}_{\boldsymbol{\upgamma}}  \subseteq \widetilde{\mathscr{U}}^{\textsc{ff}}_{\boldsymbol{\upgamma}} $ holds. Moreover, if $\boldsymbol{\upgamma}\equiv \gamma\in [0,1]$, the two sets coincide.
	\end{proposition}

	\begin{remark}\label{rem:InclusionIsStrict}
		\begin{itemize}
			\item[(i)] When transitioning from a fixed coefficient $\gamma$ to its functional form $\boldsymbol{\upgamma}$, the utility space $\widetilde{\mathscr{U}}^{\textsc{ff}}_{\boldsymbol{\upgamma}}$ emerges as a natural generalization of $\widetilde{\mathscr{U}}^*_{\gamma}$. This resembles the extension observed in the MFSD framework, where the integral condition is adapted to incorporate the functional form of $\boldgamma{\upgamma}$. A closer examination of the $\boldsymbol{D}_\pm(\boldsymbol{\upgamma})$ property reveals that $\boldsymbol{\upgamma}$ controls the maximal possible increase in the one-sided derivatives based on location, i.e., $u_+(y)/u_+(x) \le 1/\boldsymbol{\upgamma}(y)$, where its non-decreasing nature permits a greater increase at lower values of the real line. This implies that steeper local convexities are allowed at lower values. Although $\boldgamma{\upgamma}$ has similar interpretations in the class $\widetilde{\mathscr{U}}^{\textsc{mf}}_{\boldsymbol{\upgamma}}$, its control in $\widetilde{\mathscr{U}}^{\textsc{ff}}_{\boldsymbol{\upgamma}}$ is more localized, permitting utility functions to be more non-concave. 
			In fact,  as discussed in Remark \ref{rem:example}, we can construct an insightful example of such a function where $u \in \widetilde{\mathscr{U}}^{\textsc{ff}}_{\boldsymbol{\upgamma}}$ but $u \notin  \widetilde{\mathscr{U}}^{\textsc{mf}}_{\boldsymbol{\upgamma}}$. In general, we have $\widetilde{\mathscr{U}}^{\textsc{ff}}_{\boldsymbol{\upgamma}}\setminus \widetilde{\mathscr{U}}^{\textsc{mf}}_{\boldsymbol{\upgamma}}\neq \emptyset$ as soon as $\boldgamma{\upgamma}\not\equiv \gamma \in [0,1]$.

			\item[(ii)] One advantage of the utility class $\widetilde{\mathscr{U}}^{\textsc{ff}}_{\boldsymbol{\upgamma}}$ is that for a given utility function $u$, determining whether $u \in \widetilde{\mathscr{U}}^{\textsc{ff}}_{\boldsymbol{\upgamma}}$ is relatively straightforward compared to  $\widetilde{\mathscr{U}}^{\textsc{mf}}_{\boldsymbol{\upgamma}}$. The inclusion relation in Proposition \ref{prop:mf_one_sided} allows us to develop tools that easily check whether a given function $u \in \widetilde{\mathscr{U}}^{\textsc{ff}}_{\boldsymbol{\upgamma}}$ also belongs to $\widetilde{\mathscr{U}}^{\textsc{mf}}_{\boldsymbol{\upgamma}}$. In Section \ref{sec:economical}, we introduce the notion of partial greediness, which not only provides meaningful decision theoretic interpretations but also facilitates this verification.

			\item[(iii)] The bounds of the upward jumps in the derivative $u^\prime$ at kink points are depend on the specific location of the kink. For example, let $x^*_1<x^*_2$ be two points where $u$ has kinks, then
			\begin{equation*}
				\frac{u_+^\prime(x^*_1)}{u_-^\prime(x^*_1)} \le   \frac{1}{\boldgamma{\upgamma}(x^*_1) }   \quad  \text{ and }   \quad  \frac{u_+^\prime(x^*_2)}{u_-^\prime(x^*_2)} \le \frac{1}{\boldgamma{\upgamma}(x^*_2)}.
			\end{equation*}
			Such a feature is not available in the context of $\widetilde{\mathscr{U}}^*_\gamma$  and the ratio $1/\gamma$ provides a global bound independent of the location of kinks.
		\end{itemize}
	\end{remark}

	Next, we prove a "if and only if" theorem for the FFSD that establishes connection with the integral condition:
	\begin{equation} 
		\label{eq:integral_condition_FFSD}
		\int_{-\infty}^t \frac{1}{\boldgamma{\upgamma}(x)}(F(x)-G(x))_- \, dx \le \int_{-\infty}^t (F(x)-G(x))_+ \, dx, \quad \forall t \in \mathbb{R}.
	\end{equation}
	
	But first, we have the following remark concerning the well-definedness of the integral condition.
	\begin{remark}\label{rem:just_before_the_proof_FFSD} 
		For a given function $\boldsymbol{\upgamma}$ and distribution functions $F$ and $G$, the integral condition \eqref{eq:integral_condition_FFSD} is well-defined if the integral on the left-hand side finite for every $t\in \R$. First, $(F(x)-G(x))_-$ is bounded and non-negative, assuming $\boldsymbol{\upgamma}: \mathbb{R} \rightarrow (0,1]$ addresses this. Moreover, we place the following assumptions:
		(i) the distribution functions have finite means,
		(ii) the first crossing point between the distributions, $x_1^*$ exists where $(F(x)-G(x))_- = 0$ for all $x \leq x_1^*$. If they never cross, then \eqref{eq:integral_condition_FFSD} holds trivially, allowing the statements to directly follow from the FSD. Therefore, for a given non-decreasing function $\boldsymbol{\upgamma}: \mathbb{R} \rightarrow (0,1]$, we have $$\int_{-\infty}^{t}\frac{1}{\boldgamma{\upgamma}(x)}(F(x)-G(x))_- \, dx\le\int_{x^*_1}^{t}\frac{1}{\boldgamma{\upgamma}(x^*_1)}(F(x)-G(x))_-dx,$$ for all $t\in \R$ by the monotonicity of $\boldsymbol{\upgamma}$.
		
	\end{remark}

	Now, we begin by establishing one direction of the "if and only if" theorem.
	
	\begin{lemma}\label{lem:FFSD:base_type} Let $\boldgamma{\upgamma}: \mathbb{R} \rightarrow (0,1]$ be an arbitrary non-decreasing right-continuous function with finitely many jumps.  
		Let $F$ and $G$ be two arbitrary distribution functions crossing finitely many times. For every $t \in \R$, define the continuous (utility) function $u_{\textsc{ff},t}^{\prime}$ via its right-derivative as 
		\begin{equation}\label{eq:FFSD_basetpe}
			u_{\textsc{ff},t}^{\prime}(x)=
			\begin{cases}
				1 & \text { if } G(x) \le F(x) \text { and } x \leq t \\ 
				1/\boldgamma{\upgamma}(x) & \text { if } F(x) < G(x) \text { and } x \leq t \\ 
				0 & \text { if } t<x.
			\end{cases}
		\end{equation}
		Then, $u_{\textsc{ff},t} \in\widetilde{\mathscr{U}}^{\textsc{ff}}_{\boldsymbol{\upgamma}}$ for every $t \in \R$.
	\end{lemma}
	
	\begin{proposition}\label{prop:FFSD:Utility_Implies_IntegralCondition}
		Let $\boldgamma{\upgamma}: \mathbb{R} \rightarrow (0,1]$ be an arbitrary non-decreasing right-continuous function with finitely many jumps.  
		Let $F$ and $G$ be two arbitrary distribution functions with \textbf{finite  means} and crossing finitely many times. Then, relation  
		\begin{equation}\label{eq:FFSD:Extension_UtilityCondition}
			\E_F[u_{\textsc{ff},t}]  \le \E_G \left[u_{\textsc{ff},t}\right],  \quad \forall \, t \in \R
		\end{equation}
		implies that the integral condition \eqref{eq:integral_condition_FFSD} holds.
	\end{proposition}
	
	We are now prepared to present the main theorem.
	
	\begin{theorem}\label{thm:MFSD:iff}
		Let $ \boldsymbol{\upgamma}: \mathbb{R} \to (0,1] $ be a non-decreasing, right-continuous function with a finite number of discontinuities. Consider two distribution functions $ F $ and $ G $ that intersect only finitely many times. Then, the following statements are equivalent: 
		\begin{itemize}
			\item[(a)] 	$ F \preceq^{\textsc{ff}}_{\boldsymbol{\upgamma}\textsc{-SD}} G $.
			
			\item[(b)] Integral condition \eqref{eq:integral_condition_FFSD} holds, that is $$\int_{-\infty}^t \frac{1}{\boldgamma{\upgamma}(x)}(F(x)-G(x))_- \, dx \le \int_{-\infty}^t (F(x)-G(x))_+ \, dx \quad \forall t \in \mathbb{R}.$$
		\end{itemize}
	\end{theorem}

	We conclude this section with the following remark.
	
	\begin{remark}
		\begin{itemize}
			\item[(i)] It is evident that $ \widetilde{\mathscr{U}}^*_{\thickbarr{\boldsymbol{\upgamma}}} \subseteq \widetilde{\mathscr{U}}^{\textsc{mf}}_{\boldsymbol{\upgamma}} \subseteq \widetilde{\mathscr{U}}^{\textsc{ff}}_{\boldsymbol{\upgamma}} \subseteq \widetilde{\mathscr{U}}^{*}_{\barbeloww{\boldsymbol{\upgamma}}}$, and these inclusions are strict as soon as $ \boldsymbol{\upgamma} $ is a non-constant function. Therefore, FFSD is stronger than MFSD. In the terminology of Section \ref{sec:MFSDUniverse}, the inclusion relation $ \mathscrbf{E}_{\textsc{ff}}(\boldsymbol{\upgamma}) \subseteq \mathscrbf{E}_{\textsc{mf}} (\boldsymbol{\upgamma})$ for the universe of distributions is directly verified by the integral conditions \eqref{eq:Extension_gammaSD_IntegralCondition} and \eqref{eq:integral_condition_FFSD}. 
			
			\item[(ii)] There is a particularly interesting aspect here that deserves to be pointed out. Both MFSD and FFSD interpolate between FSD and SSD, but with distinct features. MFSD, thanks to its integral condition \eqref{eq:Extension_gammaSD_IntegralCondition}, offers a form of local interpolation that is straightforward to interpret. However, when we turn to the corresponding utility class $ \mathscr{U}^{\textsc{mf}}_{\boldsymbol{\upgamma}} $, the interpolation is less obvious. Determining which utility functions belong to this class becomes challenging in terms of both interpretation and verification. In contrast, the interpolation for FFSD is clearly visible in the utility class $ \widetilde{\mathscr{U}}^{\textsc{ff}}_{\boldsymbol{\upgamma}} $ and allows for easy verification of whether a given utility function belongs to it.  However, the integral condition \eqref{eq:integral_condition_FFSD} lacks a clear interpretation of the interpolation between FSD and SSD, unlike what is observed in MFSD.  
			
		\end{itemize}
	\end{remark}

	\subsection{FFSD: Examples}\label{sec:FFSD_examples}

	In this section, we present examples that highlight the key differences between the MFSD and FFSD frameworks. Thanks to the inclusion $\widetilde{\mathscr{U}}^{\textsc{mf}}_{\boldsymbol{\upgamma}} \subseteq \widetilde{\mathscr{U}}^{\textsc{ff}}_{\boldsymbol{\upgamma}}$ given in Proposition \ref{prop:mf_one_sided}, it is clear that all example utility functions presented in Section \ref{sec:utilityclass} for $\mathscr{U}^{\textsc{mf}}_{\boldsymbol{\upgamma}}$ are also belong to $\widetilde{\mathscr{U}}^{\textsc{ff}}_{\boldsymbol{\upgamma}}$, provided they are both-sided differentiable. Non-differentiable functions outside the MFSD utility class also exist, but we focus on those that are both realistic and practically significant. Specifically, we concentrate on the family of piecewise continuously differentiable (PCD) functions, which are particularly relevant for modelling decision making behaviour in applications.

	\begin{proposition}\label{prop:piece_wise} Let $u:\R\to \R$ be a piecewise continuously differentiable function. Then, we have the following:   
		\begin{itemize}
			\item[(a)] $u$ both-sided differentiable on $\R$, i.e. it fulfils $\boldsymbol{D}_\pm$ property.
			
			\item[(b)] Consider a non-decreasing function $\boldsymbol{\upgamma}:\R \to [0,1]$, which is continuous at each sub-division point $\{x_1,\dots,x_n\}$ of $u$. Then, the derivative-relation conditions, $$(i) \,\, 0 \le \boldsymbol{\upgamma}(y) u'_+(y) \le u'_+(x), \, \forall  x\le y \quad (ii)\, \,  0 \le \boldsymbol{\upgamma}(y) u'_-(y) \le u'_-(x), \, \forall  x\le y$$
			are equivalent.
		\end{itemize}
	\end{proposition}
	
	The Proposition \ref{prop:piece_wise} item (a) establishes that PCD functions, indeed belong to  $\widetilde{\mathscr{U}}^{\textsc{ff}}_{\boldsymbol{\upgamma}}$ as soon as they satisfies derivative-relation property for an appropriately given function  $\boldsymbol{\upgamma}$. It is also possible to utilize conditions on opposite sides, such as $$ 0 \le \boldsymbol{\upgamma}(y) u'_+(y) \le u'_-(x), \, \forall  x\le y.$$ Note that the condition above is particularly useful in scenarios where the utility function exhibits upward jumps in its derivative, as discussed in item (iv) of Remark \ref{rem:InclusionIsStrict}. On the other hand, they are not necessarily included in $\widetilde{\mathscr{U}}^{\textsc{mf}}_{\boldsymbol{\upgamma}}$, as its definition requires that $u$ be expressible as a linear combination of functions within $\widetilde{\mathscr{U}}^{\text{Union}}_{\boldsymbol{\upgamma}}$. To further illustrate this distinction, we provide the counterpart of Example \ref{ex:MFSDUtilityStrictSubsetFSD} in the FFSD framework. This example also highlights the key structural differences between this utility space and $\widetilde{\mathscr{U}}^*_{\gamma}$.

	\begin{example}  
		The purpose of this example is to provide a parametrized family $\mathscr{U}_{\Theta}$ so that the only possibility that $\mathscr{U}_{\Theta} \subseteq \widetilde{\mathscr{U}}^*_{\gamma}$ for some constant $\gamma \in [0,1]$ is to have $\gamma=0$ (hence, corresponding to the utility class of the FSD). However, one can find a suitable non-decreasing function $\boldsymbol{\upgamma}$ so that 
		\begin{equation*}
			\mathscr{U}_{\Theta} \subseteq \widetilde{\mathscr{U}}^{\textsc{ff}}_{\boldsymbol{\upgamma}} \neq \widetilde{\mathscr{U}}_{0} \quad \text{ and more importantly  } \quad \mathscr{U}_{\Theta}  \not\subseteq  \widetilde{\mathscr{U}}^{\textsc{mf}}_{\boldsymbol{\upgamma}}.
		\end{equation*}
		Let $\Theta=(0,1)$ and consider a function $\widetilde{\boldsymbol{\upgamma}} : \Theta \to (0,1)$ with $\lim_{\theta \to 0}\widetilde{\boldsymbol{\upgamma}}(\theta)=0$. For every $\theta \in \Theta$, we define the following PCD utility function $u_\theta$ via its right-derivative as 
		\begin{equation*}
			u'_\theta(x)=\begin{cases}
				u_{\theta,1}^\prime(x) & \mbox{ if }  -\infty< x < \theta,\\
				x &  \mbox{ if }  \theta\le x  < \sqrt{\theta}  ,\\
				u_{\theta,2}^\prime(x) & \mbox{ if }  \sqrt{\theta}\le x < \infty,\\
			\end{cases}
		\end{equation*}
		where $u_{\theta,1}^\prime$ and $u_{\theta,2}^\prime$ are arbitrary non-negative and non-increasing continuous functions with $\inf_{x\in \R}u_{\theta,1}^\prime(x)=\sqrt{\theta}$ and $\sup_{x\in \R} u_{\theta,2}^\prime(x)=\theta$. Thus, for a given $\theta$, it is straightforward to see that $u_\theta \in \widetilde{\mathscr{U}}^{*}_{\gamma}$ for some constant $\gamma$ (i.e, we have $ 0 \le  \gamma  u'_{\theta}(y) \le u'_{\theta} (x), \,\forall x\le y $) \textbf{if and only if} the condition 
		\begin{equation}\label{eq:example_theta_condition}
			\sqrt{\theta} \ge \gamma 
		\end{equation} 
		holds true.  Let  $\mathscr{U}_{\Theta } : =\left \{       u_\theta \, :\, \theta \in \Theta          \right \}$. Now, observe that, for some constant $\gamma$, the validity of the set inclusion $\mathscr{U}_{\Theta} \subseteq \mathscr{U}^*_{\gamma}$ dictates that $\gamma=0$ must be. This conclusion directly follows from \eqref{eq:example_theta_condition}  when $\theta\to 0$. Hence, the only scenario where $u_\theta$ for every $\theta\in \Theta$ included within a single utility space associated with the fractional $(1+\gamma)$\textsc{-SD} corresponds to the utility class  $\widetilde{\mathscr{U}}_{0} = \widetilde{\mathscr{U}}^*_{\gamma=0}$. Next, consider the continuos non-decreasing function 	
		\begin{align*}
			\boldgamma{\upgamma} (x)  =
			\begin{cases}
				0   & \mbox{ if } x < 0 ,\\
				x  &\mbox{ if }  0 \le x \le 1,\\
				1    & \mbox { if }    x  >1.
			\end{cases}	
		\end{align*}	
		We first show that $\mathscr{U}_{\Theta} \subseteq \widetilde{\mathscr{U}}^{\textsc{ff}}_{\boldsymbol{\upgamma}}$. In order to do this, note that, for each  $\theta \in \Theta$, the condition 
		\begin{equation}\label{eq:example_theta_condition2}
			\theta \ge u'_{\theta}(y) \boldgamma{\upgamma}(y) = y \times y , \quad \forall y \in [\theta, \sqrt{\theta})
		\end{equation}
		is both necessary and sufficient for the validity of $u_\theta \in \widetilde{\mathscr{U}}^{\textsc{ff}}_{\boldsymbol{\upgamma}}$ in which obviously holds true.  Lastly, we show that $\mathscr{U}_{\Theta}  \not\subseteq  \widetilde{\mathscr{U}}^{\textsc{mf}}_{\boldsymbol{\upgamma}}$. Fix $\theta \in \Theta$. First, when $y\approx\sqrt{\theta}$ (from left) and  $x=\theta< y$, it follows from \eqref{eq:example_theta_condition2} that
		\begin{equation}\label{eq:example_theta_condition3}
			\frac{1}{\boldgamma{\upgamma}(y)}=\frac{1}{\sqrt{\theta}}=\frac{u_\theta'(y)}{u_\theta'(x)}.
		\end{equation}
		Next, we discuss that $u_\theta \notin \widetilde{\mathscr{U}}^{\text{Union}}_{\boldsymbol{\upgamma}}$. Note that $u_\theta$ is not globally concave function and hence $u_\theta \notin \widetilde{\mathscr{U}}^{+\infty}_{\thickbarr{\boldsymbol{\upgamma}} =1}$. Now, by contradiction,  assume that there exists $t\in \R $ such that $u_\theta \in \widetilde{\mathscr{U}}^t_{\boldsymbol{\upgamma}(t)}$.  Therefore, based on the equality in \eqref{eq:example_theta_condition3}, this is only possible when $t=\sqrt{\theta}$. However, this leads to a contradiction, as the derivative of the subsequent concave segment,  $u'_{2,\theta}$ of $u_\theta$  is not necessarily equal to 0 after $\sqrt{\theta}$. Hence, $u_\theta$ can only belong to $ \widetilde{\mathscr{U}}^{\textsc{mf}}_{\boldsymbol{\upgamma}}$ if it can be represented as a convex combination of functions that belong to $\widetilde{\mathscr{U}}^{\text{Union}}_{\boldsymbol{\upgamma}}$. For a moment, assume that $u_\theta = \lambda_1 v_1 + \lambda_2 v_2$ where $v_1 \in \widetilde{\mathscr{U}}^{t = \sqrt{\theta}}_{\boldsymbol{\upgamma}(t) =\sqrt{\theta}}$ with $\frac{v_1'(\sqrt{\theta})}{v_2'(\theta)}=\frac{1}{\sqrt{\theta}}$ and $v_2 \in \widetilde{\mathscr{U}}^*_{1}$.  Then, from the mediant inequity \eqref{eq:mediant_inq} we obtain

		\begin{equation*}
			\frac{u_\theta'(\sqrt{\theta})}{u_\theta'(\theta)}=\frac{\lambda_1v_1(\sqrt{\theta})+\lambda_2v_2(\sqrt{\theta})}{\lambda_1v_1(\theta)+\lambda_2v_2(\theta)}=\frac{\lambda_1\sqrt{\theta}+\lambda_2v_2(\sqrt{\theta})}{\lambda_1\theta+\lambda_2v_2(\theta)}<\frac{1}{\sqrt{\theta}}
		\end{equation*}
		since $\frac{v_2(\sqrt{\theta})}{v_2(\theta)}\le 1$ for any $v_2\in \widetilde{\mathscr{U}}^*_{1}$ by definition in which violates \eqref{eq:example_theta_condition3}. The general case can be discussed in a similar way relying on mediant inequality and the facts that any linear combination should contain at least one piece from the utility space $\widetilde{\mathscr{U}}^{\sqrt{\theta}}_{\sqrt{\theta}}$ and another piece of $\widetilde{\mathscr{U}}^{*}_{1}$. 
		
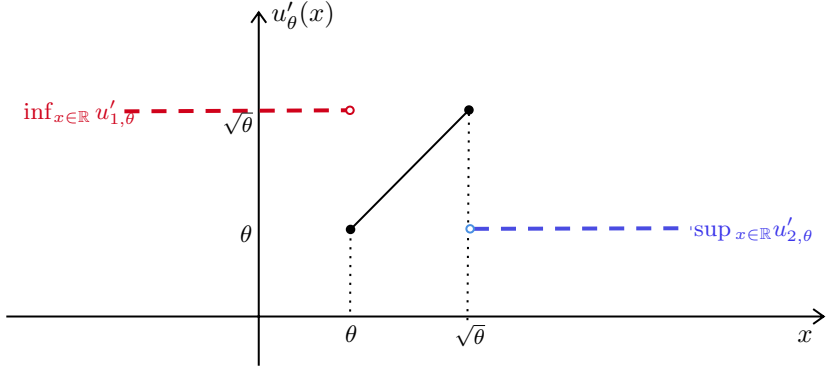
\begin{figure}[H]
			
			\centering
			
			\tikzset{every picture/.style={line width=0.75pt}} 
			
			\begin{tikzpicture}[x=0.75pt,y=0.75pt,yscale=-1,xscale=1,scale=0.75]
				
				\draw  (21.5,238.8) -- (565.5,238.8)(189.5,34.8) -- (189.5,271.8) (558.5,233.8) -- (565.5,238.8) -- (558.5,243.8) (184.5,41.8) -- (189.5,34.8) -- (194.5,41.8)  ;
				\draw  [dash pattern={on 0.84pt off 2.51pt}]  (250.6,180.4) -- (250.3,238.5) ;
				\draw    (250.6,180.4) -- (327.75,101.98) ;
				\draw [shift={(329.4,100.3)}, rotate = 314.53] [color={rgb, 255:red, 0; green, 0; blue, 0 }  ][line width=0.75]      (0, 0) circle [x radius= 3.35, y radius= 3.35]   ;
				\draw [shift={(250.6,180.4)}, rotate = 314.53] [color={rgb, 255:red, 0; green, 0; blue, 0 }  ][fill={rgb, 255:red, 0; green, 0; blue, 0 }  ][line width=0.75]      (0, 0) circle [x radius= 3.35, y radius= 3.35]   ;
				\draw  [dash pattern={on 0.84pt off 2.51pt}]  (329.4,102.8) -- (328.4,241.3) ;
				\draw [color={rgb, 255:red, 208; green, 2; blue, 27 }  ,draw opacity=1 ][line width=1.5]  [dash pattern={on 5.63pt off 4.5pt}]  (100.1,101.2) -- (245.65,100.51) ;
				\draw [shift={(247.7,100.5)}, rotate = 359.73] [color={rgb, 255:red, 208; green, 2; blue, 27 }  ,draw opacity=1 ][line width=1.5]      (0, 0) circle [x radius= 3.05, y radius= 3.05]   ;
				\draw [color={rgb, 255:red, 74; green, 76; blue, 226 }  ,draw opacity=1 ][fill={rgb, 255:red, 52; green, 54; blue, 222 }  ,fill opacity=1 ][line width=1.5]  [dash pattern={on 5.63pt off 4.5pt}]  (328.9,180) -- (477.5,179.6) ;
				\draw [shift={(328.9,180)}, rotate = 359.85] [color={rgb, 255:red, 74; green, 76; blue, 226 }  ,draw opacity=1 ][fill={rgb, 255:red, 74; green, 76; blue, 226 }  ,fill opacity=1 ][line width=1.5]      (0, 0) circle [x radius= 3.05, y radius= 3.05]   ;
				
				\draw (243.6,242.6) node [anchor=north west][inner sep=0.75pt]  [font=\small]  {$\theta $};
				\draw (175.1,175.4) node [anchor=north west][inner sep=0.75pt]  [font=\small]  {$\theta $};
				\draw (315.6,242.4) node [anchor=north west][inner sep=0.75pt]  [font=\footnotesize]  {$\sqrt{\theta }$};
				\draw (161.6,104.9) node [anchor=north west][inner sep=0.75pt]  [font=\footnotesize]  {$\sqrt{\theta }$};
				\draw (14,93.2) node [anchor=north west][inner sep=0.75pt]  [font=\small,color={rgb, 255:red, 208; green, 2; blue, 27 }  ,opacity=1 ]  {$\inf_{x\in \mathbb{R}} u'_{1,\theta }$};
				\draw (196,25.4) node [anchor=north west][inner sep=0.75pt]    {$u'_{\theta }( x)$};
				\draw (546,246.4) node [anchor=north west][inner sep=0.75pt]    {$x$};
				\draw (480.2,169.2) node [anchor=north west][inner sep=0.75pt]  [font=\small,color={rgb, 255:red, 59; green, 53; blue, 234 }  ,opacity=1 ]  {$\sup{}_{x\in \mathbb{R}} u'_{2,\theta }$};

			\end{tikzpicture}
			\caption{Plot of $u_\theta'$}

		\end{figure}

	\end{example}
	We conclude with the following remark.
	\begin{remark}\label{rem:example}
		Apart from the examples above, we can also construct a non-decreasing function $\boldsymbol{\upgamma}$ and a function $u \in \widetilde{\mathscr{U}}^{\textsc{ff}}_{\boldsymbol{\upgamma}}$ such that $u \notin \widetilde{\mathscr{U}}^{\textsc{mf}}_{\boldsymbol{\upgamma}}$. Additionally, there exist distribution functions $F$ and $G$ with $F \preceq^{\textsc{mf}}_{(1+\boldsymbol{\upgamma})\textsc{-SD}} G$, for which the constructed function $u$ violates the non-negativity condition $\E_G[u] - \E_F[u] \geq 0$. While the construction of these functions and distributions is straightforward, it is omitted here for brevity (see \cite[Example 4.2.2.]{thesis}). This confirms that the class $\widetilde{\mathscr{U}}^{\textsc{ff}}_{\boldsymbol{\upgamma}}$ cannot serve as (although as one may naturally expect) as a utility class for MFSD.
	\end{remark}

	\section{MFSD \& FFSD: Economical aspects}\label{sec:economical}
	
	In this section, our primary focus is on examining the decision theoretic aspects of both the FFSD and MFSD frameworks. Additionally, we introduce a novel concept of partial greediness, which not only offers insights from a economics perspective but also provides a straightforward criterion for determining whether a given utility class belonging to $\widetilde{\mathscr{U}}^{\textsc{ff}}_{\boldsymbol{\upgamma}}$ also belongs to $\widetilde{\mathscr{U}}^{\textsc{mf}}_{\boldsymbol{\upgamma}}$.
	
	\subsection{Partial Greediness}\label{sec:greediness}
	
	In this section, we introduce the novel notion of partial greediness of a utility function at an arbitrary given point $x\in \R$ that provides a meaning to the function $\boldgamma{\upgamma}$ from a decision making perspective in the both multi-fractional and functional fractional stochastic dominance frameworks. The notion of partial greediness at the location $x\in \R$ characterizes the greediness behaviour of decision makers over half-open intervals $(x,\infty)$ rather than the entire real line as in \cite{Greediness}.  They define the index of global greediness (non-concavity) for a strictly increasing function $u:\R\to \R$ as follows: 
	\begin{equation}\label{eq:global_greediness}
		G_u:=\sup _{x_1<x_2 \leq x_3<x_4}\left(\frac{u\left(x_4\right)-u\left(x_3\right)}{x_4-x_3} / \frac{u\left(x_2\right)-u\left(x_1\right)}{x_2-x_1}\right).
	\end{equation}

	It is known that always  $G_u \ge 1$ and $G_u =1$ if and only if $u$ is a concave function. Moreover, when $u$ is differentiable, it is true that $G_u = \sup_{x \le y} u'(y)/ u'(x)$ (see \cite{Greediness}). In this regard, the index $G_u$ \textbf{globally} determines the maximal magnitude of the increase in the slope of the utility function (marginal utility) over its entire domain. Upon closer examination of the set condition of the utility space $\mathscr{U}^*_{\gamma}$, as provided in \eqref{eq:MullerUtilitySpacesNonSmooth}, it becomes evident that $\gamma$ serves as an upper bound for the level of greediness (non-concavity) $G_u$ for all $u \in \mathscr{U}^*_\gamma$, established by the inequality $G_u \leq 1/\gamma$. Inspired from \cite{Greediness}, we define a more local version of the greediness index as the following.

	\begin{definition}[\textbf{Discrete partial greediness}] \label{def:DiscreteLocalGreediness}
		Let $u$ be a strictly increasing function. The discrete partial greediness at location $x \in \R$ is defined as:
		
		\begin{equation}\label{eq:DiscreteLocalGreediness}
			G^d_{\text{par}}(u;x):=\sup_{x < x_1<x_2\le x_3<x_4}\left(\frac{u(x_4)-u(x_3)}{x_4-x_3}/\frac{u(x_2)-u(x_1)}{x_2-x_1}\right).
		\end{equation}
	\end{definition}

	\begin{proposition}\label{prop:DiscreteLocalGreedinessProperties}
		\begin{itemize}
			\item[(a)] Clearly, $G^d_{\text{par}}(u;\cdot) \le G_u$ where the global index of the greediness $G_u$ is given by \eqref{eq:global_greediness}.
			\item[(b)] Function $x \mapsto G^d_{\text{par}}(u;x)$ is non-increasing, i.e. $G^d_{\text{par}}(u;x_2) \le G^d_{\text{par}}(u;x_1)$ for $x_1 \le x_2$. Moreover, $G^d_{\text{par}}(u;x) \to G_u$ as $x \to -\infty$.
			\item[(c)] For every $x$, $G^d_{\text{par}}(u;x) \ge 1$ and, $G^d_{\text{par}}(u;x)=1$ if and only if $u$ is concave on $(x,\infty)$. In addition, $u$ is concave on $\R$ if and only if $G^d_{\text{par}}(u;x)=1$ for every $x\in\R$.
			\item[(d)] Function $x \mapsto G^d_{\text{par}}(u;x)$ is right continuous.
		\end{itemize}
	\end{proposition}

	\begin{remark}\label{rem:DiscreteLocalGreediness}
		\begin{itemize}
			\item[(i)] From a mathematical perspective, $G^d_{\text{par}}(u;x)$ is a measure of  non-concavity over half-open intervals $(x, \infty)$.
			\item[(ii)] In the context of decision theory, $G_u$ serves as a global measure, capturing the maximal proportional increase in a decision maker's valuation of an additional cent as wealth transitions from lower to higher values. Hence, $G_u$ takes into account all possible values of wealth. This makes it a natural notion for measuring the global greediness of a decision maker over $\R$. Building on the concept of $G_u$, for a fixed wealth level $x\in \R$, the measure $G^d_{\text{par}}(u;x)$ quantifies the maximum amplification of a decision maker's valuation within a specific region — specifically when their wealth exceeds $x$, i.e. for $(x,\infty)$. We direct the reader to \cite{maoCharacterizationsRiskAversion2019} for related concepts of local greediness within the framework of cumulative prospect theory.
			\item[(iii)] The discrete partial greediness $G^{d}_{\text{par}}$ can be well-defined for functions not necessary being strictly increasing. For example, it fully makes sense to talk about $G^{d}_{\text{par}}(u;\cdot)$ for functions $u \in \widetilde{\mathscr{U}}^{\textsc{ff}}_{\boldsymbol{\upgamma}}$ which contains functions which allows for non-strictly increasing segments. We will discuss it in a great details later in Remark \ref{rem:ContinuousLocalGreedinessConcaveGammaSppce}. 
		\end{itemize}
		
	\end{remark}

	Although the discrete partial greediness $G^d_{\text{par}}$ provides us with a measure of greediness attached to the location however it is not well suited to work with functions of our interest in the space $\widetilde{\mathscr{U}}^{\textsc{ff}}_{\boldsymbol{\upgamma}}$. This is due to the fact that functions in the latter space are defined via conditions on the one-sided derivatives. Therefore, in line with the approach followed at the beginning of this chapter, we introduce a new concept of local nature that is more suitable for the  $\widetilde{\mathscr{U}}^{\textsc{ff}}_{\boldsymbol{\upgamma}}$ structure. We present the following results in a more general context, specifically by requiring continuity and one-sided differentiability, instead of both-sided differentiability.
	


	\begin{definition}[\textbf{Continuous partial greediness}] \label{def:ContinuousLocalGreediness}
		Let $u$ be a strictly increasing function with right-derivatives exist at each point $x \in \R$, we define the continuous partial greediness at the location $x$ as follows
		\begin{equation}\label{eq:ContinuousLocalGreediness}
			G^c_{\text{par}}(u;x):= \sup_{x < y < z} \frac{u'_+(z)}{u'_+(y)}.
		\end{equation}
		The same definition applies to left-differentiable functions by utilizing left-derivatives.
	\end{definition}

	Next, we give the following proposition stating that two notions of discrete and continuous partial greediness coincide in more general framework.

	\begin{proposition}\label{prop:LocalGreedinessCoincide}
		Let $u$ be a continuous, strictly increasing and either left or right-derivatives exist at each point.   Then $$G^d_{par}(u,x)=G^c_{par}(u,x), \quad \text{ for all }   x \in \R.$$ 
	\end{proposition}

	\begin{remark}\label{rem:ContinuousLocalGreedinessAgreesTwoSidedDerivatives}
		\begin{itemize}	
			\item[(i)] There is no ambiguity in Definition \ref{def:ContinuousLocalGreediness} in the sense that when function $u$ is both-sided differentiable, then it immediately follows from the proof above that $G^{c}_{\text{par}} (u;\cdot)$ is independent of the choice of one-sided derivatives.	
			\item[(ii)]From the perspective of the minimal characterization of concavity via one-sided derivatives, $G^c_{\text{par}}(u;x)$ also serves as a measure of non-concavity, i.e. $G^c_{\text{par}}(u;x)=1$ if and only if $u$ is concave on $(x,\infty)$.
			
			\item[(iii)] It is worth to mention that the continuous partial greediness is invariant under any linear transformation.
		\end{itemize}
	\end{remark}

	Hereafter, in our framework, Proposition \ref{prop:LocalGreedinessCoincide} allows us to use notation $G_{\text{par}}(u;\cdot)$ interchangeably without any ambiguity.

	\begin{proposition} \label{prop:ContinuousLocalGreedinessImpliesCancavity}
		Let $u_1,\dots,u_n$ be $n$ strictly increasing continuous functions with either left or right-derivatives exist at each point $x \in \R$.  Then, for non-negative constants $\lambda_1,\dots, \lambda_n$, the following statements hold:
		\begin{itemize}
			\item[(a)] For every x, we have  $	G_{\text{par}} (\lambda_1 u_1+ \cdots+\lambda_n u_n ;x) \le \max_{1 \le k \le n}  G_{\text{par}}(u_k;x)$.
			\item[(b)] For every $x$, $G_{\text{par}}(\cdot;x)$ is a sub-linear functional in the sense that 
			\begin{equation*}
				G_{\text{par}} (\lambda_1 u_1+ \cdots+\lambda_n u_n;x) \le \sum_{k=1}^{n} G_{\text{par}}(u_k;x). 
			\end{equation*}	
		\end{itemize}
	\end{proposition}

	\begin{remark}\label{rem:ContinuousLocalGreedinessConcaveGammaSppce} In this remark, we delve into both notions of partial greediness within the context of the (utility) space $\widetilde{\mathscr{U}}^{\textsc{ff}}_{\boldsymbol{\upgamma}}$, where the assumption of strict increasing functions may not hold in general. For example, note that from the derivative-relation condition given in \eqref{eq:D_pm(gamma)Property} 
		\begin{equation} \label{eq:rem_set_condition}
			0\le \boldsymbol{\upgamma}  (y)  u_{-}' (y) \le u_{-}'(x), \, \quad \forall \, x\le y,	
		\end{equation}
		one can easily see that	if $u_-'(x)=0$, then $u_-'(y)=0$ for all $y>x$. Therefore, when  considering the ratio $u_-'(y)/u_-'(x)$, $ x\le y$, the only problematic scenario we may encounter is when we have a $0/0$ case. Hence, the convention $0/0=1$ removes all the ambiguities. Moreover, this does not affect the measurement of non-concavity, as the lower bound, $1$ is reached if (only if) the function is concave and constant functions are concave. In addition, due to convention above, Proposition \ref{prop:ContinuousLocalGreedinessImpliesCancavity} remains valid for the functions chosen from the utility space $\widetilde{\mathscr{U}}^{\textsc{ff}}_{\boldsymbol{\upgamma}}$. We will make use of this several instance later on in Section \ref{sec:MFSDUtilitySpaceLocalGreediness}.
	\end{remark}

	\begin{corollary}\label{cor:LocalGreedinessConcaveGammaSapce}
		Let $u \in \widetilde{\mathscr{U}}^{\textsc{ff}}_{\boldsymbol{\upgamma}}$. Then, for every $x$, $G^d_{par}(u;x)=G^c_{par}(u;x)=: G_{\text{par}}(u;x)$ and moreover,  
		\begin{equation}\label{eq:UpperBoundLocalGreediness}
			G_{\text{par}} (u;x) \le  \frac{1}{\boldsymbol{\upgamma}(x)}.
		\end{equation}
	\end{corollary}

	In the following remark, we highlight the decision theoretic implications of the $\boldsymbol{\upgamma}$ function.
	
	\begin{remark}\label{rem:LocalGreedinessGamma}
		\begin{itemize}
			\item[(i)]Analogous to the fixed $\gamma$ coefficient in fractional SD controlling the \textbf{global} index of greediness/non-concavity, in our context, both in the multi-fractional and functional fractional SD frameworks, the function $\boldsymbol{\upgamma}$ controls the \textbf{partial} greediness or non-concavity as per Corollary \ref{cor:LocalGreedinessConcaveGammaSapce}.
			\item[(ii)] From a mathematical point of view, the relation \eqref{eq:UpperBoundLocalGreediness} provides local upper bounds on the maximal steepness of the non-concave segments, depending on the location $x$. Under the assumption that $\boldgamma{\upgamma}$ is a non-decreasing, the utility function $u \in \widetilde{\mathscr{U}}^{\textsc{mf}}_{\boldsymbol{\upgamma}}\subseteq \widetilde{\mathscr{U}}^{\textsc{ff}}_{\boldsymbol{\upgamma}}$  is allowed to have less steeper non-concave segments as one moves to the right hand side of the real axis. This is perfectly in line with the local interpolation property of MFSD (see Example \ref{ex:local_interpolation}). Essentially, the non-decreasing feature of function  $\boldsymbol{\upgamma}$ enforces a stronger dominance (closer to FSD) on the lower
			values of the real line, while the dominance progressively weakens (closer to SSD) as we move towards the higher values on the real line.
			\item[(iii)] From a decision theoretic perspective, the bounding relation \eqref{eq:UpperBoundLocalGreediness} is meaningful and consistent with the realistic behaviour of decision makers. That is, the upper bound of partial greediness at wealth level $x$, i.e. the maximal possible proportional increase in valuing an additional cent as wealth grows higher than $x$, decreases as the wealth level $x$ increases. In other words, as the decision maker becomes wealthier, the maximum satisfaction he can get from an additional cent decreases.
			
		\end{itemize}
		
	\end{remark}

	\subsection{\texorpdfstring{$\widetilde{\mathscr{U}}^{\textsc{\fontsize{5}{8}\selectfont{MF}}}_{\boldsymbol{\upgamma}}$}{U-MF-gamma} and the partial greediness  } \label{sec:MFSDUtilitySpaceLocalGreediness}
	In this section, we provide several tools  in terms of the partial greediness $G_{\text{par}}(u;\cdot)$ to verify a given function $u \in \widetilde{\mathscr{U}}^{\textsc{ff}}_{\boldsymbol{\upgamma}}$ belongs to the multi-fractional utility space $\widetilde{\mathscr{U}}^{\textsc{mf}}_{\boldsymbol{\upgamma}}$. The underlying principle is based on the observation stated in Remark \ref{rem:InclusionIsStrict}, which notes that for a fixed  $\boldsymbol{\upgamma}$, the space $\widetilde{\mathscr{U}}^{\textsc{ff}}_{\boldsymbol{\upgamma}}$ encompasses more non-concave functions compared to $\mathscr{U}^{\textsc{mf}}_{\boldsymbol{\upgamma}}$.   Therefore, for a given function $u$, we employ the partial greediness to evaluate its non-concavity and subsequently determine whether it belongs to $\widetilde{\mathscr{U}}^{\textsc{mf}}_{\boldsymbol{\upgamma}}$. 
	First, we start with the following lemma.
	
	\begin{lemma} \label{lem:ExtremeComponentMustExist}
		Let $\boldsymbol{\upgamma}:\R \to [0,1]$ be an arbitrary non-decreasing function. Let $u=\sum_{k=1}^{n} \lambda_k u_k \in \widetilde{\mathscr{U}}^{\textsc{mf}}_{\boldsymbol{\upgamma}}$, $n\in \N$, $\lambda_k \ge 0$ and $u_k \in \widetilde{\mathscr{U}}^{\text{Union}}_{\boldsymbol{\upgamma}}$, $k=1,\dots,n$. Assume that there exists $x_0 \in \R$ such that $\boldsymbol{\upgamma}(x_0) \neq 0$, and moreover 
		
		\begin{equation}\label{eq:ExtremeComponentMustExist}
			G_{\text{par}}(u;x_0) = \frac{1}{\boldsymbol{\upgamma}(x_0)}.
		\end{equation}	
		Then, there exists an index $1 \le j \le n$ so that
		\begin{equation*}
			G_{\text{par}}(u;x_0) = G_{\text{par}}(u_j;x_0) =\max_{1 \le k\le n} G_{\text{par}}(u_k;x_0) =\frac{1}{\boldsymbol{\upgamma}(x_0)}.
		\end{equation*}	
	\end{lemma}

	The main message of Lemma \ref{lem:ExtremeComponentMustExist} is that for $u \in \widetilde{\mathscr{U}}^{\textsc{mf}}_{\boldsymbol{\upgamma}}$, $G_{\text{par}}(u;x)$ touches the upper bound given in \eqref{eq:UpperBoundLocalGreediness} at $x_0$ if and only if $u$ has a extreme component belonging to $\widetilde{\mathscr{U}}^{x_0}_{\boldsymbol{\upgamma}(x_0)} \subsetneq \widetilde{\mathscr{U}}^{\text{Union}}_{\boldsymbol{\upgamma}}$. We proceed with the following proposition, which asserts that for any $u \in \widetilde{\mathscr{U}}^{\textsc{mf}}_{\boldsymbol{\upgamma}}$, the upper bound for greediness (non-concavity) is reached only when $\boldgamma{\upgamma}$ is a constant function around that point. The usefulness of this proposition becomes evident in the subsequent corollary, stating that if $G_{\text{par}}(u;x)$ reaches the upper bound at some point and $\boldgamma{\upgamma}$ is not constant around that point, then $u$ cannot belong to $\widetilde{\mathscr{U}}^{\textsc{mf}}_{\boldsymbol{\upgamma}}$. It is important to note that such conditions are not required for functions in $\widetilde{\mathscr{U}}^{\textsc{ff}}_{\boldsymbol{\upgamma}}$.

	\begin{proposition}\label{prop:GammamustBeConstant}
		Let $\boldsymbol{\upgamma}:\R \to [0,1]$ be an arbitrary non-decreasing function.  Assume that $u \in \widetilde{\mathscr{U}}^{\textsc{mf}}_{\boldsymbol{\upgamma}}$ and  there exists $x_0 \in \R$ such that $\boldsymbol{\upgamma}(x_0) \neq 0$, and moreover 
		
		\begin{equation}\label{eq:GammamustBeConstant}
			G_{\text{par}}(u;x_0) = \frac{1}{\boldsymbol{\upgamma}(x_0)}.
		\end{equation}	
		Then, there exists an open interval $I=I_{x_0} = (x_0,x_0+\delta)$ for some $\delta >0$ such that $\boldsymbol{\upgamma}$ is a constant function on $I$.
	\end{proposition}

	\begin{corollary}\label{cor:BigHammerCor1}
		Let $\boldsymbol{\upgamma}:\R \rightarrow [0,1]$ be an arbitrary non-decreasing function such that there exists an open subset $U \subseteq \R$ where $\boldsymbol{\upgamma}$ is strictly increasing on $U$.  Assume that $u \in \widetilde{\mathscr{U}}^{\textsc{ff}}_{\boldsymbol{\upgamma}}$ and there exists $x_0 \in U$ such that 
		\begin{equation}\label{eq:gloc_equal_x_0}
			G_{\text{par}}\left(u; x_0\right)=\frac{1}{\boldgamma{\upgamma}(x_0)}.
		\end{equation}
		Then,  $u \notin\widetilde{\mathscr{U}}^{\textsc{mf}}_{\boldsymbol{\upgamma}}$.
	\end{corollary}
	
	Next, we present the main result of this section, which states that if, for a given $u \in \widetilde{\mathscr{U}}^{\textsc{ff}}_{\boldsymbol{\upgamma}}$, the partial greediness $G_{\text{par}}\left(u; x\right)$ touches the upper bound for at least two distinct value of function $\boldgamma{\upgamma}$, then $u \not\in \widetilde{\mathscr{U}}^{\textsc{mf}}_{\boldsymbol{\upgamma}}$. This follows because, for $u \in \widetilde{\mathscr{U}}^{\textsc{mf}}_{\boldsymbol{\upgamma}}$, once $G_{\text{par}}\left(u; x\right)$ reaches its upper bound at a certain point $x$ due to the extreme component (as per Lemma \ref{lem:ExtremeComponentMustExist}), $u$ must remain constant thereafter. If not, the linear structure of $\widetilde{\mathscr{U}}^{\textsc{mf}}_{\boldsymbol{\upgamma}}$ implies that combining this extreme component with other functions that are not constant after $x$—i.e., those unable to exhibit as steep non-concave segments—would detract the degree of non-concavity of $u$ at $x$. Conversely, by the definition of the $\boldsymbol{D}_\pm(\boldsymbol{\upgamma})$ property, this does not apply to the functions within $\widetilde{\mathscr{U}}^{\textsc{ff}}_{\boldsymbol{\upgamma}}$.

	\begin{theorem}\label{thm:BigHammer}
		Let $\boldsymbol{\upgamma}:\R \to [0,1]$ be an arbitrary non-decreasing function. Assume that $u \in \widetilde{\mathscr{U}}^{\textsc{ff}}_{\boldsymbol{\upgamma}}$ and  there exist $x_0, x_1 \in \R$ such that $0 < \boldsymbol{\upgamma}(x_0) \neq \boldsymbol{\upgamma}(x_1)<1$. Let
		
		\begin{equation}\label{eq:BigHammer}
			G_{\text{par}}(u;x_k) = \frac{1}{\boldsymbol{\upgamma}(x_k)}, \quad k=0,1.
		\end{equation}
		Then, $u \notin \widetilde{\mathscr{U}}^{\textsc{mf}}_{\boldsymbol{\upgamma}}$.
	\end{theorem}

	We conclude this section with the following corollary that adds the last piece to the overall framework. From Proposition \ref{prop:GammamustBeConstant} we already know that for a given $u$, if $G_{\text{par}}(u;x_0)$ reaches the upper bound around a point where $\boldgamma{\upgamma}$ is strictly increasing, then $u$ cannot belong to $\widetilde{\mathscr{U}}^{\textsc{mf}}_{\boldsymbol{\upgamma}}$. Furthermore, this corollary establishes that when $\boldgamma{\upgamma}$ is a constant function almost everywhere, and $G_{\text{par}}(u;x_0)$ touches the upper bound for more than one value of $\boldgamma{\upgamma}$, then $u$ does not lie in $\widetilde{\mathscr{U}}^{\textsc{mf}}_{\boldsymbol{\upgamma}}$.

	\begin{corollary}\label{cor:BigHammerCor2}
		Let $\boldsymbol{\upgamma}:\R \to [0,1]$ be a piecewise step function of the form given in \eqref{eq:GammaStepFunction}. Assume that $u \in \widetilde{\mathscr{U}}^{\textsc{ff}}_{\boldsymbol{\upgamma}}$ and  there exist $x_i \in (t_{i-1},t_i)$ and $x_j \in (t_{j-1},t_j)$, where $1 \le i \neq j \le n$ such that 
		
		\begin{equation*}
			G_{\text{par}} (u;x_k) = \frac{1}{\boldsymbol{\upgamma}(x_k)}, \quad k=i,j.
		\end{equation*}	
		Then, $u\notin \widetilde{\mathscr{U}}^{\textsc{mf}}_{\boldsymbol{\upgamma}}$. 
	\end{corollary}

		\subsection{Average risk aversion}
		
		In the preceding sections, we explored the implications of the function $\boldgamma{\upgamma}$ from an economics perspective, emphasizing non-differentiable utility functions. Now, our focus shifts to differentiable utility functions, specifically strictly increasing twice-differentiable utility functions within $\widetilde{\mathscr{U}}^{\textsc{ff}}_{\boldsymbol{\upgamma}}$. In this context, function $\boldgamma{\upgamma}$ can be interpreted as a somewhat local lower bound to the average risk aversion (or upper bound to risk lovingness) behavior of a decision maker. Recall the well known \textit{Arrow–Pratt measure} of absolute risk aversion (ARA) coefficient:
		\begin{equation*}
			r_u(x) = r(x): =-\frac{u^{\prime\prime}(x)}{u^\prime(x)}, \quad x\in \R.
		\end{equation*}
		Then, one can easily deduce a local lower bounds on the \textit{average absolute risk aversion coefficient} as 
		\begin{equation}\label{eq:lowerboundARA_y}
			\int_x^y -\frac{u^{\prime\prime}(z)}{u^\prime(z)}dz= \left[-\ln(u^\prime(y))+\ln(u^\prime(x))\right]=\ln\bigg(\frac{u^\prime(x)}{u^\prime(y)}\bigg) \geq \ln(\boldsymbol{\upgamma}(y)), \quad \forall x\le y.
		\end{equation}	
		When $\boldsymbol{\upgamma}(y)=0$, the above estimate does not provide any insight as the quantity $\ln (\boldsymbol{\upgamma}(y)) = -\infty$.  Obviously, when $u$ is fully concave the latter inequality always is in place (one side is negative and the other side is positive). However, in more interesting cases namely when utility function $u$ allows for some local non-concavity, the relation   \eqref{eq:lowerboundARA_y} indicates that function $\boldsymbol{\upgamma}$ controls (on average) how much function $u$ can deviate of being locally risk averse. Hence, for lower values of wealth $y$, both utility spaces $\widetilde{\mathscr{U}}^{\textsc{ff}}_{\boldsymbol{\upgamma}}$ and $\widetilde{\mathscr{U}}^{\textsc{mf}}_{\boldsymbol{\upgamma}}$ allows the utility functions to represent more risk loving behaviour due to the non-decreasing property of $\boldsymbol{\upgamma}$. This is consistent with \cite{FriedmanSavage} and \cite{Markowitz} models where investors are more risk averse (buying insurance) when they have more wealth than their present wealth. Looking at it from the viewpoint of decision theory, this implies that both FFSD and MFSD orders distributions such that not only do all risk averse decision makers agree on the ordering, but also less risk averse ones, who are willing to take some risks based on their wealth level. This is in contrast to fractional SD, where the degree of risk behavior remains independent of wealth. Thus, the value of $\boldgamma{\upgamma}(y)$ can be interpreted as the smallest average risk aversion degree (or maximum risk lovingness) at wealth level $y$, required to ensure agreement on the ordering between risk averse and those who are risk loving to some extent.\\

		Another interesting observation is that when function $\boldsymbol{\upgamma}$ is continuous and by a direct application of the mean value theorem, in addition, we can observe that: for every $x$ and every $y$ in a tiny neighbourhood of $x$ that 
		
		\begin{equation*}
			\frac{\log \left(  \boldsymbol{\upgamma} (x)\right )   }{y-x}\lessapprox r_u(x), \quad x \neq y.
		\end{equation*}
		The latter heuristic observation interprets that function $\boldsymbol{\upgamma}$ at each instance controls the non-concavity of utility function $u$ through the ARA coefficient.  Similarly, one can discuss the relative absolute risk aversion coefficient  $R_{u}(x)=-x\frac{u^{\prime\prime}(x)}{u^\prime(x)}$ as well. \\
		
		In conclusion, it is evident that both the MFSD and FFSD methods utilize extra information regarding the position (wealth) of non-concave segments in the utility function $u$. Consequently, this approach enables a more comprehensive and realistic representation of decision makers' behavior, aligning with the Expected Utility theory. 

		\section{Functional almost stochastic dominance}\label{sec:func_ASD}
		
		In this section, we extend the concept of Almost Stochastic Dominance (ASD), introduced in \cite{almost_SD}. This extension employs a combined concave/ convex approach, paralleling the methodologies discussed in \cite{muller} within the framework of FFSD. Before proceeding, let us first recall the definition of the $\varepsilon$\textsc{-ASD} utility space consisting of smooth functions,
		\begin{equation*}
			\mathscr{U}_{\varepsilon\textsc{-asd}}  = \Bigg\{ u :\R \to \R \in C^{1}(\R) :   \,     
			0 \le u^\prime(x) \le \inf_{x\in \R}u^\prime(x) \left[\frac{1-\varepsilon}{\varepsilon}\right],  \, \forall   x \Bigg\}   
		\end{equation*}
		where $\varepsilon\in (0,1/2)$. Inspired by  the derivative-relation conditions given in \eqref{eq:D_pm(gamma)Property}, we obtain the following natural functional extension of the utility set $\mathscr{U}_{\varepsilon\textsc{-asd}}$ through  
		\begin{align}
			&(i)\, 0 \leq u'_-(x) \leq \inf_{x \in \R} u'_-(x) \left[\frac{1 - \boldsymbol{\upepsilon}(x)}{\boldsymbol{\upepsilon}(x)}\right], \quad \forall x, \label{eq:almost_(i)} \\
			&(ii)\, 0 \leq u'_+(x) \leq \inf_{x \in \R} u'_+(x) \left[\frac{1 - \boldsymbol{\upepsilon}(x)}{\boldsymbol{\upepsilon}(x)}\right], \quad \forall x \label{eq:almost_(ii)}
		\end{align}  
		where $\boldsymbol{\upepsilon}:\R \to (0,1/2)$.  Accordingly, we define the following property.
		
		\begin{definition}[$\boldsymbol{D}^{\textsc{asd}}_\pm(\boldsymbol{\upepsilon})$\textbf{-property}]\label{def:D_pm(epsion)Property} Let $\boldsymbol{\upepsilon}:\R \to (0,1/2)$ be an arbitrary function. We say that both-sided differentiable,  function $u:\R \to \R$ satisfies in $\boldsymbol{D}^{\textsc{asd}}_\pm(\boldsymbol{\upepsilon})$ if it is both-sided differentiable and at least one of the conditions given in \eqref{eq:almost_(i)} and \eqref{eq:almost_(ii)} holds.
		\end{definition}
		
		Now, we are ready for the formal definition.
		
		\begin{definition}[\textbf{Functional  $\boldsymbol{\upepsilon}$\textsc{-ASD}}] \label{def:fucnitonal_ASD} Let $\boldsymbol{\upepsilon}:\R \to (0,1/2)$ be an arbitrary function. Let $F$ and $G$ be two distribution functions. We say that $G$ dominates $F$ in the sense of the functional  $\boldsymbol{\upepsilon}$\textsc{-ASD}, denote by $F \preceq_{\boldsymbol{\upepsilon}\textsc{-ASD}} G$ if 
			\begin{equation}\label{eq:functional_ASD_utiltiy_contion}
				\E_F  \left[  u \right]  \le \E_G \left[ u \right], \quad \forall \, u \in \widetilde{\mathscr{U}}^{\textsc{asd}}_{\boldsymbol{\upepsilon}}
			\end{equation}	
			where \begin{equation}\label{eq:functional_asd_utility_class}
				\widetilde{\mathscr{U}}^{\textsc{asd}}_{\boldsymbol{\upepsilon}}   := \Big\{   u: \R \to \R  \, :\,    \boldsymbol{D}^{\textsc{asd}}_\pm(\boldsymbol{\upepsilon}) \text{ property holds}    \Big\}                
			\end{equation}	
			provided that expectations exist.
		\end{definition}
		
		Following the approach used in the proof of Proposition \ref{prop:mf_one_sided}, we can demonstrate that setting $\boldsymbol{\upepsilon}\equiv \varepsilon \in (0,1/2)$ results in the equality $\widetilde{\mathscr{U}}^{\textsc{asd}}_{\boldsymbol{\upepsilon}} = \widetilde{\mathscr{U}}^*_{\varepsilon\textsc{-asd}}$. Here, $\mathscr{U}^*_{\varepsilon\textsc{-asd}}$ represents the set of non-differentiable utility functions associated with $\varepsilon$-ASD. Furthermore, a close examination of conditions \eqref{eq:almost_(i)} and \eqref{eq:almost_(ii)} shows that the function $\boldsymbol{\upepsilon}$ controls the maximal possible jump in the one-sided derivatives at a given location $x$. This control is highly localized, meaning its impact does not extend to values other than $x$, unlike other orders introduced so far. For instance, at those points where $\boldsymbol{\upepsilon}(x)\approx 0$, utility functions in $\widetilde{\mathscr{U}}^{\textsc{asd}}_{\boldsymbol{\upepsilon}}$ may exhibit greater jumps in their one-sided derivatives compared to points where $\boldsymbol{\upepsilon}(x)\approx \frac{1}{2}$, at which the  one-sided derivatives are constrained to be closer to their infimum. Also, note that any $u \in \widetilde{\mathscr{U}}^{\textsc{asd}}_{\boldsymbol{\upepsilon}}$ is either a constant function or strictly increasing.  It is evident that, for any chosen function $\boldsymbol{\upepsilon}:\mathbb{R} \to (0,1/2)$, the set $\widetilde{\mathscr{U}}^{\textsc{asd}}_{\boldsymbol{\upepsilon}}$ is non-empty and it is easy to verify whether a given utility function belong to it. Moreover, it is always possible to choose an appropriate $\boldsymbol{\upepsilon}$ for specific purposes, without restricting the selection to solely monotone functions. Next, we present the "if and only if" theorem along with the corresponding integral condition for a given $\boldsymbol{\upepsilon}:\mathbb{R} \to (0,1/2)$, as follows:
		\begin{equation}\label{eq:functional_ASD_integral}
			\int_{-\infty}^\infty \frac{1}{\boldsymbol{\upepsilon}(x)} (F(x)-G(x))_-dx \le \int_{-\infty}^\infty  |(F(x)-G(x))|dx.
		\end{equation}
		
		Before proceeding, we give the following remark, which highlights a few important points.
		
		\begin{remark} \begin{itemize}
				\item[(i)] We give the proof without assuming monotonicity on $\boldsymbol{\upepsilon}$ and avoid arguments relying on the first crossing point of given distribution functions. To ensure the integral condition is meaningful and avoid potential technical difficulties, we consider a restricted subset of pairs of distribution functions.  For a given $\boldsymbol{\upepsilon}:\mathbb{R} \to (0,1/2)$, we define:
				\begin{equation*}
					\operatorname{L}^1(\boldsymbol{\upepsilon}):=\left\{F,G\in \mathscrbf{D}(\R):\int_{-\infty}^{\infty}\left|\frac{1}{\boldsymbol{\upepsilon}(x)} (F(x)-G(x))_-\right|dx<\infty\right\}.
				\end{equation*} 
				\item[(ii)] The integral condition \eqref{eq:functional_ASD_integral} can be equivalently represented as:
				\begin{equation}\label{eq:functional_ASD_integral_alternative}
					\int_{-\infty}^\infty \frac{1-\boldsymbol{\upepsilon}(x)}{\boldsymbol{\upepsilon}(x)} (F(x)-G(x))_-dx \le \int_{-\infty}^\infty  (F(x)-G(x))_+dx.
				\end{equation}

				\item[(iii)] Building on the approach detailed in item (ii) of Remark \ref{rem:just_before_the_proof_FFSD} for FFSD, we prove one direction of the theorem under the additional assumptions regarding both $\boldgamma{\upepsilon}$ and the crossing points of the given distribution functions.

				\item[(iv)] To focus on essential aspects of the proof and avoid unnecessary discussion, we assume $\E_F[|u|],\E_G[|u|]<\infty$ for all $u \in \widetilde{\mathscr{U}}^{\textsc{asd}}_{\boldsymbol{\upepsilon}}$.
			\end{itemize}	
		\end{remark}
		
		\begin{theorem}\label{thm:FASD_iff} Let $\boldsymbol{\upepsilon}:\mathbb{R} \to (0,1/2)$ be an arbitrary right-continuous function with finitely many jumps and one-sided limits at each point. Let $F$ and $G$ be distribution functions in $\operatorname{L}^1(\boldsymbol{\upepsilon})$ crossing finitely many times. Then, $F \preceq_{\boldsymbol{\upepsilon}\textsc{-ASD}} G$ if and only if the integral condition \eqref{eq:functional_ASD_integral} holds.
		\end{theorem}

		We conclude that the utility class $\widetilde{\mathscr{U}}^{\textsc{asd}}_{\boldsymbol{\upepsilon}}$ admits an integral condition, leading to the functional $\boldsymbol{\upepsilon}$\textsc{-ASD} as a meaningful ordering relation. This extension could be beneficial for broader applications and interpretations in various contexts, especially in scenarios where traditional $\varepsilon$\textsc{-ASD} falls short.

		\section{Conclusion}\label{sec:conclusion}
		
		In this paper, we propose a two-fold generalization of fractional stochastic dominance by replacing the fixed parameter $\gamma$ with a function $\boldsymbol{\upgamma}:\mathbb{R} \rightarrow [0,1]$. This leads to the introduction of two novel families, MFSD and FFSD. These generalizations effectively interpolate between FSD and SSD, addressing the limitations of both. Unlike fractional SD, which offers only global interpolation, our approach enables more informative local interpolation. This distinction allows MFSD and FFSD to rank distributions that are otherwise indistinguishable under fractional SD.  Moreover, the corresponding utility classes consist of functions with local non-concavities, where the steepness of the convexities is controlled by the function  $\boldgamma{\upgamma}$.\\
		
		Another contribution of our framework is the introduction of partial greediness, which captures the evolution of greediness behavior with changes in wealth. As wealth increases, decision makers exhibit diminishing greediness, reflected in the non-increasing bounds defined by $\boldsymbol{\upgamma}$. Partial greediness also serves as a practical criterion for identifying whether a utility function falls under MFSD or FFSD. Additionally, we extend the same analysis to risk taking behaviour through the bounding relationship of average risk aversion, offering insights that fractional SD cannot provide due to its reliance on a fixed $\gamma$. Lastly, we provide a generalized family of FASD that extends the results of ASD.\\
		
		We introduced a novel framework for understanding choice under uncertainty, offering a more comprehensive and flexible approach than traditional methods. Our extension presents a unique methodology that has contributed to the literature, opening up new avenues for research and practical applications across various fields. Future studies could investigate the application of these generalizations in specific economic models, such as portfolio optimization and behavioural economics, and explore their implications for decision making under uncertainty. There is also significant potential for applications in actuarial science, particularly in risk assessment, where the ability to compare risk distributions is critical. 
		

\addcontentsline{toc}{section}{Appendices}

\section*{Appendix}

\appendix

\addtocontents{toc}{\protect\setcounter{tocdepth}{0}}
\section{{Construction of Example \ref{ex:3}}}

First, $F \preceq^{\textsc{mf}}_{ (1+ \boldgamma{\upgamma})\textsc{\textsc{-SD}}} G$ but $F \npreceq_{(1+\gamma)\textsc{\textsc{-SD}}} G$. We discuss separately the following two possible cases:

\noindent \textbf{Case 1:}  There is a left-side neighbourhood $N_\delta(t_0) = (t_0-\delta,t_0)$, $\delta >0$ such that function $\boldgamma{\upgamma} (t)=\boldgamma{\upgamma}(t_0)$ for all $t \in N_\delta (t_0)$. Choose $\ell < \delta$ small enough (in the sense that $\boldsymbol{\upgamma} (t_0) - \ell > \gamma$). Now, consider the square with the length $\ell$ and with the right-up corner at $(t_0,\boldsymbol{\upgamma}(t_0))$. Let's denote the square with $S(\ell)$. Next, there are two natural numbers $M, N \in \N$ so that $\gamma <  M/N  <  \boldgamma{\upgamma}(t_0)$ (rational numbers are dense in $\R$).  Without loosing of generality, we can assume that $(M+N)/MN < \ell^{-1}$ (otherwise we work with $KM$, and $KN$ with large $K$ integer. This keeps the ratio unchanged however note that $KN +KM/(KM)(KN) = (1/K) \times (M+N)/(MN) \to 0$ as $K \to \infty$).   Now, we equally divide $S(\ell)$ into $(MN)^2$ little squares, each with a side length of $\ell/(MN)$ and denote it as $\ell^*$. Let's also denote the $t$-coordinate of the left-bottom corner of $S(\ell)$ by $t^* $ such that $t^* + 2\ell^* \in N_\delta(t_0)$ and define 		
\begin{align*}
	F(t)   =& 
	\begin{cases}
		0   & \quad \quad \quad \mbox { if }    t <t^*\\
		N \ell^*  & \quad \quad \quad \mbox{ if }   t^* \le t < t^* + 2\ell^* \\
		1 & \quad \quad \quad \mbox { if }   t \ge t^* + 2\ell^*
	\end{cases}	\\
	G(t)  =&
	\begin{cases}
		0 & \mbox { if }  t <t^* + \ell^*\\
		(M+N) \ell^* & \mbox { if }   t^* + \ell^* \le t < t^* + 2\ell^* \\
		1 & \mbox { if }  t \ge t^* + 2\ell^* .
	\end{cases}	
\end{align*} 
Then, we have 

\begin{equation*}
	A_+:= \int_{t^*}^{t^* + \ell^*} \left(  F(x) - G(x) \right)_+ dx = N (\ell^*)^2\text{ and }\,
	A_- : = \int_{t^* + \ell^*}^{t^* + 2\ell^*}  \left(  F(x) - G(x) \right)_- dx = M (\ell^*)^2.
\end{equation*}	
Therefore, we have 
$F \preceq^{\textsc{mf}}_{ (1+ \boldgamma{\upgamma})\textsc{-SD}} G$ but $F \not\preceq_{(1+\gamma)\textsc{\textsc{-SD}}} G$ since
$$\gamma < \frac{A_-}{A_+}= \frac{M}{N} <  \boldsymbol{\upgamma}(t^*+\ell^*)=\boldsymbol{\upgamma}(t^*+2\ell^*)=\boldsymbol{\upgamma} (t_0).$$

\begin{figure}[H]

	\tikzset{every picture/.style={line width=0.75pt}} 
	
	\begin{tikzpicture}[x=0.75pt,y=0.75pt,yscale=-0.81,xscale=0.87]
		
		\draw  (25,245.17) -- (337.23,245.17)(56.22,123) -- (56.22,258.74) (330.23,240.17) -- (337.23,245.17) -- (330.23,250.17) (51.22,130) -- (56.22,123) -- (61.22,130)  ;
		\draw    (56.98,201.49) -- (253.48,201.74) ;
		\draw  [dash pattern={on 0.84pt off 2.51pt}]  (213.01,148.81) -- (213,224.43) -- (213,243.5) ;
		\draw    (136.01,148.81) -- (213.01,148.81) ;
		\draw   (138.01,264.2) .. controls (138.05,268.87) and (140.4,271.18) .. (145.07,271.13) -- (165.57,270.94) .. controls (172.24,270.88) and (175.59,273.18) .. (175.64,277.85) .. controls (175.59,273.18) and (178.9,270.82) .. (185.57,270.75)(182.57,270.78) -- (206.07,270.56) .. controls (210.74,270.51) and (213.05,268.16) .. (213,263.49) ;
		\draw  [dash pattern={on 0.84pt off 2.51pt}]  (136.01,148.81) -- (136,224.43) -- (136,243.5) ;
		\draw    (184.01,149.15) -- (184.01,173.15) ;
		\draw    (184.01,173.15) -- (212.01,173.15) ;
		\draw   (211.67,148.81) .. controls (211.67,148.08) and (212.27,147.48) .. (213.01,147.48) .. controls (213.74,147.48) and (214.34,148.08) .. (214.34,148.81) .. controls (214.34,149.55) and (213.74,150.15) .. (213.01,150.15) .. controls (212.27,150.15) and (211.67,149.55) .. (211.67,148.81) -- cycle ;
		\draw  [line width=1.5] [line join = round][line cap = round] (213.78,149.24) .. controls (213.65,148.84) and (212.91,148.8) .. (212.53,148.99) ;
		\draw  [dash pattern={on 4.5pt off 4.5pt}] (165.51,123.48) -- (233.2,123.48) -- (233.2,191.17) -- (165.51,191.17) -- cycle ;
		\draw  [dash pattern={on 4.5pt off 4.5pt}]  (233.2,122.63) -- (410.89,64.31) ;
		\draw  [dash pattern={on 4.5pt off 4.5pt}]  (233.2,191.17) -- (413.14,244.31) ;
		\draw    (420.42,220.37) -- (569.43,220.38) ;
		\draw    (419.93,71.88) -- (568.93,71.9) ;
		\draw    (419.93,71.88) -- (420.42,220.37) ;
		\draw    (568.93,71.9) -- (569.43,220.38) ;
		\draw  [fill={rgb, 255:red, 0; green, 0; blue, 0 }  ,fill opacity=1 ] (565.77,71.9) .. controls (565.77,70.15) and (567.19,68.74) .. (568.93,68.74) .. controls (570.68,68.74) and (572.09,70.15) .. (572.09,71.9) .. controls (572.09,73.64) and (570.68,75.06) .. (568.93,75.06) .. controls (567.19,75.06) and (565.77,73.64) .. (565.77,71.9) -- cycle ;
		\draw   (421.08,231.45) .. controls (421.17,236.12) and (423.54,238.41) .. (428.21,238.32) -- (430.29,238.28) .. controls (436.96,238.16) and (440.33,240.43) .. (440.42,245.1) .. controls (440.33,240.43) and (443.62,238.04) .. (450.29,237.92)(447.29,237.97) -- (452.38,237.88) .. controls (457.05,237.79) and (459.34,235.42) .. (459.25,230.75) ;
		\draw [line width=1.5]    (420.83,151.18) -- (500.01,151.56) ;
		\draw   (460.83,230.45) .. controls (460.97,235.12) and (463.37,237.38) .. (468.04,237.24) -- (470.76,237.15) .. controls (477.42,236.95) and (480.82,239.18) .. (480.96,243.85) .. controls (480.82,239.18) and (484.08,236.75) .. (490.75,236.55)(487.75,236.64) -- (493.47,236.46) .. controls (498.13,236.32) and (500.39,233.92) .. (500.25,229.25) ;
		\draw [color={rgb, 255:red, 208; green, 2; blue, 27 }  ,draw opacity=1 ][line width=1.5]    (459.25,122.55) -- (500.01,123.04) ;
		\draw  [dash pattern={on 0.84pt off 2.51pt}]  (459.25,122.55) -- (459.25,220.38) ;
		\draw  [color={rgb, 255:red, 0; green, 0; blue, 0 }  ,draw opacity=0 ][fill={rgb, 255:red, 208; green, 2; blue, 27 }  ,fill opacity=0.27 ] (420.64,151.56) -- (459.06,151.56) -- (459.06,220.75) -- (420.64,220.75) -- cycle ;
		\draw  [color={rgb, 255:red, 0; green, 0; blue, 0 }  ,draw opacity=0 ][fill={rgb, 255:red, 74; green, 144; blue, 226 }  ,fill opacity=0.3 ] (459.06,123.04) -- (500.01,123.04) -- (500.01,151.56) -- (459.06,151.56) -- cycle ;
		\draw [color={rgb, 255:red, 208; green, 2; blue, 27 }  ,draw opacity=1 ][line width=1.5]    (420.42,220.37) -- (459.25,220.38) ;
		\draw  [fill={rgb, 255:red, 255; green, 255; blue, 255 }  ,fill opacity=1 ] (497.01,151.56) .. controls (497.01,149.9) and (498.35,148.56) .. (500.01,148.56) .. controls (501.66,148.56) and (503.01,149.9) .. (503.01,151.56) .. controls (503.01,153.22) and (501.66,154.56) .. (500.01,154.56) .. controls (498.35,154.56) and (497.01,153.22) .. (497.01,151.56) -- cycle ;
		\draw  [dash pattern={on 0.84pt off 2.51pt}]  (500.01,154.56) -- (500,219) ;
		\draw  [color={rgb, 255:red, 208; green, 2; blue, 27 }  ,draw opacity=1 ][fill={rgb, 255:red, 255; green, 255; blue, 255 }  ,fill opacity=1 ] (456.25,220.38) .. controls (456.25,218.72) and (457.59,217.38) .. (459.25,217.38) .. controls (460.91,217.38) and (462.25,218.72) .. (462.25,220.38) .. controls (462.25,222.03) and (460.91,223.38) .. (459.25,223.38) .. controls (457.59,223.38) and (456.25,222.03) .. (456.25,220.38) -- cycle ;
		\draw  [color={rgb, 255:red, 208; green, 2; blue, 27 }  ,draw opacity=1 ][fill={rgb, 255:red, 255; green, 255; blue, 255 }  ,fill opacity=1 ] (497.01,123.04) .. controls (497.01,121.38) and (498.35,120.04) .. (500.01,120.04) .. controls (501.66,120.04) and (503.01,121.38) .. (503.01,123.04) .. controls (503.01,124.7) and (501.66,126.04) .. (500.01,126.04) .. controls (498.35,126.04) and (497.01,124.7) .. (497.01,123.04) -- cycle ;
		\draw  [color={rgb, 255:red, 208; green, 2; blue, 27 }  ,draw opacity=1 ][fill={rgb, 255:red, 208; green, 2; blue, 27 }  ,fill opacity=1 ] (456.25,122.55) .. controls (456.25,120.89) and (457.59,119.55) .. (459.25,119.55) .. controls (460.91,119.55) and (462.25,120.89) .. (462.25,122.55) .. controls (462.25,124.21) and (460.91,125.55) .. (459.25,125.55) .. controls (457.59,125.55) and (456.25,124.21) .. (456.25,122.55) -- cycle ;
		\draw   (425,77) -- (482.75,77) -- (482.75,107.25) -- (425,107.25) -- cycle ;
		\draw [color={rgb, 255:red, 208; green, 2; blue, 27 }  ,draw opacity=1 ][line width=1.5]    (451.75,85.8) -- (477,86) ;
		\draw [color={rgb, 255:red, 0; green, 0; blue, 0 }  ,draw opacity=1 ][line width=1.5]    (451.25,98.55) -- (476.5,98.75) ;
		\draw  [color={rgb, 255:red, 0; green, 0; blue, 0 }  ,draw opacity=1 ][fill={rgb, 255:red, 0; green, 0; blue, 0 }  ,fill opacity=1 ] (417.64,151.56) .. controls (417.64,149.9) and (418.98,148.56) .. (420.64,148.56) .. controls (422.3,148.56) and (423.64,149.9) .. (423.64,151.56) .. controls (423.64,153.22) and (422.3,154.56) .. (420.64,154.56) .. controls (418.98,154.56) and (417.64,153.22) .. (417.64,151.56) -- cycle ;
		\draw  [dash pattern={on 0.84pt off 2.51pt}]  (184.01,173.15) -- (184.8,246.6) ;
		
		\draw (41.5,195.65) node [anchor=north west][inner sep=0.75pt]    {$\gamma $};
		\draw (206,245.4) node [anchor=north west][inner sep=0.75pt]  [font=\footnotesize]  {$t_{0}$};
		\draw (198,133.9) node [anchor=north west][inner sep=0.75pt]  [font=\scriptsize]  {$\boldsymbol{\upgamma }( t_{0})$};
		\draw (109.5,278.08) node [anchor=north west][inner sep=0.75pt]  [font=\small]  {$N_{\delta }( t_{0}) =( t_{0} -\delta ,t_{0})$};
		\draw (194.5,175.55) node [anchor=north west][inner sep=0.75pt]  [font=\tiny]  {$\ell $};
		\draw (177,158.71) node [anchor=north west][inner sep=0.75pt]  [font=\tiny]  {$\ell $};
		\draw (187.42,156.77) node [anchor=north west][inner sep=0.75pt]  [font=\scriptsize]  {$S( \ell )$};
		\draw (482,50.11) node [anchor=north west][inner sep=0.75pt]  [font=\normalsize]  {$S( \ell )$};
		\draw (545.67,53.73) node [anchor=north west][inner sep=0.75pt]  [font=\scriptsize]  {$( t_{0} ,\boldsymbol{\upgamma }( t_{0}))$};
		\draw (412.83,220.04) node [anchor=north west][inner sep=0.75pt]  [font=\footnotesize]  {$t^{*}$};
		\draw (392.78,144.69) node [anchor=north west][inner sep=0.75pt]  [font=\tiny]  {$N\ell^*$};
		\draw (437.11,250) node [anchor=north west][inner sep=0.75pt]  [font=\tiny]  {$\ell^*$};
		\draw (476.86,249) node [anchor=north west][inner sep=0.75pt]  [font=\tiny]  {$\ell^*$};
		\draw (370.08,114.94) node [anchor=north west][inner sep=0.75pt]  [font=\tiny,color={rgb, 255:red, 208; green, 2; blue, 27 }  ,opacity=1 ]  {$( M+N)\ell^*$};
		\draw (428.91,176.18) node [anchor=north west][inner sep=0.75pt]    {$A_{+}$};
		\draw (469.91,127.45) node [anchor=north west][inner sep=0.75pt]    {$A_{-}$};
		\draw (61.23,109.39) node [anchor=north west][inner sep=0.75pt]  [font=\normalsize]  {$\boldsymbol{\upgamma }( t)$};
		\draw (328.4,247.39) node [anchor=north west][inner sep=0.75pt]    {$t$};
		\draw (426,79.15) node [anchor=north west][inner sep=0.75pt]  [font=\scriptsize]  {$G( t)$};
		\draw (426.5,91.9) node [anchor=north west][inner sep=0.75pt]  [font=\scriptsize]  {$F( t)$};
		\draw (180.5,245.4) node [anchor=north west][inner sep=0.75pt]  [font=\footnotesize]  {$t^{*}$};

	\end{tikzpicture}
\end{figure}

\textbf{Case 2:}  Function $\boldsymbol{\upgamma}$ is not flat at any left-neighbourhood of $t_0$. If $\boldsymbol{\upgamma}$ is left-continuous at $t_0$, then the construction boils down to that of the previous case with two natural numbers $M, N \in \N$ where $\gamma <  M/N  <  \inf_{t\in N_\delta(t_0) }\boldgamma{\upgamma}(t)$ so that we have 
$$\gamma < \frac{A_-}{A_+}= \frac{M}{N} <  \inf_{t\in N_\delta(t_0) }\boldgamma{\upgamma}(t)\le \boldsymbol{\upgamma}(t^*+\ell^*)\le \boldsymbol{\upgamma}(t^*+2\ell^*) \le \boldsymbol{\upgamma} (t_0).$$

\begin{figure}[H]

	\tikzset{every picture/.style={line width=0.75pt}} 
	
	\begin{tikzpicture}[x=0.75pt,y=0.75pt,yscale=-0.8,xscale=0.78]
		
		\draw  (7,198.67) -- (319.23,198.67)(38.22,76.5) -- (38.22,212.24) (312.23,193.67) -- (319.23,198.67) -- (312.23,203.67) (33.22,83.5) -- (38.22,76.5) -- (43.22,83.5)  ;
		\draw    (38.98,168.49) -- (235.48,168.74) ;
		\draw  [dash pattern={on 0.84pt off 2.51pt}]  (195.01,102.81) -- (195,178.43) -- (195,197.5) ;
		\draw    (118.01,134.81) -- (195.01,134.81) ;
		\draw   (120.01,218.2) .. controls (120.05,222.87) and (122.4,225.18) .. (127.07,225.13) -- (147.57,224.94) .. controls (154.24,224.88) and (157.59,227.18) .. (157.64,231.85) .. controls (157.59,227.18) and (160.9,224.82) .. (167.57,224.75)(164.57,224.78) -- (188.07,224.56) .. controls (192.74,224.51) and (195.05,222.16) .. (195,217.49) ;
		\draw  [dash pattern={on 0.84pt off 2.51pt}]  (118.01,102.81) -- (118,178.43) -- (118,197.5) ;
		\draw    (167.01,135.65) -- (167.01,159.65) ;
		\draw    (167.01,159.65) -- (195.01,159.65) ;
		\draw   (193.67,102.81) .. controls (193.67,102.08) and (194.27,101.48) .. (195.01,101.48) .. controls (195.74,101.48) and (196.34,102.08) .. (196.34,102.81) .. controls (196.34,103.55) and (195.74,104.15) .. (195.01,104.15) .. controls (194.27,104.15) and (193.67,103.55) .. (193.67,102.81) -- cycle ;
		\draw  [line width=1.5] [line join = round][line cap = round] (195.78,103.24) .. controls (195.65,102.84) and (194.91,102.8) .. (194.53,102.99) ;
		\draw    (118.01,134.81) .. controls (158.01,104.81) and (155.01,132.95) .. (195.01,102.95) ;
		\draw  [fill={rgb, 255:red, 0; green, 0; blue, 0 }  ,fill opacity=1 ] (116.8,134.81) .. controls (116.8,134.15) and (117.34,133.61) .. (118.01,133.61) .. controls (118.67,133.61) and (119.21,134.15) .. (119.21,134.81) .. controls (119.21,135.48) and (118.67,136.02) .. (118.01,136.02) .. controls (117.34,136.02) and (116.8,135.48) .. (116.8,134.81) -- cycle ;
		\draw  (332,199.99) -- (644.23,199.99)(363.22,77.82) -- (363.22,213.56) (637.23,194.99) -- (644.23,199.99) -- (637.23,204.99) (358.22,84.82) -- (363.22,77.82) -- (368.22,84.82)  ;
		\draw    (362.98,157.81) -- (620.16,158.43) ;
		\draw    (549.33,105.02) -- (549.33,129.02) ;
		\draw    (521.33,129.02) -- (549.33,129.02) ;
		\draw  [fill={rgb, 255:red, 0; green, 0; blue, 0 }  ,fill opacity=1 ] (522.36,105.82) .. controls (522.73,104.7) and (522.14,103.5) .. (521.02,103.12) .. controls (519.91,102.74) and (518.7,103.34) .. (518.33,104.45) .. controls (517.95,105.57) and (518.55,106.77) .. (519.66,107.15) .. controls (520.77,107.53) and (521.98,106.93) .. (522.36,105.82) -- cycle ;
		\draw   (518.35,174.43) .. controls (518.35,173.24) and (519.32,172.27) .. (520.51,172.27) .. controls (521.7,172.27) and (522.66,173.24) .. (522.66,174.43) .. controls (522.66,175.62) and (521.7,176.58) .. (520.51,176.58) .. controls (519.32,176.58) and (518.35,175.62) .. (518.35,174.43) -- cycle ;
		\draw [color={rgb, 255:red, 0; green, 0; blue, 0 }  ,draw opacity=1 ]   (520.34,105.13) -- (594.99,104.93) ;
		\draw    (363.22,199.99) .. controls (403.22,169.99) and (475.11,192.04) .. (518.35,174.43) ;
		\draw  [dash pattern={on 0.84pt off 2.51pt}]  (520.34,105.13) -- (520.61,157.58) ;
		\draw  [dash pattern={on 0.84pt off 2.51pt}]  (520.51,176.58) -- (520.64,198.5) ;
		\draw  [dash pattern={on 0.84pt off 2.51pt}]  (520.61,157.58) -- (520.51,172.27) ;
		\draw   (519.61,217.7) .. controls (519.66,222.37) and (522.01,224.68) .. (526.67,224.63) -- (547.17,224.44) .. controls (553.84,224.38) and (557.19,226.68) .. (557.24,231.35) .. controls (557.19,226.68) and (560.5,224.32) .. (567.17,224.25)(564.17,224.28) -- (587.67,224.06) .. controls (592.34,224.01) and (594.65,221.66) .. (594.6,216.99) ;
		\draw  [dash pattern={on 0.84pt off 2.51pt}]  (594.99,104.93) -- (594.98,180.55) -- (594.98,199.62) ;
		\draw  [dash pattern={on 0.84pt off 2.51pt}]  (167.01,159.65) -- (167.4,200) ;
		
		\draw (24,161.9) node [anchor=north west][inner sep=0.75pt]    {$\gamma $};
		\draw (188,199.4) node [anchor=north west][inner sep=0.75pt]  [font=\footnotesize]  {$t_{0}$};
		\draw (181.5,87.9) node [anchor=north west][inner sep=0.75pt]  [font=\scriptsize]  {$\boldsymbol{\upgamma }( t_{0})$};
		\draw (91.5,232.58) node [anchor=north west][inner sep=0.75pt]  [font=\small]  {$N_{\delta }( t_{0}) =( t_{0} -\delta ,t_{0})$};
		\draw (177.25,160.8) node [anchor=north west][inner sep=0.75pt]  [font=\tiny]  {$\ell $};
		\draw (160,145.21) node [anchor=north west][inner sep=0.75pt]  [font=\tiny]  {$\ell $};
		\draw (170.42,143.27) node [anchor=north west][inner sep=0.75pt]  [font=\scriptsize]  {$S( \ell )$};
		\draw (41,137.9) node [anchor=north west][inner sep=0.75pt]  [font=\scriptsize]  {$\inf_{\{t\in N_{\delta }( t_{0})\}}\boldsymbol{\upgamma }( t)$};
		\draw (349.75,152.72) node [anchor=north west][inner sep=0.75pt]    {$\gamma $};
		\draw (508.75,90.22) node [anchor=north west][inner sep=0.75pt]  [font=\scriptsize]  {$\boldsymbol{\upgamma }( t_{0})$};
		\draw (531.5,132.87) node [anchor=north west][inner sep=0.75pt]  [font=\tiny]  {$\ell $};
		\draw (553.34,115.1) node [anchor=north west][inner sep=0.75pt]  [font=\tiny]  {$\ell $};
		\draw (524.34,113.53) node [anchor=north west][inner sep=0.75pt]  [font=\scriptsize]  {$S( \ell )$};
		\draw (513.6,201.9) node [anchor=north west][inner sep=0.75pt]  [font=\footnotesize]  {$t_{0}$};
		\draw (491.1,232.08) node [anchor=north west][inner sep=0.75pt]  [font=\small]  {$N_{\delta }( t_{0}) =( t_{0} ,t_{0} +\delta )$};
		\draw (44.23,65.55) node [anchor=north west][inner sep=0.75pt]  [font=\normalsize]  {$\boldsymbol{\upgamma }( t)$};
		\draw (301.83,201.42) node [anchor=north west][inner sep=0.75pt]    {$t$};
		\draw (368.85,67.3) node [anchor=north west][inner sep=0.75pt]  [font=\normalsize]  {$\boldsymbol{\upgamma }( t)$};
		\draw (631.1,202.92) node [anchor=north west][inner sep=0.75pt]    {$t$};
		\draw (163.5,199.2) node [anchor=north west][inner sep=0.75pt]  [font=\footnotesize]  {$t^{*}$};

	\end{tikzpicture}
\end{figure}

Otherwise, namely that $\boldsymbol{\upgamma}(t) < \gamma$ on a left neighbourhood $N_\delta (t_0)$.  In this scenario, since function $\boldsymbol{\upgamma}$ is non-decreasing, we can assume that on a tiny right neighbourhood of $t_0$ function $\boldsymbol{\upgamma}$ is flat.  Now, we can similarly construct two distribution functions $F$ and $G$ in the very tiny square $S(\ell)$  on the right hand side of $t_0$ (the left-bottom corner is $t_0$ instead of $t^*$). Note that this time constructed $F$ and $G$ take $0$ on $(-\infty,t_0)$ thus, the non-pleasant values of $\boldsymbol{\upgamma}$ i.e. $\boldgamma{\upgamma}(t)< \gamma$ for $t<t_0$ become irrelevant.

Second, $ F \preceq_{ (1+\gamma)\textsc{-SD}} G$ but $F \npreceq_{ (1+ \boldgamma{\upgamma})\textsc{-SD}}G$. Recall that there exists (at least one) $t^0$ such that $\boldsymbol{\upgamma} (t^0) < \gamma$. Again the construction is the same as above, meaning that after $t^0$ we define $F$ and $G$ the same and discuss in a left neighbourhood of $t^0$ whether $\boldsymbol{\upgamma}$ is flat or not and continue as before.

\section{{Proofs}}

\subsection{\textbf{Proof of results in Section \ref{sec:definition}}}

\begin{proof}[B.1.1]\textbf{ Proof of Theorem \ref{thm:universal_relation}}
	(a) The inclusions are immediate consequence of the monotonicity of integral condition \eqref{eq:Extension_gammaSD_IntegralCondition} (see Remark \ref{rem:RightAfterMainDefinition} (i)). We only prove the strict inclusion case. Suppose that $\boldgamma{\upgamma}$ is not a constant function and so that there exists at least one $t\in \R$ with $\boldgamma{\upgamma}(t)>\barbelow{\boldgamma{\upgamma}}$.  Then, we can always find a constant $c\in(0,1)$ satisfying $\sqrt{c}+\sqrt{c}\,\boldgamma{\upgamma}(t)<1$. Fix such $c$ and construct the following pair of distribution functions $(F^*,G^*)$: 
	\begin{align*}
		F^*(x)&= \begin{cases}
			0   &\mbox{ if }  x< t-\sqrt{c}\\
			\sqrt{c} &\mbox { if } t-\sqrt{c} \le x < t+\sqrt{c}\\
			1 & \mbox{ if }  t+\sqrt{c} \le x 
		\end{cases}\\
		G^*(x)&= \begin{cases}
			0   &\mbox{ if }  x< t\\
			\sqrt{c}+\sqrt{c} \,  \boldgamma{\upgamma}(t) &\mbox { if } t \le x < t+\sqrt{c}\\
			1 & \mbox{ if }  t+\sqrt{c} \le x 
		\end{cases}.
	\end{align*}
	Then, $F^*$ and $G^*$ have a single crossing at $t$ and  
	\begin{equation*}
		A_+:= \int_{-\infty}^t (F^*(x)-G^*(x))_+dx= c, \, \, A_- := \int_{t}^{t+\sqrt{c}} (F^*(x)-G^*(x))_-dx= \boldgamma{\upgamma}(t)c.
	\end{equation*}
	Note that $\barbelow{\boldgamma{\upgamma}}<\frac{A_-}{A_+} =\boldgamma{\upgamma}(t)\le \boldgamma{\upgamma}(x), \,   \forall x\ge t$. Thus, $(F^*,G^*)\in \mathscrbf{E}_{\textsc{mf}}(\boldgamma{\upgamma})$ but $(F^*,G^*)\notin \mathscrbf{E}_{\textsc{f}}(\barbelow{\boldsymbol{\upgamma}})$.  This shows that the first inclusion is strict, $	\mathscrbf{E}_{\textsc{f}}(\barbelow{\boldsymbol{\upgamma}}) \subsetneq \mathscrbf{E}_{\textsc{mf}}(\boldgamma{\upgamma})$.
	Similarly, we can show that the second inclusion $ \mathscrbf{E}_{\textsc{mf}}(\boldgamma{\upgamma})\subsetneq \mathscrbf{E}_{\textsc{f}}(\thickbar{\boldsymbol{\upgamma}})$ is strict as well. For a non-constant $\boldsymbol{\upgamma}$, consider fixing $t \in \mathbb{R}$ such that $\thickbar{\boldsymbol{\upgamma}} > \boldsymbol{\upgamma}(t)$. In this case, a constant $c \in (0,1)$ can be chosen such that $\sqrt{c} + \sqrt{c}\,\thickbar{\boldsymbol{\upgamma}} < 1$. With this $c$, one can then construct the corresponding pair of distribution functions $(F^*, G^*)$ with single crossing, as above. This time, we have  $(F^*,G^*)\in \mathscrbf{E}_{\textsc{f}}(\thickbar{\boldsymbol{\upgamma}})$ but $(F^*,G^*)\notin \mathscrbf{E}_{\textsc{mf}}(\boldgamma{\upgamma})$ since $\boldgamma{\upgamma}(t)<\frac{A_-}{A_+} =\thickbar{\boldgamma{\upgamma}}$ meaning that the integral condition \eqref{eq:Extension_gammaSD_IntegralCondition} fails at $t$.  
	
	(b)  First, the set equalities in \eqref{eq:unviersal_inclusion} are obvious. Next, note that the first strict inclusion in relation \eqref{eq:unviersal_inclusion2} is a direct application of Example \ref{ex:identical_means}. Finally, we show that the second set equality in relation \eqref{eq:unviersal_inclusion2}. First, we have 
	\begin{equation*}\label{eq:proof_inc}
		\bigcup_{    \mathclap{\substack{  \boldgamma{\upgamma}:\R \to [0,1]     \\
					\text{non-decreasing} \\
					\boldgamma{\upgamma} \not\equiv 1  }  } }  \,  \mathscrbf{E}_{\textsc{mf}}(\boldgamma{\upgamma})\quad \subseteq \quad \mathscrbf{E}_{\textsc{ssd}}.
	\end{equation*}
	In order to demonstrate the converse, let $(F,G)$ be an arbitrary element of $\mathscrbf{E}_{\textsc{ssd}}$. Then,  w.l.o.g,  there exists the first crossing point $x^*\in \R$ (otherwise, we have $F \preceq_{\textsc{FSD}} G$ and the result becomes trivial) for the pair $(F,G)$ meaning that $\int_{-\infty}^{x^*} \left(F(x) - G(x) \right)_{+}dx>0$, $\int_{-\infty}^{x^*} \left(F(x) - G(x) \right)_{-}dx=0$ and $F(x^*)=G(x^*)$.  Now, we define the following function:
	\begin{equation*}\label{eq:inlusion_proof_sup}
		\boldsymbol{\upgamma}(F,G)(t):= \begin{cases}
			0 &\mbox{ if } t\le x^*\\
			\sup_{x^* \le s \leq t} \frac{\int_{x^*}^{s} \left(F(x) - G(x) \right)_{-} dx}{\int_{-\infty}^{s} \left(F(x) - G(x) \right)_{+} dx} &\mbox{ if }  x^*\le t.
		\end{cases}
	\end{equation*}
	Clearly, $\boldsymbol{\upgamma}(F,G)$ is a non-negative, non-decreasing, non-constant, continuous function and it is bounded by one.  Therefore, 
	\begin{equation*}
		\boldsymbol{\upgamma}(F,G)\in\{ \boldgamma{\upgamma}: \R \to [0,1] :  \mbox{ non-decreasing and } \boldgamma{\upgamma}\not\equiv 1 \},
	\end{equation*}
	and moreover, we have $F \preceq^{\textsc{mf}}_{ (1+\boldsymbol{\upgamma}(F,G))\textsc{-SD}} G$.  This completes the proof. 
\end{proof}

\subsection{\textbf{Proof of results in Section \ref{sec:utilityclass}}}
\begin{proof}[B.2.1]\textbf{ Proof of Lemma \ref{lem:NiceUtilityBelongClass}}
	Let $t\in \R$. First, we avoid of the trivial cases $\boldsymbol{\upgamma}(t)=0$ (since by very definition, function $u$ is monotone) as well as $\boldgamma{\upgamma}(t)=1$. Next, note that $u_t$ is a continuous piece-wise linear function so that for any $x \le t$ both left $(u_t^{\prime})_-(x)$ and right $(u_t^{\prime})_+(x)$ derivatives exist and we have $(u_t^{\prime})_\pm(x)\in \{\boldgamma{\upgamma}(t),1\}$. This in particular implies that for all $x<y\le t$, $\boldgamma{\upgamma}(t)\le\frac{u_t(y)-u_t(x)}{y-x}\le 1.$ Thus, for all $x_1<x_2\le x_3<x_4$, the quotient relation given in  \eqref{eq:OneSidedUtilityClass} holds true. Moreover, $u_t$ is a constant function on the region $(t,\infty)$. Hence, it follows that $u_t \in \mathscr{U}^{t}_{\boldgamma{\upgamma}(t)}$ for every $t$.
\end{proof}

\begin{proof}[B.2.2]\textbf{ Proof of Proposition \ref{prop:Utility_Implies_IntegralCondition}} Fix $ t \in \mathbb{R} $. The case $ \boldgamma{\upgamma}(t) = 0 $ corresponds to FSD hence the claim follows. The assumption of having finite means implies that $\E_G[|u_t|],\E_F[|u_t|]<\infty$. Employing the integration by parts, we obtain
	\begin{align*}
		\E_G\left[u_t\right]-\E_F\left[u_t\right]&=\int_\R  u_t (x) d(F-G)(x) =  \int_\R  \left(F(x) -G(x)\right)  du_t (x)\\ =&\boldgamma{\upgamma}(t) \int_{-\infty}^{t} \left(F(x)-G(x)\right)_+dx-\int_{-\infty}^{t} \left(F(x)-G(x)\right)_- dx \ge 0.\nonumber
	\end{align*}
\end{proof}	

\begin{proof}[B.2.3]\textbf{ Proof of Proposition \ref{prop:u_t_properties}}
	(a)-(d)-(e) The proof is straightforward. (b) Continuity follows from the set condition of $ \mathscr{U}^{t}_{\boldgamma{\upgamma}(t)} $. Given that $ u $ is non-decreasing, for any $ x < y $, it follows that $ u(y) - u(x) = |u(y) - u(x)| $. Let $ a, a' $ be arbitrary real numbers with $ a' < a $. Then, for any $ x, y $ such that $ a \le x < y $, we have 
	$$ 0 \le \frac{u(y) - u(x)}{y - x} \boldsymbol{\upgamma}(t) \le \frac{u(a) - u(a')}{a - a'}.$$
	From this inequality, for all $ x, y \in [a, \infty) $, we have $|u(y) - u(x)| \le K |y - x| $
	where $ K = \frac{1}{\boldsymbol{\upgamma}(t)} \frac{u(a) - u(a')}{a - a'} \ge 0 $. Consequently, $ u $ exhibits Lipschitz continuity on $ [a, \infty) $, implying it is also absolutely continuous.
	(c) For any $\boldsymbol{\upgamma}(t) = c \in [0, 1)$, we can directly construct a function that may not exhibit differentiability on both sides. Specifically, consider piecewise continuous functions where the first derivative oscillates in the neighborhood of 0, causing the upper and lower Dini derivatives from the right to be unequal. Consequently, the right-hand derivative fails to exist. For example, for $t=27$ and $\boldsymbol{\upgamma}(t)=1/3$, 
	\begin{equation*}
		f(x)=\begin{cases}
			\frac{x}{2}+\frac{3^n}{2} &\mbox{ if } 3^n\le x \le 2(3^n), n \in \{z\in Z:z\le 2\} \\
			\frac{3x}{2}- \frac{3^{n+1}}{2}&\mbox{ if } 2(3^n)\le x \le 3^{n+1}\\
			\frac{3x}{2} &\mbox{ if } x<0\\
			27&\mbox{ if } 27\le x
		\end{cases}
	\end{equation*}
	is such a function.

	\begin{figure}[H]
		\centering
		\begin{tikzpicture}
			\begin{axis}[
				axis lines=left,
				xlabel=$ x$ ,
				ylabel=$ f(x)$ ,
				xmin=0, xmax=30,
				ymin=0, ymax=30,
				samples=100,
				scale=0.7,
				]
				
				\foreach \n in {-2,-1,0,1,2} {
					\addplot[domain=3^\n:2*3^\n,blue,line width=1.2pt] {x/2 + (3^\n)/2};
					
					\addplot[domain=2*3^\n:3^(1+\n),red,line width=1.2pt] {3*x/2 - (3^\n)*3/2};
				}
				
				\addplot[domain=0:30,black,dashed,line width=1pt] {x};
				
				\addplot[domain=0:30,black,dashed,line width=1pt] {3*x/4};
				
				\addplot[domain=27:30, blue, line width=1.2pt] {27};
				
				\node[anchor=west, font=\tiny] at (axis cs:20,25) {$ y=x$ };
				\node[anchor=west, font=\tiny] at (axis cs:20,14.75) {$ y=\frac{3x}{4}$ };

			\end{axis}
		\end{tikzpicture}
	\end{figure}

	(g) Let $(t_n)$ be a sequence of real numbers diverging to infinity. First note that by inclusion relations given in statement (e), we have 
	$$\bigcap_{k=n}^{\infty} \mathscr{U}^{t_k}_{\boldsymbol{\upgamma}(t_k)}=\mathscr{U}^{t_n}_{\thickbarr{\boldsymbol{\upgamma}}}.$$
	Thus,  $\displaystyle{\liminf_{n\rightarrow \infty }\mathscr{U}^{t_n}_{\boldsymbol{\upgamma}(t_n)}}=\displaystyle{\bigcup_{n\ge1}} \mathscr{U}^{t_n}_{\thickbarr{\boldsymbol{\upgamma}}}=\mathscr{U}^{+\infty}_{\thickbarr{\boldsymbol{\upgamma}}}$.  	Similarly, we can show that $$\displaystyle{\limsup_{n\rightarrow \infty }\mathscr{U}^{t_n}_{\boldsymbol{\upgamma}(t_n)}} =\bigcap_{n\ge 1}\bigcup_{k\ge n}^{\infty} \mathscr{U}^{t_k}_{\boldsymbol{\upgamma}(t_k)} =\mathscr{U}^{+\infty}_{\thickbarr{\boldsymbol{\upgamma}}}.$$ Therefore,
	$
	\displaystyle{\limsup_{t_n\rightarrow \infty }\mathscr{U}^{t_n}_{\boldsymbol{\upgamma}(t_n)}}=\displaystyle{\liminf_{t_n\rightarrow \infty }\mathscr{U}^{t_n}_{\boldsymbol{\upgamma}(t_n)}} = \mathscr{U}^{+\infty}_{\thickbarr{\boldsymbol{\upgamma}}}$. (f) Similar to part (g). This completes the proof.
\end{proof}

\begin{proof}[B.2.4]\textbf{ Proof of Theorem \ref{thm:Main_Theorem_IFF}}
	Some parts of the proof involve technicalities that are discussed in a great care in \cite[Theorem 2*]{tesfatsion_efficiency}. IT turns out that assumptions of finite means i.e. $-\infty<\mu_F,\mu_G<\infty$ together with the well-definiteness of the quantity $\E_G[u] - \E_F[u]$ for every $u \in \mathscr{U}^{\textsc{mf}}_{\boldsymbol{\upgamma}}$ provide us with the most transparent scenario. Hence, we assume that those hypotheses are available without mentioning in each required instance. 
	
	The implication (b) to (a) is just Proposition \ref{prop:Utility_Implies_IntegralCondition}. Hence, we prove (a) implies (b). Due to the linearity of the mathematical expectation and the structure of the elements in the class $\mathscr{U}^{\textsc{mf}}_{\boldsymbol{\upgamma}}$ it is enough to show that 
	\begin{equation*}
		\mathbb{E}_F[u] \leq \mathbb{E}_G[u]  \quad \text{ for every } \quad u \in \mathscr{U}^{t}_{\boldsymbol{\upgamma}(t)}  \quad \text{ where } \quad t \in (-\infty,\infty].
	\end{equation*}  Without loss of generality, we assume the existence of a first crossing point of $ F $ and $ G $, denoted by $ x_1^* $. If no such point exists i.e. $F$ and $G$ never cross, then we are in the FSD case, and the statement holds immediately from \cite[Theorem 1*]{tesfatsion_efficiency}. First, 
	let $t=\infty$. Pick $u \in \mathscr{U}^{+\infty}_{\thickbarr{\boldsymbol{\upgamma}}} = \mathscr{U}^{*}_{\thickbarr{\boldsymbol{\upgamma}}}$. Now, observe that the integral condition \eqref{eq:Extension_gammaSD_IntegralCondition} clearly yields that the integral condition 
	\begin{equation*}
		\int_{-\infty}^t  (F(x)-G(x))_{-} d x \leq \thickbarr{\boldsymbol{\upgamma}} \int_{-\infty}^t(F(x)-G(x))_{+} d x, \quad \forall t\in \mathbb{R},
	\end{equation*}
	that is $F \preceq_{ (1+ \thickbarr{\boldsymbol{\upgamma}})\textsc{-SD}} G$ and  therefore, the result directly follows from \cite[Theorem 2.4]{Muller}. Next, let $t \in \mathbb{R}$ and $u \in \mathscr{U}^{t}_{\boldsymbol{\upgamma}(t)}$ be arbitrary. Before proceeding further we emphasize the following points: 
	\begin{itemize}
		\item[(i)] As $u$ is arbitrary, we shall the positivity requirement:
		\begin{equation}\label{eq:main_proof_we_need_to_show_stronger}
			\int_{-\infty}^s \left( F(x)-G(x) \right) d u(x)\ge 0 \textbf{ for all } s\leq t.
		\end{equation}
		\item[(ii)]  Without loss of generality, we assume $  \boldsymbol{\upgamma}(t) \neq 0 $ and $ x_1^* < t $. Absent these conditions, we once again are in the FSD framework on the interval $(-\infty,t]$.	
	\end{itemize}
	Hereafter, by the positivity of the integrals, 
	$ F \preceq^{\textsc{mf}}_{(1+\boldgamma{\upgamma})\textsc{-SD}} G $
	implies
	\begin{equation}\label{eq:proof_alternative integral}
		\int_{-\infty}^{s} \left( F(x) - G(x) \right)_{-}dx \le \boldgamma{\upgamma}(t) \int_{-\infty}^{s} \left( F(x) - G(x) \right)_{+} dx, \quad \forall s\leq t
	\end{equation}
	for any $t\in \R$, we will follow the similar arguments provided in the proof of Theorem 2 in \cite{efficiency}. For a more detailed discussion, see also the proof of Theorem 2* in \cite{tesfatsion_efficiency}. \\

	
	First, for any $s\le t$, we can write \eqref{eq:main_proof_we_need_to_show_stronger} as
	\begin{align}
		&\int_{-\infty}^s \left(F(x)-G(x) \right)d u(x)&\\ =&\int_{-\infty}^s \left(F(x)-G(x)\right)_+ du(x)-\int_{-\infty}^s \left(F(x)-G(x)\right)_- du(x)\nonumber \\
		= &\int_{-\infty}^s \left(F(x)-G(x)\right)_+ du(x)-\int_{x_1^*}^s \left(F(x)-G(x)\right)_- du(x)\label{eq:proof_expected_value}
	\end{align}
	where the last equality follows from the fact that we have $\left(F(x)-G(x)\right)_-=0$ for all $ x \le x_1^* $. Now, define the following transformation $T:x_1^*< s\le t \mapsto T(s)$ as
	\begin{equation} \label{eq:proof_transformation_equality}
		T(s):=\inf\left\{ s^\prime : \boldgamma{\upgamma}(t)\int_{-\infty}^{s^\prime} \left( F(x)-G(x)\right)_+dx = \int_{x_1^*}^{s} \left(F(x)-G(x)\right)_-dx \right\}
	\end{equation}
	such that 
	\begin{equation}\label{eq:proof_derivative_equality} \boldgamma{\upgamma}(t)\int_{-\infty}^{T(s)} \left(F(x)-G(x)\right)_+dx = \int_{x_1^*}^{s} \left(F(x)-G(x)\right)_-dx,\quad \forall s\leq t.\end{equation}
	
	We have the following observations:	
	\begin{itemize}
		
		\item[(i)] Denote 
		\begin{equation*}
			A_+(s)=\boldgamma{\upgamma}(t) \int_{-\infty}^{s} \left(F(x) - G(x)\right)_+dx  \text{ and }  A_-(s)=\int_{x^*_1}^{s} \left(F(x) - G(x)\right)_-dx.
		\end{equation*}
		Then,
		\begin{itemize}
			\item[(a)] both $ A_+(s) $ and $ A_-(s) $ are non-negative, \textbf{non-decreasing} and continuous functions 
			\item[(b)] given the integral condition in \eqref{eq:proof_alternative integral} holds,  it is clear that $ A_+(s) \ge A_-(s) $ for all $ x^*_1<s \le t $.
		\end{itemize}
		Therefore, combination of (a) and (b) ensures that for any $ x^*_1<s \le t $, there exists $T(s) \leq s $ such that the equality in \eqref{eq:proof_derivative_equality} , i.e. $$A_+(T(s)) = A_-(s)$$ is achieved. Also, note that $T_+(x^*_1)=\lim_{s\to x^*_1}T(s)=-\infty$.
		
		\item[(ii)] Let $D_+ = \{x^*_1 < s \leq t : F(s) \geq G(s)\}$ and $D_- = \{s \leq t : F(s) < G(s)\}$, so that we have $D_+ \cup D_- = \{s : x^*_1 < s \leq t\}$. Then, for any $s, s' \in D_-$ we have 
		$$s > s'\implies A_-(s) > A_-(s')\implies T(s) > T(s')$$ where the last relation is a direct consequence of the equality in  \eqref{eq:proof_derivative_equality}. In a similar way, one can show that $T$ is constant on each segment (between crossing points) of $D_+$. It follows that $T$ differentiable almost everywhere on $D_+\cup D_-$.  Therefore, differentiating equation \eqref{eq:proof_derivative_equality}, one obtains:
		\begin{equation}\label{eq:proof_derivative_T(s)}
			\boldgamma{\upgamma}(t)  \left(   F(T(s))-G(T(s))  \right)_+ T'(s)= \left(  F(s)-G(s) \right)_- 
		\end{equation}
		for almost all $x^*_1<s\leq t$.
		Observe that, we have $T'(s)=0$ for almost all $s\in D_+$ and $T'(s)>0$ for almost all $s\in D_-$. 
	\end{itemize}
	Now, we turn back to our main statement \eqref{eq:main_proof_we_need_to_show_stronger}. Substituting \eqref{eq:proof_derivative_T(s)} into the integrand of the second integral on the right-hand side of \eqref{eq:proof_expected_value}, we get	\begin{align*}
		&\int_{-\infty}^s \left(F(x)-G(x)\right) d u(x)\\= &\int_{-\infty}^s \left(F(x)-G(x)\right)_+ du(x)- \boldgamma{\upgamma}(t)\int_{(x_1^*,s]} \left(F(T(x))-G(T(x))\right)_+ T'(x) du(x)
	\end{align*}
	where the last integral in the equation should be evaluated as a Lebesgue-Stieltjes integral. Now, for every $s \le t$, let $S:=\{x:x^*_1<x\leq s\}\cap D_-$. First, we have 
	\begin{align*}
		&\boldgamma{\upgamma}(t)\int_{(x^*_1,s]=S \cup D_+}  \left(  F(T(x))-G(T(x)) \right) _+ T'(x) du(x)\\=&\boldgamma{\upgamma}(t)\int_{S}  \left(  F(T(x))-G(T(x)) \right) _+ T'(x) du(x),
	\end{align*}
	since $T'(x)=0$ a.e. on $D_+$. Recall that $ u $ is a non-decreasing function, hence $u'$ exists except on a set with a Lebesgue measure zero. Moreover, it follows from Proposition \ref{prop:u_t_properties} that $u$ is absolutely continuous on $(x^*_1,s]$. Therefore, we have
	\begin{align*}
		&\boldgamma{\upgamma}(t)\int_{S}  \left(  F(T(x))-G(T(x)) \right) _+ T'(x) du(x)\\=&\int_{S}  \left(  F(T(x))-G(T(x)) \right) _+ T'(x)\boldgamma{\upgamma}(t) u'(x)dx.
	\end{align*} Recall that $T(x)\le x$ for all $x$ and  $\boldgamma{\upgamma}(t)u'(y)\le u'(x)$ holds for $x\le y$ whenever $u'$ exists hence, 
	\begin{align*}
		&\int_{S}  \left(  F(T(x))-G(T(x)) \right) _+ T'(x)\boldgamma{\upgamma}(t) u'(x)dx\\ \le &\int_{S} \left(   F(T(x))-G(T(x))\right)_+ T^\prime(x) u'(T(x))dx\\
		= &\int_{S} \left(   F(T(x))-G(T(x))\right)_+ du(T(x)).
	\end{align*}
	Furthermore, recall that $ T $ is strictly increasing on $ D_- $. Consequently, $ T $ is also strictly increasing on $ S $.  Using the change of variable $ z = T(x) $, we can then deduce 
	\begin{equation*}
		\int_{S} \left(   F(T(x))-G(T(x))\right)_+ du(T(x))=\int_{T(S)} \left( F(z)-G(z) \right)_+du(z).
	\end{equation*}
	Clearly, we have $ T(S) \subseteq (T_+(x^*_1),T(s)] $ since $ S \subseteq (x^*_1,s] $. Moreover, the Stieltjes measure $du$ is non-negative, as $ u $ is non-decreasing. Thus, 
	\begin{equation*}
		\int_{T(S)} \left( F(z)-G(z) \right)_+du(z) \le \int_{(T_+(x^*_1),T(s)]} \left( F(z)-G(z) \right)_+du(z).
	\end{equation*}
	Ultimately, we arrive at
	\begin{equation*}
		\int_{-\infty}^s \left( F(x)-G(x) \right)_-du(z) \le \int_{(T_+(x^*_1),T(s)]} \left( F(z)-G(z) \right)_+du(z).
	\end{equation*}
	Next, integrating the above conclusion back into \eqref{eq:proof_expected_value}, we get
	\begin{align*}
		&\int_{-\infty}^s  \left(F(x)-G(x)\right) d u(x)\\\ge& \int_{-\infty}^s \left(F(x)-G(x)\right)_+ du(x)-\int_{(T_+(x^*_1),T(s)]} \left(F(x)-G(x)\right)_+ du(x)\nonumber\\
		=& \int_{(-\infty,s]} \left(F(x)-G(x)\right)_+ du(x)-\int_{(T_+(x^*_1),T(s)]} \left(F(x)-G(x)\right)_+ du(x)\nonumber\\
		=&\int_{[T(s),s]} \left(F(x)-G(x)\right)_+ du(x) \ge 0.\label{eq:T(s)_inequlity_proof}
	\end{align*}
	This completes the proof.

\end{proof}

\subsection{\textbf{Proof of results in Section \ref{sec:mathematical_aspects}}}
\begin{proof}[B.3.1] \textbf{ Proof of Proposition  \ref{prop:utility_class_properties}}
	(a) For any given non-decreasing function $\boldsymbol{\upgamma}:\mathbb{R} \to [0,1]$, the set $\mathscr{U}^{*}_{\thickbarr{\boldgamma{\upgamma}}} \subseteq \mathscr{U}^{\textsc{mf}}_{\boldsymbol{\upgamma}}$ contains all linear, non-decreasing functions. (b) Given the linear structure of $\mathscr{U}^{\textsc{mf}}_{\boldsymbol{\upgamma}}$, it is sufficient to show that for any $t \in \mathbb{R}$, if $u \in \mathscr{U}^t_{\boldsymbol{\upgamma}(t)}$, then $u_c \in \mathscr{U}^t_{\boldsymbol{\upgamma}(t)}$ for $c \ge 0$. Fix $t \in \mathbb{R}$ and $c \ge 0$. Then, it can be directly verified that $u_c$ satisfies the set condition on the inequality of quotients given in  \eqref{eq:OneSidedUtilityClass}. Moreover, if $u(x) = u(t)$ for all $x \ge t$, then $u_c(x) = u(x + c) = u(t)=u_c(t)$ for all $x \ge t$. Therefore, $u_c \in \mathscr{U}^t_{\boldsymbol{\upgamma}(t)}$. Note that if $c < 0$, then the latter claim is not necessarily true.
\end{proof}

\begin{proof}[B.3.2] \textbf{ Proof of Proposition  \ref{prop:basic_properties}}
	
	(a) This is a direct implication of the integral condition given in \eqref{eq:Extension_gammaSD_IntegralCondition}.
	(b) If $F \preceq^{\textsc{mf}}_{ (1+ \boldgamma{\upgamma})\textsc{-SD}} G$, then by letting as $t \to \infty $ the integral condition \ref{eq:Extension_gammaSD_IntegralCondition}, we obtain that
	\begin{equation*}
		\thickbar{\boldgamma{\upgamma}}	\int_{-\infty}^\infty (F(x)-G(x))_+dx - \int_{-\infty}^\infty (F(x)-G(x))_-dx \ge 0.
	\end{equation*}
	Since $\bar{\boldgamma{\upgamma}} \leq 1$, it follows that $\mu_G - \mu_F \ge 0$.
	(c) Obvious. (d) For every non-negative constant $c$, random variable $X+c$ has cumulative distribution function $F_c$, defined by $F_c(t) := F(t-c)$. Since $F$ is a non-decreasing function, it follows that $F(t) \geq F(t-c)$ for all $t \in \mathbb{R}$, hence $F \preceq_{\textsc{FSD}} G$ and obviously   $F \preceq^{\textsc{mf}}_{ (1+ \boldgamma{\upgamma})\textsc{-SD}}  F_c$. (e) Similarly, if $a\leq b$ then $\delta_{a}  \preceq_{\textsc{FSD}} \delta_{b}$.

\end{proof}

\begin{proof}[B.3.3] \textbf{ Proof of Proposition \ref{prop:StabilityProperties}}
	(a) This is a direct application of  part (b) of Proposition \ref{prop:utility_class_properties}. (b)  If $X \preceq^{\textsc{mf}}_{ (1+ \boldgamma{\upgamma})\textsc{-SD}} Y$, then by part (a) we have $[X+Z|Z=z] \preceq^{\textsc{mf}}_{ (1+ \boldgamma{\upgamma})\textsc{-SD}} [Y+Z|Z=z]$ holds for all $z \in \text{supp}(Z) \subseteq [0,\infty)$. Hence, $\E[u(X+Z)|Z=z]\le \E[u(Y+Z)|Z=z]$, $\forall u\in \mathscr{U}^{\textsc{mf}}_{\boldsymbol{\upgamma}}$. Then, for any $u\in \mathscr{U}^{\textsc{mf}}_{\boldsymbol{\upgamma}}$ we have
	\begin{equation*}
		\E[u(X+Z)]=\E[\E[u(X+Z)|Z=z]]\le \E[\E[u(Y+Z)|Z=z]]=\E[u(Y+Z)].
	\end{equation*} 
	
	(c) Similarly, for any $u\in \mathscr{U}^{\textsc{mf}}_{\boldsymbol{\upgamma}}$ we have
	\begin{equation*}
		\mathbb{E}[u(X)]=\mathbb{E}[\mathbb{E}[u(X) \mid \Theta]] \leq \mathbb{E}[\mathbb{E}[u(Y) \mid \Theta]]=\mathbb{E}[u(Y)].
	\end{equation*}
	(d) This is just a direct application of the integral condition \eqref{eq:Extension_gammaSD_IntegralCondition}.
\end{proof}
\begin{proof}[B.3.4] \textbf{ Proof of Proposition \ref{prop:gamma_MFgamma}}
	(a) Let $v\in\mathscr{U}^{\textsc{mf}}_{\boldsymbol{\upgamma}}$ and define $u(x):=v(\phi(x))$. Then, it is sufficient to show $u \in \mathscr{U}^{\textsc{mf}}_{\boldsymbol{\upgamma}\times \gamma}$ for all $v$ and for all $\phi \in \mathscr{U}^{*}_{\boldsymbol{\upgamma}}$ with $\phi(x)\ge x$. Due to the linear structure of $\mathscr{U}^{\textsc{mf}}_{\boldsymbol{\upgamma}}$ without loss of generality we can assume $v\in \mathscr{U}^t_{\boldsymbol{\upgamma}(t)}$. Then, for any $x_1<x_2\le x_3<x_4$ \footnote{ We assume $\frac{u(x_4)-u(x_3)}{x_4-x_3}>0$ (=0 case is trivial).} we have
	\begin{align*}
		(\boldgamma{\upgamma}(t)\times\gamma)\frac{u(x_4)-u(x_3)}{x_4-x_3}&=\boldsymbol{\upgamma}(t)\frac{v(\phi(x_4))-v(\phi(x_3))}{\phi(x_4)-\phi(x_3)}  \times \gamma \frac{\phi(x_4)-\phi(x_3)}{x_4-x_3}\\
		&\le \frac{v(\phi(x_2))-v(\phi(x_1))}{\phi(x_2)-\phi(x_2)} \frac{\phi(x_2)-\phi(x_1)}{x_2-x_1}\\
		&=\frac{u(x_2)-u(x_1)}{x_2-x_1}.
	\end{align*}
	In addition, for every $x \ge t$,	we have that $u(x) =v (\phi(x)) \ge v (x) =v(t)$. Hence, $u\in \mathscr{U}^t_{\boldsymbol{\upgamma}(t)\times \gamma} \subseteq \mathscr{U}^{\textsc{mf}}_{\boldsymbol{\upgamma}\times \gamma}.$
	(b) Similar to the previous case, it is simple to observe that for every $\psi \in \mathscr{U}^{\textsc{mf}}_{\boldsymbol{\upgamma}}$ and every $v \in \mathscr{U}^{*}_{\gamma}$, function $u = v \circ \psi \in \mathscr{U}^{\textsc{mf}}_{\gamma \times \boldsymbol{\upgamma}}$.
	
\end{proof}

\subsection{\textbf{Proof of results in Section \ref{sec:FFSD_definition}}}

\begin{proof}[B.4.1] \textbf{ Proof of Proposition \ref{prop:one_is_enough}} It suffices to demonstrate that $0 \le \gamma u'_+(y) \le u'_+(x)$ for all $x \le y$ implies
	
	$$ \gamma\frac{u(x_4)-u(x_3)}{x_4-x_3} \le \frac{u(x_2)-u(x_1)}{x_2-x_1} $$
	
	for all $x_1<x_2\le x_3 <x_4$. Consider such four arbitrary points, and by  \cite[Lemma  5.3.7]{rajwadeSurprisesCounterexamplesReal2007}, there exist $x \in (x_1, x_2)$ and $y \in (x_3, x_4)$ such that
	
	$$ u_{+}'(x) \leq \frac{u(x_2)-u(x_1)}{x_2-x_1} \quad \text { and } \quad \frac{u(x_4)-u(x_3)}{x_4-x_3} \leq u_{+}'(y) . $$
	
	For each $\gamma \in [0,1]$, we then obtain:
	
	$$ 0 \leq \frac{u(x_4)-u(x_3)}{x_4-x_3} \gamma \leq u_{+}'(y) \gamma \leq u_{+}'(x) \leq \frac{u(x_2)-u(x_1)}{x_2-x_1}. $$
	
	A similar argument holds for the left derivatives. This completes the proof.
\end{proof}

\begin{proof}[B.5.2] \textbf{ Proof of Proposition \ref{prop:mf_one_sided}} 
	Due to the linear structure of the elements in 	$\widetilde{\mathscr{U}}^{\textsc{mf}}_{\boldsymbol{\upgamma}}$, it is enough to show that $\widetilde{\mathscr{U}}^t_{\boldsymbol{\upgamma}(t)}, \widetilde{\mathscr{U}}^{\infty}_{\thickbarr{\boldsymbol{\upgamma}}}\subseteq \widetilde{\mathscr{U}}^{\textsc{ff}}_{\boldsymbol{\upgamma}}$ for every real $t$. Let $u \in \widetilde{\mathscr{U}}^t_{\boldsymbol{\upgamma}(t)}$. Then, we have $ 0\le \boldsymbol{\upgamma} (t) u'_+(y) \le u'_+(x)$ for all $x \le y$ thanks to $\boldsymbol{D}_\pm$ property. This implies that $ 0\le \boldsymbol{\upgamma} (y) u'_+(y) \le u'_+(x)$ for all $x \le y$ is true since $\boldsymbol{\upgamma}$ is non-decreasing, and $u'_+(x)=0$ for each $x \ge t$. The argument for $t=\infty$ is the same. When $\boldsymbol{\upgamma}$ is a constant function, i.e., $\boldsymbol{\upgamma} \equiv \gamma$, the other direction $\widetilde{\mathscr{U}}^{\textsc{ff}}_{\gamma} \subseteq \widetilde{\mathscr{U}}^{\textsc{mf}}_{\gamma}$ can be directly inferred from the proof of Proposition \ref{prop:one_is_enough} by noting that $\boldsymbol{D}_\pm(\boldgamma{\upgamma})$ reduces to $\gamma u_+'(y)\le u_+'(x)$ and $\widetilde{\mathscr{U}}^{\textsc{mf}}_{\gamma}=\widetilde{\mathscr{U}}^{*}_{\gamma}$.  The analogous statement for left-sided derivatives follows in a similar manner.
\end{proof}

\begin{proof}[B.4.2] \textbf{ Proof of Lemma \ref{lem:FFSD:base_type}} Fix $ t \in \mathbb{R} $. If $t \le x^*_1$ (first crossing point), then $ u_{\textsc{ff},t}' $ is a non-decreasing linear function with as constant segment. The claim directly follows. Let $ t \ge x^*_1 $. Then, $ u_{\textsc{ff},t}^{\prime}$ is continuous almost everywhere with possible discontinuities at the crossing points of $ F $ and $ G $ or due to jumps in the $ \boldsymbol{\upgamma} $ function. Denote the set of potential discontinuity points as $ \{x^*_1, x_2, \dots, x_n\} $ then $ u_{\textsc{ff},t} $ is continuously differentiable on every $ (x_i, x_{i+1}) $ and differentiable from both sides at every point, including the discontinuity points. Next, we show that $ u_{\textsc{ff},t} $ also satisfies derivative-relation condition.  Fix $ y $. If $ y \geq t $, then the $ \boldsymbol{D}_\pm(\boldsymbol{\upgamma}) $ condition immediately follows, as $ \boldsymbol{\upgamma} $ is non-negative. Consider the case when $ y \leq t $. 
	To verify the $ \boldsymbol{D}_\pm(\boldsymbol{\upgamma}) $ condition via right-derivatives, four distinct cases need to be considered for all $ x \leq y $: (i) $\boldgamma{\upgamma}(y)\le \frac{1}{\boldgamma{\upgamma}(x)}$, (ii) $\frac{1}{\boldgamma{\upgamma}(y)} \boldgamma{\upgamma}(y)\le 1$, (iii) $\frac{1}{\boldgamma{\upgamma}(y)} \boldgamma{\upgamma}(y)\le \frac{1}{\boldgamma{\upgamma}(x)} $ and (iv) $ \boldgamma{\upgamma}(y)\le 1$. All four cases hold true as $ \boldsymbol{\upgamma} $ takes values in the interval $(0, 1]$. Hence, the claim holds. If $ \boldsymbol{\upgamma} $ were left-continuous, then defining $ u_{\textsc{ff},t} $ via left-derivatives would lead to the same conclusion. This completes the proof.
\end{proof}

\begin{proof}[B.4.3] \textbf{Proof of Proposition  \ref{prop:FFSD:Utility_Implies_IntegralCondition}} Employing the integration by parts formula, we obtain
	\begin{align*}
		&\E_G\left[u_{\textsc{ff},t}\right]-\E_F\left[u_{\textsc{ff},t}\right]=\int_{\R} \left(F(x)-G(x)\right) du_{\textsc{ff},t}(x)\\
		=&\int_{-\infty}^{t} \left(F(x)-G(x)\right)_+dx-\int_{-\infty}^{t}\frac{1}{\boldgamma{\upgamma}(x)} \left(F(x)-G(x)\right)_- dx \ge 0.\nonumber
	\end{align*}
\end{proof}

\begin{proof}[B.4.4] \textbf{Proof of Theorem   \ref{thm:MFSD:iff}} The proof follows directly by replicating the steps in the proof of Theorem \ref{thm:Main_Theorem_IFF} [B.2.4]. Define transformation  $T:x^*_1< t \mapsto T(t)$ as
	\begin{equation*} \label{eq:proof_transformation_equality_CV}
		T(t):=\inf \left\{t^{\prime}: \int_{-\infty}^{t^{\prime}}(F(x)-G(x))_{+} d x=\int_{x^*_1}^t \frac{1}{\boldgamma{\upgamma}(x)}(F(x)-G(x))_{-} d x\right\}
	\end{equation*} where $x^*_1$ is the first crossing point.
	Then,
	\begin{equation*}
		\boldgamma{\upgamma}(t)  \left(   F(T(t))-G(T(t))  \right)_+ T'(t)= \left(  F(t)-G(t) \right)_- \text { for almost all }  x^*_1< t.
	\end{equation*}
	It follows that \begin{align*}
		&\E_G[u]-\E_F[u]=\int_{-\infty}^t \left( F(x)-G(x) \right) d u(x)\\
		=&\int_{-\infty}^t \left(F(x)-G(x)\right)_+ du(x)- \int_{(x_1^*,t]}\boldgamma{\upgamma}(x) \left(F(T(x))-G(T(x))\right)_+ T'(x) du(x).
	\end{align*}
	Then, the set condition  $ \boldsymbol{\upgamma}(y)u'(y) \leq u'(x) $ for all $ x \leq y $ where $ u' $ exists implies that  
	\begin{align*}
		&\int_{S}\boldgamma{\upgamma}(x) \left(   F(T(x))-G(T(x))\right)_+ T^\prime(x) u'(x) dx\\
		\le &\int_{S} \left(   F(T(x))-G(T(x))\right)_+ T^\prime(x) u'(T(x))dx\\
		= &\int_{S} \left(   F(T(x))-G(T(x))\right)_+ du(T(x))
	\end{align*}
	given that $S=\{x:x^*_1<x\leq t\}\cap \{t:x^*_1< t \}$. Using the change of variable $ z = T(x) $, we get 
	\begin{align*}
		\int_{S} \left(   F(T(x))-G(T(x))\right)_+ du(T(x))=&\int_{T(S)} \left( F(z)-G(z) \right)_+du(z)\\ \le &\int_{(T_+(x^*_1),T(t)]} \left( F(z)-G(z) \right)_+du(z).
	\end{align*}
	In turn, this suggests
	\begin{equation*}
		\int_{-\infty}^t \left( F(x)-G(x) \right)_-du(z) \le \int_{(T_+(x^*_1),T(t)]} \left( F(z)-G(z) \right)_+du(z).
	\end{equation*} Ultimately, we get
	\begin{align*}
		&\int_{-\infty}^t  \left(F(x)-G(x)\right) d u(x)\\\ge& \int_{-\infty}^t \left(F(x)-G(x)\right)_+ du(x)-\int_{(T_+(x^*_1),T(t)]} \left(F(x)-G(x)\right)_+ du(x)\nonumber\\
		=& \int_{(-\infty,t]} \left(F(x)-G(x)\right)_+ du(x)-\int_{(T_+(x^*_1),T(t)]} \left(F(x)-G(x)\right)_+ du(x)\nonumber\\
		=&\int_{[T(t),t]} \left(F(x)-G(x)\right)_+ du(x) \ge 0.\label{eq:T(s)_inequlity_proof_CV}
	\end{align*}
	
\end{proof}

\subsection{\textbf{Proof of results in Section \ref{sec:FFSD_examples}}}

\begin{proof}[B.5.1] \textbf{ Proof of Proposition \ref{prop:piece_wise}} 
	(a) By definition, the function $ u $ is differentiable on each sub-interval $ (x_i, x_{i+1}) $. The existence of the right and left derivatives at the subdivision points is ensured by the existence of one-sided limits of the derivative at those points. (b) Consider the PCD function $ u $, which satisfies condition (i) for a given $ \boldsymbol{\upgamma} $. The statement is clearly true at the differentiability points of $ u $, i.e., for every $ x, y \in (x_i, x_{i+1}) $. Additionally, due to the continuity of $ \boldsymbol{\upgamma} $ at each $ x_i $, the statement also holds at the limiting points.
	
\end{proof}
\subsection{\textbf{Proof of results in Section \ref{sec:greediness}}}
\begin{proof}[B.6.1] \textbf{ Proof of Proposition \ref{prop:DiscreteLocalGreedinessProperties}} 
	Items (a) and (b) are obvious. Next, we prove (c). Since $u$ is strictly increasing, hence for every given $x$, there is a point $y = y_x \in (x ,\infty)$ that $u$ is differentiable at $y$. Hence, by letting $x_1 \to y^-$ and $x_4 \to y^+$ we arrive to the claim.   The validity of the second statement is just rely on the basic definition of concave function in terms monotonicity of quotients.  Lastly, we show part (d). Fix $x_0 \in \R$. Let $x_n \to x^+_0$. Then, based on parts (b), (c) and moving to a subsequence, we have $ \lim_{n\to \infty} G^d_{\text{par}}(u;x_n) \le G^d_{\text{par}}(u;x_0)$. Now, let $\varepsilon >0$ be arbitrary. Then, there exist four points $x_0 < x_1 < x_2 \le x_3 < x_4$ such that 
	
	\begin{equation*}
		G^d_{\text{par}} (u;x_0 ) - \varepsilon < \frac{u(x_4)-u(x_3)}{x_4-x_3}/\frac{u(x_2)-u(x_1)}{x_2-x_1}.
	\end{equation*}	  
	On the other hand side, there exists $n_0 \in \N$ such that $x_n < x_1$ for every $n \ge n_0$. Therefore, $$\sup_{n \ge n_0} G^d_{\text{par}}(u;x_n) \ge  \frac{u(x_4)-u(x_3)}{x_4-x_3}/\frac{u(x_2)-u(x_1)}{x_2-x_1}.$$ Hence, by part (b), we obtain $\lim_{n\to \infty} G^d_{\text{par}}(u;x_n)  > G^d_{\text{par}} (u;x_0 ) - \varepsilon$. Letting $\varepsilon \to 0$ we achieve the claim. 
	
\end{proof}

\begin{proof}[B.6.2] \textbf{ Proof of Proposition \ref{prop:LocalGreedinessCoincide}} 
	It is clear that $G^c_{\text{par}}(u;x)\le G^d_{\text{par}}(u;x)$.  Thus, to complete the proof, it remains to show $G^c_{par}(u;x)\ge G^d_{par}(u;x)$. 	We assume that $u$ is right-differentiable at any point. The case of left-differentiable is the same. Let $\varepsilon>0$. Then, by the definition of $G^d_{par}(u,x)$, there exist four points, $x \le x_1<x_2\le x_3<x_4$ such that $$G:=\left(\frac{u(x_4)-u(x_3)}{x_4-x_3}/\frac{u(x_2)-u(x_1)}{x_2-x_1}\right) > G^d_{par}(u;x)-\varepsilon.$$ 
	On the other hand, there exist $c_1, c_2\in (x_1,x_2)$ and $c_3,c_4 \in (x_3,x_4)$ such that
	\begin{equation*}
		u'_+(c_1)\le	\frac{u(x_2)-u(x_1)}{x_2-x_1}\le u'_+(c_2) 	\quad \text{ and } \quad  
		u'_+(c_3)\le	\frac{u(x_4)-u(x_3)}{x_4-x_3}\le u'_+(c_4)
	\end{equation*}
	Hence,
	$$G^c_{\text{par}}(u;x)\ge \sup_{x < y<z}\frac{u'_+(z)}{u'_+(y)} \ge \frac{u'_+(c_4)}{u'_+(c_1)}\ge G > G^d_{par}(u,x)-\varepsilon.$$  By letting $\varepsilon \to 0$,  we complete the proof. 
\end{proof}

\begin{proof}[B.6.3] \textbf{ Proof of Proposition \ref{prop:ContinuousLocalGreedinessImpliesCancavity}} 
	(a) It is a direct application of the generalized mediant inequality. (b) It is also a direct application of part (a) and the fact that $ \max\{ a,b \}\le a+b$ for every two non-negative real numbers. 
\end{proof}

\begin{proof}[B.6.4] \textbf{ Proof of Corollary \ref{cor:LocalGreedinessConcaveGammaSapce}} 
	(a) It is a direct application of the generalized mediant inequality. (b) It is also a direct application of part (a) and the fact that $ \max\{ a,b \}\le a+b$ for every two non-negative real numbers. 
\end{proof}

\subsection{\textbf{Proof of results in Section \ref{sec:MFSDUtilitySpaceLocalGreediness}}}

\begin{proof}[B.7.1] \textbf{ Proof of Lemma  \ref{lem:ExtremeComponentMustExist}} 
	Recall that by Proposition \ref{prop:ContinuousLocalGreedinessImpliesCancavity} and Corollary \ref{cor:LocalGreedinessConcaveGammaSapce}, we have $$G_{\text{par}}(u;x_0)\le \max_{1 \le k \le n}	G_{\text{par}}(u_k;x_0)$$ and $G_{\text{par}}(u_k;x_0) \le \frac{1}{\boldgamma{\upgamma}(x_0)}$ for each $k=1,...,n$. Now, relation \eqref{eq:ExtremeComponentMustExist} yields that 
	\begin{equation*}
		\frac{1}{\boldsymbol{\upgamma}(x_0)} = G_{\text{par}}(u;x_0) \le \max_{1 \le k \le n} 	G_{\text{par}}(u_k,x_0) \le \frac{1}{\boldsymbol{\upgamma}(x_0)}.
	\end{equation*}
	Hence, the result follows.
\end{proof}

\begin{proof}[B.7.2] \textbf{ Proof of Proposition  \ref{prop:GammamustBeConstant}} 
	If $\boldsymbol{\upgamma}(x_0)=1$, then it is obvious that $\boldsymbol{\upgamma}$ is a constant function over the interval $(x_0,\infty)$. Hence, hereafter, we assume that $\boldsymbol{\upgamma}(x_0)<1$. Next, let $u$ takes the form $u=\sum_{k=1}^{n} \lambda_k u_k \in \widetilde{\mathscr{U}}^{\textsc{mf}}_{\boldsymbol{\upgamma}}$, $n\in \N$, $\lambda_k \ge 0$ and $u_k \in \widetilde{\mathscr{U}}^{\text{Union}}_{\boldsymbol{\upgamma}}$, $k=1,...,n$. Now, by Lemma \ref{lem:ExtremeComponentMustExist}, there exists an index $1 \le j \le n$ so that 
	\begin{equation*}
		\frac{1}{\boldsymbol{\upgamma}(x_0)} = G_{\text{par}}(u;x_0) =	G_{\text{par}}(u_j;x_0).
	\end{equation*}
	Recall that $u_j \in \widetilde{\mathscr{U}}^{\text{Union}}_{\boldsymbol{\upgamma}}$, so there exists $t\in [- \infty,\infty]$ with $u_j \in \widetilde{\mathscr{U}}^{t}_{\boldsymbol{\upgamma}(t)}$. The case $t=-\infty$ obviously not possible. If $t=\infty$, then clearly $\boldsymbol{\upgamma}$ is a constant function over the interval $(x_0,\infty)$. Now, let $t \in \R$. If $t \le x_0$, then we must have $G_{\text{par}}(u_j;x_0)=1$ that contradicts our assumption. Therefore, $t> x_0$ must be. By definition, we have 
	
	\begin{equation*}
		\frac{1}{\boldsymbol{\upgamma}(x_0)} = G_{\text{par}}(u_j;x_0) \le  \frac{1}{\boldsymbol{\upgamma}(t)} \le  \frac{1}{\boldsymbol{\upgamma}(x_0)}. 
	\end{equation*}	
	Since, $\boldsymbol{\upgamma}$ is non-decreasing, therefore must be a constant function over the interval $(x_0,t)$, hence the claims follows. 
\end{proof}

\begin{proof}[B.7.3] \textbf{ Proof of Theorem   \ref{thm:BigHammer}} 	We proceed by contradiction. Assume that $u \in \widetilde{\mathscr{U}}^{\textsc{mf}}_{\boldsymbol{\upgamma}}$ takes the form,
	\begin{equation*}
		u = \sum_{k=1}^{n}\lambda_k u_k, \quad \lambda_k \ge 0, \, u_k \in \widetilde{\mathscr{U}}_{\boldgamma{\upgamma}}^{\text {Union }}, n \in \N.
	\end{equation*}	
	Now, Lemma \ref{lem:ExtremeComponentMustExist} implies that there at least exist two indexes $1 \le j_0, j_1 \le n$, $j_0 \neq j_1$ such that 
	
	$$G_{\text{par}}(u_{j_k},x_k)=\frac{1}{\boldgamma{\upgamma}(x_k)}, \quad k=0,1.$$
	
	Let's assume that  $x_0 < x_1$ (hence, by our assumption $\boldsymbol{\upgamma}(x_0) < \boldsymbol{\upgamma}(x_1)$). Let $A_0 = \{ 1\le k \le n \, : \, G_{\text{par}} (u_k;x_0)= \frac{1}{\boldsymbol{\upgamma}(x_0)}\}$. Note that $j_0 \in A_0$ and  let

	\begin{equation*}
		\mathcal{E}_0 = \{  \varepsilon >0 \, :\, 	G_{\text{par}}(u_{j_0},x_0)=\frac{1}{\boldsymbol{\upgamma}(x_0)}> \max_{    k \in A^c_0}G_{\text{par}}(u_k,x_0) + \varepsilon \}.
	\end{equation*}
	Next, for $\varepsilon \in \mathcal{E}_0$, by definition, there exist four points with $x_0 < x^\varepsilon_1<x^\varepsilon_2\le x^\varepsilon_3<x^\varepsilon_4$ such that
	\begin{align*}
		&\frac{u(x^\varepsilon_4)-u(x^\varepsilon_3)}{x^\varepsilon_4-x^\varepsilon_3}/\frac{u(x^\varepsilon_2)-u(x^\varepsilon_1)}{x^\varepsilon_2-x^\varepsilon_1} =:\frac{u[x^\varepsilon_4,x^\varepsilon_3]}{u[x^\varepsilon_2,x^\varepsilon_1]}\\>&\underbrace{G_{\text{par}}(u,x_0)}_{=1/\boldsymbol{\upgamma}(x_0)\text{(By assumption)}} - \quad \varepsilon  \, > \,  \max_{k \in A^c_0}G_{\text{par}}(u_k,x_0)\ge \frac{u_k[x^\varepsilon_4,x^\varepsilon_3]}{u_k[x^\varepsilon_2,x^\varepsilon_1]}
	\end{align*}
	for every $k\in A^c_0$.  This means that (with $\bar{x}^\varepsilon = (x^\varepsilon_1,x^\varepsilon_2,x^\varepsilon_3,x^\varepsilon_4)$),
	
	\begin{equation*}
		u[\bar{x}^\varepsilon]:=	\frac{u[x^\varepsilon_4,x^\varepsilon_3]}{u[x^\varepsilon_2,x^\varepsilon_1]}=\frac{\sum_{k=1}^{n}\lambda_k u_k[x^\varepsilon_4,x^\varepsilon_3]}{\sum_{k=1}^{n}\lambda_k u_k[x^\varepsilon_2,x^\varepsilon_1]}  >   u_k[  \bar{x}^\varepsilon ],      \quad  \forall k\in A^c_0.
	\end{equation*}
	On the other hand, by the well-known mediant inequality we have 
	$	u_k [\bar{x}^\varepsilon] > u [\bar{x}^\varepsilon]$,  $k \in A_0$. This leads to the following final inequality that for every index $k \in A_0$, 
	\begin{equation*}
		\frac{1}{\boldsymbol{\upgamma}(x_0)}\ge 	u_k [\bar{x}^\varepsilon] > u [\bar{x}^\varepsilon]> G_{\text{par}}(u;x_0) - \varepsilon, 
	\end{equation*}
	Let $\varepsilon \to 0$ to obtain that

	\begin{equation*}
		\lim_{\varepsilon \to 0} u_k [\bar{x}^\varepsilon]  = \lim_{\varepsilon \to 0}  u [\bar{x}^\varepsilon] = \frac{1}{\boldsymbol{\upgamma}(x_0)}, \quad k \in A_0.
	\end{equation*}
	Let $u_{A_0} = \sum_{k \in A_0} \lambda_k u_k$, and similarly $u_{A^c_0} = \sum_{k \in A^c_0} \lambda_k u_k$.  First, by the mediant inequality $ \lim_{\varepsilon\to 0} u_{A_0} [\bar{x}^{\varepsilon}] = \frac{1}{\boldsymbol{\upgamma}(x_0)}$. This, implies that,
	\begin{align*}\label{eq:equality_limit}
		&\frac{1}{\boldsymbol{\upgamma}(x_0)}	= \lim_{\varepsilon \to 0} u_{A_0} [\bar{x}^\varepsilon] =  \lim_{\varepsilon \to 0} u [ \bar{x}^\varepsilon]=   \lim_{\varepsilon \to 0}
		\frac{   u_{A_0}[x^\varepsilon_4,x^\varepsilon_3]   + u_{A^c_0}  [x^\varepsilon_4,x^\varepsilon_3]    }{u_{A_0}[x^\varepsilon_2,x^\varepsilon_1]   + u_{A^c_0}  [x^\varepsilon_2,x^\varepsilon_1]  }\\
		&= \lim_{\varepsilon \to 0}  
		\Bigg[        u_{A_0}[\bar{x}^{\epsilon}]  \left(     \frac{ u_{A_0}[x^\varepsilon_2,x^\varepsilon_1]}{u_{A_0}[x^\varepsilon_2,x^\varepsilon_1]   + u_{A^c_0}  [x^\varepsilon_2,x^\varepsilon_1]  }    \right)   +    u_{A^c_0}[\bar{x}^{\epsilon}]  \left(     \frac{ u_{A^c_0}[x^\varepsilon_2,x^\varepsilon_1]}{u_{A_0}[x^\varepsilon_2,x^\varepsilon_1]   + u_{A^c_0}  [x^\varepsilon_2,x^\varepsilon_1]  }    \right)  \Bigg]	
	\end{align*}
	First, note that all the four terms in the RHS above are bounded sequences, and hence after passing to a suitable subsequence we arrive into 
	
	\begin{equation}\label{eq:LookingForFuckingContradiction}
		\frac{1}{\boldsymbol{\upgamma}(x_0)}	=  \frac{1}{\boldsymbol{\upgamma}(x_0)}	 \alpha_1 + \beta \alpha_2\end{equation}
	where we know that $ 0 \le  \beta < \frac{1}{\boldsymbol{\upgamma}(x_0)}$, $\alpha_1, \alpha_2 \in [0,1]$ with $\alpha_1 + \alpha_2 =1$, and $ \beta : = \lim_{\varepsilon \to }  u_{A^c_0} [\bar{x}^{\varepsilon}]$. First, we show that $ \beta \neq 0$. By contradiction, assume that $ \beta =0$.  	First, we claim that all the $x^\varepsilon_4 < x_1$ for sufficiently small $\varepsilon$.  Recall that we have $G_{\text{par}}(u;x_0) = \max_{k} G_{\text{par}}(u_k;x_0) = G_{\text{par}}(u_{j_0};x_0)$. Assume that $u_{j_0} \in \widetilde{\mathscr{U}}^{t}_{\boldsymbol{\upgamma} (t)}$. If $t=\infty$, this immediately implies that $\boldsymbol{\upgamma}(x_0) = \thickbarr{\boldsymbol{\upgamma}} $ that contradict $\boldsymbol{\upgamma}(x_0) < \boldsymbol{\upgamma}(x_1)$. So $t<\infty$. Now if $t > x_1$, then we would have $G_{\text{par}}(u_{j_0;x_0}) \le \frac{1}{\boldsymbol{\upgamma}(t)} < \frac{1}{\boldsymbol{\upgamma}(x_1)} < \frac{1}{\boldsymbol{\upgamma}(x_0)}$ that's again provides us with a contradiction. Therefore, $t < x_1$.  If this is the case that $x^\varepsilon_4 < x_1$, we are done otherwise since $u_{j_0}$ is a flat function after $t$ one can let $x^\varepsilon_4 =t$. If $x^\varepsilon_3 >t$ for infinitely many, then this implies that $G_{\text{par}}(u_{j_0};x_0)=1$ that contradicts. Therefore, we have $x^\varepsilon_4 \le t$ for sufficiently small $\varepsilon$.  Now, assumption $ \beta=0$ yields that $u_{j_1}[x^\varepsilon_4,x^\varepsilon_3] \to 0$. So, $G_{\text{par}}(u_{j_1};x_1)=1$ that contradicts. Now, in order to have \eqref{eq:LookingForFuckingContradiction}, we must have $\alpha_2 =0$. This immediately implies that  $u_{j_1}[x^\varepsilon_2,x^\varepsilon_1] \to 0$.  Now, since $(x^\varepsilon_1,x^\varepsilon_2)$ is a bounded sequence, after passing to yet another suitable subsequence, we can assume that $(x^\varepsilon_1,x^\varepsilon_2) \to (x^0_1,x^0_2)$ as $\varepsilon \to 0$. 
	\begin{itemize}
		\item[(i)] If $x^0_2=x^0_1=x^0$, then \begin{equation*}
			\lim_{\varepsilon \to 0}u_{j_1}[x^\varepsilon_2,x^\varepsilon_1]=(u_{j_1})'_+(x^0)=0
		\end{equation*} 
		\item[(ii)] If $x^0_2>x^0_1$ then 
		\begin{equation}
			\lim_{\varepsilon \to 0}u_{j_1}[x^\varepsilon_2,x^\varepsilon_1]= u_{j_1}(x^0_2)-u_{j_1}(x^0_1)=0
		\end{equation}
	\end{itemize}
	In both cases, it implies that $u_{j_1}$ is a constant function after $x_1$. This, in turn, implies that $
	G_{\text{par}}(u_{j_1};x_1)=1$ that provides us with contradiction. This concludes the proof.
\end{proof}

\subsection{\textbf{Proof of results in Section \ref{sec:func_ASD}}}

\begin{proof}[B.8.1] \textbf{ Proof of Theorem   \ref{thm:FASD_iff}} Sufficiency follows from the  Lemma \ref{lem:FFSD:base_type} and Proposition \ref{prop:FFSD:Utility_Implies_IntegralCondition}.  Define the continuous utility function via its right-derivative as  
	\begin{equation*}
		u_{\textsc{asd}}^{\prime}(x)= 
		\begin{cases}
			1 & \text { if } G(x) \le F(x)\\ 
			\frac{1-\boldsymbol{\upepsilon}(x)}{\boldsymbol{\upepsilon}(x)} & \text { if } F(x) < G(x).
		\end{cases}
	\end{equation*}
	By the assumptions that both $F$ and $G$ have finitely many crossing points, and $\boldsymbol{\upepsilon}$ has finitely many jumps with existing right-sided limits everywhere, it follows similarly that $u_{\textsc{asd}}$ is differentiable on both sides. Observe that $\inf_{x\in \mathbb{R}}u_{\textsc{asd}}^{\prime}(x)=\inf_{x\in \mathbb{R}}\frac{1-\boldsymbol{\upepsilon}(x)}{\boldsymbol{\upepsilon}(x)}=1$. Hence, the derivative-relation condition directly holds and $u_{\textsc{asd}}\in \widetilde{\mathscr{U}}^{\textsc{asd}}_{\boldsymbol{\upepsilon}}$ follows. Noting that $F$ and $G$ are in $\operatorname{L}^1(\boldsymbol{\upepsilon})$, we have
	\begin{align*}
		&\E_G[u_{\textsc{asd}}]-\E_F[u_{\textsc{asd}}]\\
		=&\int_{-\infty}^\infty  (F(x)-G(x))_+dx- \int_{-\infty}^\infty \left[\frac{1-\boldsymbol{\upepsilon}(x)}{\boldsymbol{\upepsilon}(x)}\right] (F(x)-G(x))_-dx\ge 0.
	\end{align*}
	For the other direction, consider an  arbitrary $u\in \widetilde{\mathscr{U}}^{\textsc{asd}}_{\boldsymbol{\upepsilon}}$. If $u$ is a constant function then the claim directly follows. Let $u$ be strictly increasing so that $\inf_{x\in \R}u'_+(x)>0$.  Then, it is sufficient to show that
	\begin{equation*}
		\label{eq:ASD_Proof_we_want_to_show}
		\E_G[u]-\E_F[u]=\int_{\R} (F(x)-G(x))_+du(x)-\int_{\R}
		(F(x)-G(x))_-du(x)\ge 0.
	\end{equation*}
	It follows that
	\begin{align*}
		&\E_G[u]-\E_F[u]\\=&\int_{\R} (F(x)-G(x))_+u'(x)dx-\int_{\R} (F(x)-G(x))_-u'(x)dx\\
		=&\int_{\R} (F(x)-G(x))_+u_+'(x)dx-\int_{\R} (F(x)-G(x))_-u_+'(x)dx\\
		\ge&\int_{\R} \inf_{x\in \R}u'_+(x)(F(x)-G(x))_+ dx- \int_{\R} \inf_{x\in \mathbb{R}}u'_+(x)\left[\frac{1-\boldsymbol{\upepsilon}(x)}{\boldsymbol{\upepsilon}(x)}\right](F(x)-G(x))_- dx\\
		\ge &\underbrace{\inf_{x\in \R}u'_+(x)}_{>0} \underbrace{\left[\int_{\R} (F(x)-G(x))_+ dx- \int_{\R}\left[\frac{1-\boldsymbol{\upepsilon}(x)}{\boldsymbol{\upepsilon}(x)}\right](F(x)-G(x))_- dx \right]}_{ \ge 0 \text{ (by the assumption)}}\ge 0,
	\end{align*}
	where first equality follows from the fact that $u$ is differentiable on both sides, meaning $u_+$ differs from $u$ at only a countable set of points and the last inequality is the direct application the set condition of $\widetilde{\mathscr{U}}^{\textsc{asd}}_{\boldsymbol{\upepsilon}}$ given in \eqref{eq:almost_(i)}.  The same conclusion can be reached with a left-continuous function $\boldsymbol{\upepsilon}$ and by employing left-derivatives in a similar manner.
\end{proof}

	\end{document}